\PassOptionsToPackage{svgnames}{xcolor}
\documentclass[%
nofootinbib,
nobibnotes,
12pt
]{iopart}

\usepackage{tikz} 
\usepackage{tikz-feynman}
\usepackage{hhline}

\usepackage{tikz}
\usetikzlibrary{plotmarks}

\usepackage{xcolor}

\usepackage{graphicx,caption,subcaption}
\usetikzlibrary{decorations.pathmorphing}

\usepackage{dcolumn}
\usepackage{bm}

\usepackage{slashbox}

\usepackage{hyperref}

\usepackage{mwe}

\usepackage{caption}
\usepackage{subcaption}

\renewcommand{\thesection}{}

\makeatletter
\def\@seccntformat#1{\csname #1ignore\expandafter\endcsname\csname the#1\endcsname\quad}
\let\sectionignore\@gobbletwo
\let\latex@numberline\numberline
\def\numberline#1{\if\relax#1\relax\else\latex@numberline{#1}\fi}
\makeatother

\usepackage{blindtext}
\usepackage [english]{babel}
\usepackage [autostyle, english = american]{csquotes}
\MakeOuterQuote{"}
\usepackage[perpage]{footmisc}

\usepackage{titlesec}
\titleformat{\section}
  {\normalfont\sffamily\large\bfseries\color{black}}
  {\thesection}{1em}{}

\hypersetup{colorlinks=true,urlcolor=blue}

\usepackage{etoolbox}
\makeatletter
\def\@mkboth#1#2{}
\newlength\appendixwidth
\preto\appendix{\addtocontents{toc}{\protect\patchl@section}}
\newcommand{\patchl@section}{%
  \settowidth{\appendixwidth}{\textbf{Appendix }}%
  \addtolength{\appendixwidth}{1.5em}%
  \patchcmd{\l@section}{1.5em}{\appendixwidth}{}{\ddt}%
}
\makeatother

\usepackage{iopams}

\expandafter\let\csname equation*\endcsname\relax

\expandafter\let\csname endequation*\endcsname\relax

\usepackage{amsmath}

\usepackage{dcolumn}
\usepackage{epsf}
\usepackage{float}

\usepackage{tabularx}

\usepackage[final]{changes}

\usepackage{lipsum}
\definechangesauthor[color=blue]{.}
\colorlet{Changes@Color}{red}
\setremarkmarkup{(#2)}

\newtheorem{theorem}{Theorem}
\newtheorem{corollary}{Corollary}[theorem] 
\usepackage{lipsum}

\usepackage{mathtools}
\DeclarePairedDelimiter{\ceil}{\lceil}{\rceil}
\DeclarePairedDelimiter\floor{\lfloor}{\rfloor}

\begin{document}

\title[]{Necessary and Sufficient Conditions for the Validity of Luttinger's Theorem}

\author{Joshuah T. Heath \& Kevin S. Bedell}

\address{Physics Department,  Boston  College,  Chestnut  Hill, Massachusetts  02467,  USA}
\ead{heathjo@bc.edu}
\vspace{10pt}
\begin{indented}
\item[]\today
\end{indented}

\begin{abstract} \noindent Luttinger's theorem is a major result in many-body physics that states the volume of the Fermi surface is directly proportional to the particle density. In its "hard" form, Luttinger's theorem implies that the Fermi volume is invariant with respect to interactions (as opposed to a "soft" Luttinger's theorem, where this invariance is lost). Despite it's simplicity, the conditions on the fermionic self energy under which Luttinger's theorem is valid remains a matter of debate, with possible requirements for its validity ranging from particle-hole symmetry to  analyticity about the Fermi surface. In this paper, we propose the minimal requirements for the application of a hard Luttinger's Theorem to a generic fermionic system of arbitrary interaction strength {by invoking the Atiyah-Singer index theorem to quantify the topologically-robust behavior of a generalized Fermi surface}. We show that the applicability of a hard Luttinger's theorem in a $D$-dimensional system is directly dependent on the existence of a $(D-1)$-dimensional manifold of gapless chiral excitations at the Fermi level, regardless of whether the system exhibits Luttinger or Fermi surfaces (i.e., manifolds of zeroes of the Green's function and inverse Green's function, respectively).The exact form of the self-energy which guarantees validity of a hard Luttinger's theorem is derived, and agreement with current experiments, numerics, and theories are discussed. 
\end{abstract}

\vspace{2pc}

\submitto{\NJP}

\maketitle

\tableofcontents

%

\section{I. Introduction}
Of fundamental importance to physics in both the IR and UV limits is the question of whether or not macroscopic phenomena can be described by the collective behavior of indivisible, well-defined particles that obey fundamental conservation laws. In the high-energy community, such an "independent-particle" approximation (IPA)\cite{Schrieffer} has lead to the successful prediction of new particles \cite{Pauli} and ultimately the creation of the present-day Standard Model \cite{Glashow, Guralnik, Weinberg}. The low-energy effective field theory of fermionic excitations
 also relies heavily upon an IPA, as the presence of a 
 Fermi surface usually permits us to construct an isomorphism between the eigenstates of the non-interacting Fermi gas and the interacting Fermi system via either perturbative\cite{Landau1} or renormalization group\cite{Benfatto,Polchinski,Shankar1, Chitov,Shankar2} arguments.
%
%
When a particle loses its mass, the IPA breaks down, resulting in the well-known scale invariant properties of photons and gauge bosons.
%
%
%
On the contrary, the presence of free {\it massive} particles described by scale invariant physics is not predicted by the Standard Model. 
Such systems are described by an "un-particle"\cite{Georgi,Georgi2} approximation (UPA), with a continuous spectrum of mass replacing the discrete observables in the IPA \cite{Nikolic}.
This unparticle "stuff" has recently been embraced by condensed matter theorists as a possible description of the normal phase of the cuprates \cite{Varma,LeBlanc,Phillips5}, leading to the possibility of an "un-Fermi liquid" state \added{in these materials} \cite{Phillips1}.
%
%
%

In the high-energy limit, unparticles may be found experimentally by detecting a loss of energy or momentum not accounted for by conservation laws \cite{Georgi,Kingman}. Analogously, unparticles in the low-energy limit should correspond to "missing" degrees of freedom (DoF) once we turn on interactions. 
This latter scenario can be studied in a certain material by checking the applicability of Luttinger's theorem \cite{Luttinger2,Luttinger3,Luttinger1,Farid}, which 
states that the direct relation between the $D$-dimensional volume contained within the Fermi surface and the total density of particles
\begin{align}
\frac{1}{(2\pi)^D}\int_{G({\bf k},\,\omega=0)>0}d^D {\bf k}&=\frac{N}{2V}\label{eqn1}
\end{align}
is invariant with respect to the particles' interaction\footnote{
 \added{Throughout this article, $G({\bf k},\,\omega)$ is interpreted as the single-particle Green's function for single-band systems and the eigenvalues of the propagator for more complex crystalline states. In the case of the latter, the left-hand side of Eqn. \eqref{eqn1} is summed over all eigenvalues \cite{Dzyal,Phillips1}}}. {
The connection between the failure of Luttinger's theorem and an ill-defined independent-particle picture is apparent when one considers the formation of Fermi arcs in the cuprate materials La$_{2-x}$Sr$_x$CuO$_2$ \cite{Eisaki} and Bi$_2$Sr$_2$CaCu$_2$O$_{8+\delta}$ \cite{Kidd}, where ARPES measurements show a breakdown of Eqn. \eqref{eqn1} as a function of hole-doping. As some fraction of the non-interacting particle density is "lost" when interactions are turned on, one must conclude that the remaining electronic excitations must be coexisting with some "stuff" which lacks a description in terms of well-defined individual excitations \cite{Phillips_book}. Indeed, the quantum critical scaling inherent to such systems allows us to describe the transport in terms of power-law Green's functions not unlike the propagators describing unparticle stuff \cite{Phillips5,Karch}, with recent work on such "power-law liquids" explicitly showing that Luttinger's theorem breaks down for unparticle-like scaling of the Green's function \cite{Phillips4,Heath}.}

Because Luttinger's theorem is a non-perturbative theory, it is a statement that describes collective behavior beyond the vicinity of some cutoff near the Fermi surface, making it a more robust criterion of the IPA than Landau-Fermi liquid theory \cite{Krastan,Affleck,Oshikawa}. Unfortunately, the scope of when and where Luttinger's theorem is valid is somewhat unclear in the present literature \added{and has been hotly debated \cite{Farid3,Rosch2,Farid4}}, with some even claiming the very definition of the theorem is "clouded in folklore"\cite{Dzyal}. This has led to a generalization of Luttinger's theorem into "hard" and "soft" variations, with the former being defined as in Eqn. \eqref{eqn1} and the latter corresponding to those systems where the left-hand-side of Eqn. \eqref{eqn1} is equal to some fraction of the total  {\color{black} non-interacting} density, known as the "Luttinger count"\cite{Phillips2,Phillips3,Phillips1,Phillips4,Essler2}. Because independent-particle behavior is only seen in systems that satisfy a hard Luttinger's theorem with trivial Luttinger count, it has been {\color{black} widely} accepted that the IPA breaks down whenever we lack a conventional Fermi surface or particle-hole symmetry\cite{Rosch}.  	
\added{This includes materials with a "Luttinger surface" \cite{Dzyal}, which corresponds to zeroes of the interacting Green's function $G({\bf k},\,\omega)$} and {\color{black} are} proposed to violate the fundamental assumptions of a hard Luttinger's theorem\footnote{
From hereon, we refer to the "hard" version of Luttinger's theorem 
as simply Luttinger's theorem.} 
\cite{Phillips2,Phillips3,Phillips1,Phillips4}.

 In this paper, we introduce the necessary and sufficient conditions in which we can safely consider the Luttinger count in {\it any} interacting fermionic system to be synonymous with the bare particle density. In other words, we outline {\color{black}when and} where an independent particle description is valid in a many-body system of arbitrary interaction strength. {By doing so, we show explicitly that Luttinger's theorem remains valid for non-Fermi liquids beyond the Tomonaga-Luttinger liquid as long as the system remains gapless.}
{Such an analysis allows us to {\color{black}write down} the exact form of the self energy that simultaneously satisfies Luttinger's theorem while also entailing the existence of a Luttinger surface. }

\section{II. Generalization of the Fermi surface}
 Of central importance to 
Luttinger's theorem
is the preservation of a Fermi surface
\cite{Luttinger1}. 
By Fermi surface, we mean here \added{(at the bare minimum)} some boundary in phase space (i) that exactly 
overlaps with the Fermi surface of the non-interacting Fermi gas at $\{{\bf k}_F\}$ in the isotropic case, (ii) where 
$G({\bf k},\,\omega)$ changes sign, and (iii) which remains experimentally detectable for some finite interaction. 


In a simple $D$-dimensional Landau-Fermi liquid,
the presence of a discontinuity in the bare particle momentum distribution $n({\bf k})$ can be interpreted as a finite quasiparticle weight\cite{Migdal}:
\begin{align}
Z_k&=n({\bf k}_F-\delta)-n({\bf k}_F+\delta)\notag\\
&=\left(1-\frac{\partial \Re \Sigma({\bf k},\,\omega)}{\partial \omega}\right)^{-1}\bigg|_{\genfrac{}{}{0pt}{}{{\bf k}={\bf k}_F}{\hspace{-2mm}\omega=0}} \label{eq3}
\end{align}
\added{where $\Sigma({\bf k},\,\omega)$ is the retarded self energy. By definition, the presence of $0<Z_k\le 1$ results in a traditional Fermi surface, and the well-known proof of Luttinger's theorem in a Fermi liquid follows} (See Appendix A for derivation). However, 
a value of $Z_k\ge 0$ is not a strong indication for the applicability of Luttinger's theorem \cite{Sachdev}, nor is a vanishing $Z_k$ an indication of its failure \cite{Haldane}. 
A well-known example of the latter is the Tomonaga-Luttinger liquid \cite{Tomonaga,Luttinger_liquid,Mattis,Haldane_Luttinger_liquid}, where 
perturbative methods \cite{Krastan} and the Lieb-Schultz-Mattis theorem \cite{Affleck} suggest that Luttinger's theorem is preserved in 1D metals despite the clear lack of a quasiparticle weight $Z_k$. {\color{black} Unlike the case of the underdoped cuprates considered previously, an independent particle picture remains in the Tomonaga-Luttinger liquid as the number of charge degrees of freedom (the "chargons") in the interacting system are always equal to the number of electrons in the 1D Fermi gas \cite{Krastan}.}

\added{From the g-ology construction, the distribution function for the Tomonaga-Luttinger liquid near the Fermi points becomes} \cite{Mattis, Matt}:
\begin{align}
{\color{black} n_{\pm}({\bf k})=
\frac{1}{2}-C_1 |{\bf k} \mp {\bf k}_F|^\alpha\, \textrm{sgn} (\pm {\bf k}-{\bf k}_F)-C_2 ({\bf k}\mp {\bf k}_F)}
\end{align}
where $C_1$, $C_2$, and $\alpha$ are positive constants \cite{Matt} {\color{black} and the sign $\pm$ denote right and left moving excitations, respectively}. \added{The Fermi surface at $\{{\bf k}_F\}$ is then replaced by the set of Fermi points where the $m$th derivative of the bare distribution function becomes singular \cite{Krastan}:}
\begin{align}
\{{\bf k}_F\}=\left\{
\forall\, {\bf k},\,\exists\, m \in \mathbb{N}_1:\,\frac{d^m n({\bf k})}{dk^m}\rightarrow \infty
\right\}\label{eq4}
\end{align}
 Because the momentum distribution of the Landau-Fermi liquid also exhibits a singularity in the $m=1$ derivative at $\{{\bf k}\}=\{{\bf k}_F\}$, it \added{is tempting to say} that \added{the legitimacy of} Eqn. \eqref{eq4} \added{for some $m$} might be a nearly "universal" feature of systems that obey Luttinger's theorem. \added{If this turns out to be the case, the necessary requirements on the Green's function and hence the self-energy for the case of a trivial Luttinger count could be deciphered.}

\added{The primary goal of this paper is to expand upon the work of Blagoev and Bedell, and ultimately to show} that a variant of Eqn. \eqref{eq4} is indeed a universal feature of all systems that obey Luttinger's theorem. {
In other words, {\bf we want to explicitly show that there exists some generalization of the Fermi surface in a generic, fermionic many-body system that guarantees Luttinger's theorem to be preserved}. Much as we can extract the behavior of the self energy in a Landau-Fermi liquid by imposing $0<Z_k\le 1$, proving that an equation such as Eqn. \eqref{eq4} is required for a system to obey Luttinger's theorem will then allow us to readily extract the behavior of the self energy in {\it any} system that obeys Luttinger's theorem. Ultimately, the calculation of a self energy that guarantees a trivial Luttinger count (even in the presence of a Luttinger surface) is the peripheral objective of this paper.

We can summarize the first goal of this paper with the following proposal:
\begin{theorem}
In a $D$-dimensional fermionic system, the topological index of the 
{\color{black} generating functional for all two-point Green's functions}
takes on integer value for all conventional Fermi surfaces. 
%
%
\end{theorem}
}

 We begin {our generalization of the Fermi surface} by recalling the Kadanoff-Baym functional for some general interacting fermionic system \cite{Cornwall,Luttinger3,Baym1,Baym2,Tremblay2,Polonyi}: 
\begin{align}
\Gamma[G]\approx 
\Phi[G]-\textrm{Tr}[(G_0^{-1}-G^{-1})G]+\textrm{Tr}[\log (-G)]\label{eq5}
\end{align}

\noindent where 
 $\Phi$ is the Luttinger-Ward functional, defined as the sum of all skeleton diagrams: \cite{Luttinger3,Tremblay2}:

\begin{align}
\Phi[G]&=
\begin{tikzpicture}[baseline=5.5ex,scale=0.7,scale=0.6, every node/.style={transform shape,scale=0.6}]
\begin{feynman}
\vertex(a);
\vertex[right=of a](b);
\vertex[right=of b](c);
\vertex[below=1.1 mm of c](c1);
\vertex[right=of c](d);
\vertex[right=of d](e);
\vertex[above=of c](f);
\vertex[above right=of f](g);
\vertex[above left=of f](h);
\vertex[above left=of h](h1);
\vertex[below left=of h](h2);
\vertex[below right=of g](g1);
\vertex[above right=of g](g2);
\diagram*{
(g)--[photon](h),
(g)--[quarter right](g1),
(g1)--[fermion, half right](g2),
(g2)--[quarter right](g),
(h)--[quarter right](h1),
(h1)--[fermion, half right](h2),
(h2)--[quarter right](h),
};
\end{feynman}
\end{tikzpicture}\,+\,
\begin{tikzpicture}[baseline=5.5ex,scale=0.7,scale=0.6, every node/.style={transform shape,scale=0.6}]
\begin{feynman}
\vertex(a);
\vertex[right=of a](b);
\vertex[right=of b](c);
\vertex[below=1.1 mm of c](c1);
\vertex[right=of c](d);
\vertex[right=of d](e);
\vertex[above=of c](f);
\vertex[above right=of f](g);
\vertex[above left=of f](h);
\diagram*{
(g)--[photon](h),
(g)--[half right,fermion](h),
(h)--[half right,fermion](g),
};
\end{feynman}
\end{tikzpicture}
\,+\,.\,.\,.
\end{align}

\noindent {\color{black} The Kadanoff-Baym functional is fully derived in Appendix B, and can be considered {\color{black} the full two-point irreducible (2PI)} effective quantum} action for the fermionic many-body state. {\color{black} For reasons that will soon be apparent, we want to connect the above expression to the partition function; i.e., the generating functional for the two-point Green's functions $G$. Defining $J$ and $K$ as the one-particle and two-particle sources, the partition function $Z[G]$ can be written as \cite{Hagen,Rammer} }
\begin{align}
{\color{black} Z[G]=e^{i W[J,\,K]}}
\end{align}
{\color{black} where $W[J,\,K]$ is the quantum action. Performing a double-Legendre transformation, we can connect the quantum action $W[J,\,K]$ and the 2PI effective action $\Gamma[G]$ via the following expression \cite{Hagen,Rammer}:}
\begin{align}
{\color{black} \Gamma[G]=W[J,\,K]-\psi^\dagger J-\frac{1}{2}\psi^\dagger K\psi-\frac{i}{2}\textrm{Tr}(KG)}
\end{align}

{\color{black} \noindent This allows us to write the generating functional in the form \cite{Rammer, Gasenzer,Gasenzer2}} 
%
\begin{align}
{\color{black} Z[G]
=\widetilde{Z}
\exp\left\{
i\left(
\Phi[G]-\textrm{Tr}[\Sigma G]+\textrm{Tr}[\log (-G)]\label{eq7}
\right)
\right\}}
\end{align}
\noindent {\color{black} where $\widetilde{Z}$ is dependent on any interaction-dependent constants and the sources $J$ and $K$. As the physical result corresponds to $J,\,K\rightarrow 0$ \cite{Rammer}, their dependence is of little concern to this work. Note that in a classical Bose gas, $\widetilde{Z}$ would also include a classical contribution from a non-zero vacuum expectation value. However, from Pauli exclusion we know that $\langle \psi\rangle=0$, and therefore we exclude a "classical" component to the effective 2PI action\footnote{We thank Thomas Gasenzer for clarifying this point.}. }

{\color{black} We now want to see how a Fermi surface manifests itself in the generating functional $Z$}. By definition, a Fermi surface exists when $G^{-1}({\bf k},\,\omega)=0$. 
\added{The second term in the above can be simplified via}

\begin{align}
\Sigma G&=(G_0^{-1}-G^{-1})G\notag\\
&=\frac{\omega-(\epsilon_k-\mu)}{\omega-(\epsilon_k-\mu)-\Sigma}-1\notag\\
&=\frac{\Sigma}{\omega-(\epsilon_k-\mu)-\Sigma}\notag\\
&=\frac{1}{(G_0\Sigma)^{-1}-1}\label{eq10}
\end{align}

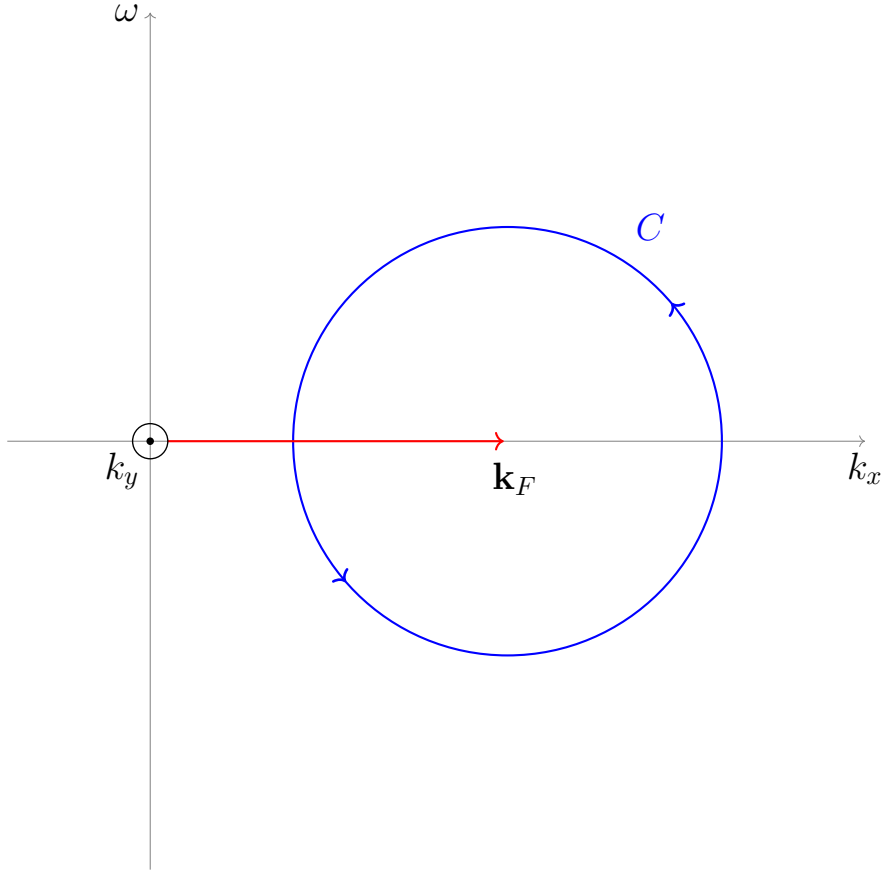
\begin{figure}
\begin{center}
\begin{tikzpicture}[scale=1.9,decoration={markings,
mark=at position 2cm with {\arrow[line width=1pt]{>}},
mark=at position 11cm with {\arrow[line width=1pt]{>}}
}
]
\draw[help lines,->] (-1,0) -- (5,0) coordinate (xaxis);
\draw[help lines,->] (0,-3) -- (0,3) coordinate (yaxis);

\path[draw,blue, line width=0.8pt,postaction=decorate] (4,0) node[below] {} -- (4,0) node[below] {} arc (0:360:1.5); 

\node (A) at (.045, 0) {};
\node[label=below:{\large \color{black} ${\bf k}_F$}] (B) at (2.55, 0) {};
\draw[->,thick,red]
  (A) edge (B) ;


\node[below] at (xaxis) {\large $k_x$};
\node[left] at (yaxis) {\large $\omega$};
\node[below left] {\large $k_y$};
\node at (3.5,1.5) {\large {\color{blue} $C$}};
\node at (0,0) {\LARGE $\odot$};
\end{tikzpicture}
\end{center}
\caption{{\color{black} Visualization of the contour $C$ taken by the path $\ell$ in the definition of the winding number Eqn. \eqref{eq9} for a simple 2D Fermi liquid, as introduced by Volovik \cite{Volovik2}. In this orientation, the $\omega$--$k_x$ axis is in the plane of the page, with the $k_y$ axis coming out of the page. By definition, the contour is taken about the Fermi momentum ${\bf k}_F$. For a 3D Fermi liquid, the contour becomes a 2D manifold in a 4D Fourier space.}}
\label{Fig1}
\end{figure}

\noindent \added{where, in all lines of the above, a trace over indices is implied {\color{black} and, in the second to-last line, we assume $\Sigma\not=0$.} {\color{black}As we are concerned about the value of the partition function in the vicinity of the Fermi momentum, $\Sigma\sim G^{-1}$, implying that $\Sigma G\sim (G_0^{-1} G-1)^{-1}$. For a conventional Fermi liquid, the interacting Green's function is proportional to the quasiparticle weight in close proximity to the Fermi momentum, and hence $\Sigma G\sim -1$. For the case of a non-Fermi liquid, $\Sigma$ is divergent, yielding the same result. Therefore, Eqn. \eqref{eq10} remains well-defined regardless of the fermionic system we consider.  }
%
Similar behavior is seen in the Luttinger-Ward functional $\Phi[G]$, which we will assume to be well-behaved and finite. {\color{black}Note that the analytic behavior of the Luttinger-Ward functional near the Fermi surface is intimately tied to the analytic behavior of the self energy, a concept explored later in this article as well as in the work of Phillips et. al. \cite{Phillips2,Phillips3,Phillips5,Phillips4}}. 

This leaves us to consider the behavior of $\log(-G)$. Ignoring the negative, this term becomes divergent at the Fermi surface, as $G^{-1}({\bf k},\,\omega)=0$ at such a boundary by definition. If we assume the other contributions are well behaved { (i.e., if we assume that the Luttinger-Ward functional doesn't diverge near the Fermi surface)}, we can then simplify the above if we restrict the functional to ${\bf k}$-points in the direct vicinity of the Fermi surface:}

\begin{align}
{\color{black} Z[G]\approx \widetilde{Z} e^{\textrm{Tr}[\log(G)]}}
\end{align}
\added{This phase on the generating functional can by quantified by a winding number:}
\begin{align}
\mathcal{N}&=\frac{1}{2\pi i}\oint_{C} d\ell\, e^{-\textrm{Tr}[\log(G(\ell))]}\frac{d}{d\ell}e^{
\textrm{Tr}[\log(G(\ell))]
}\notag\\
&=\frac{1}{2\pi i}\oint_C d\ell e^{-\textrm{Tr}[\log(G(\ell))]}e^{\textrm{Tr}[\log (G(\ell))]}\frac{d}{d\ell} \textrm{Tr}[\log(G(\ell))\notag\\
&=\frac{1}{2\pi i} \textrm{Tr} \oint_C d\ell \frac{d}{d\ell} \log(G(\ell))\label{eq9}
\end{align}
{\color{black}where the path $\ell$ in the full frequency/momentum space is taken over a contour $C$ which encloses the manifold $\{\omega,\,{\bf k}\}=\{0,\,{\bf k}_F\}$ (see Fig. \ref{Fig1}).}
{\color{black} As an example, for a 2D Fermi liquid, the contour $C$ is a one-dimensional line that winds about the 1D Fermi surface in the three-dimensional space $\{\omega,\,k_x,\,k_y\}$. For a 3D Fermi liquid, the contour $C$ is then a two-dimensional manifold that winds about the 2D Fermi surface in the 4D space $\{\omega,\,k_x,\,k_y,\,k_z\}$.}
%
%
{\color{black} It should then be clear that the phase in Eqn. \eqref{eq9}} defines a covering map $f(\ell)
=e^{iw(\ell)}$, where $f:S^1\rightarrow U(1)$ in the simplified $D=2$ fermionic system
 characterized by the homotopy class $\mathcal{N}$ given above.
\noindent When this winding number $\mathcal{N}\not =0$, 
 then the system supports a Fermi surface, {\color{black}as the contour winding number (by definition) is non-zero when the Green's function has singularities}. \added{More specifically, $\mathcal{N}=1$ when a single-band system supports solutions $\{\omega,\,{\bf k}\}=\{0,\,{\bf k}_F\}$ where $G^{-1}({\bf k},\,\omega)\rightarrow 0$, while $\mathcal{N}\in \mathbb{N}_1$ when a multi-band system obeys the same conditions \cite{Mudry}}. {\color{black}If the fermionic system lacks a Fermi surface (or, as we will see, a Luttinger surface), then the Green's function lacks a singularity at the Fermi momentum, and the winding number vanishes away as the propagator remains analytic throughout the entirety of Fourier space.}
 
It is important to note that a non-zero value of the topological index Eqn. \eqref{eq9}
 is equivalent to the topological invariant introduced by Volovik \cite{Volovik2,Volovik1} to provide a robust definition of the Fermi surface for Landau-Fermi liquids, Tomonaga-Luttinger liquids, and marginal Fermi liquids \cite{Varma, Ruckenstein}. Because such a definition was inspired by the analogous topological singularities in superfluid $^{3}$He-A (known as "boojums"\cite{Mermin1,Mermin2}), we will refer to \added{the $(D-1)$-dimensional manifolds characterized by non-zero winding number $\mathcal{N}$} as "snarks" for conciseness\footnote{From the last stanza of Lewis Carroll's {\it The Hunting of the Snark}: "He had softly and suddenly vanished away--/
   For the Snark was a Boojum, you see."}.
   
\added{In Volovik's original argument, the existence of $\mathcal{N}\not=0$ is a direct result of the singularity in the interacting Green's function at the Fermi level. However, simple manipulation of Volovik's term given above yields a non-zero winding number for Luttinger surface solutions, where the Green's function itself has zeroes:}

\begin{align}
\mathcal{N}&=\frac{1}{2\pi i}\textrm{Tr}\oint_C d\ell G^{-1}(\ell)\frac{d}{d\ell} G(\ell)\notag\\
&=
\frac{1}{2\pi i} \textrm{Tr} \left\{ G^{-1}(\ell)G(\ell)\right\} \bigg|_{C}-\frac{1}{2 \pi i } \textrm{Tr} \oint_C G(\ell) \frac{d}{d\ell} G^{-1}(\ell)
\end{align}
\added{As long as we assume the Green's function is holomorphic in the vicinity of the Fermi surface, the first integral disappears, and we are left with }
\begin{align}
\mathcal{N}&=\frac{1}{2 \pi i } \textrm{Tr} \oint_{C'} G(\ell) \frac{d}{d\ell} G^{-1}(\ell)\label{eq12}
\end{align}
where we have changed the handedness of our contour from $C$ to $C'$. {From hereon, we assume the handedness of the contour which defines the topological indices Eqns. \eqref{eq9} and \eqref{eq12} is taken such that $\mathcal{N}\ge 0$. In the presence of a multi-band system (where a sum over the eigenvalues of the fermionic propagator is implied in the formula for the winding number), each contour in the sum is similarly taken such that each value in the sum is positive. This leads to the following corollary to the theorem on the previous page: 
 
\begin{corollary}
The topological index of the $D$-dimensional{\color{black} generating functional for all two-point Green's functions} cannot distinguish between the presence of a $(D-1)$-dimensional Fermi surface and a $(D-1)$-dimensional Luttinger surface.
%
\end{corollary}}

The above follows from basic calculus, and predicts a non-zero solution for the winding number $\mathcal{N}$ for all Luttinger surfaces with a well-behaved Luttinger-Ward functional. However, the existence of such solutions is not predicted by Volovik's original argument, which is directly dependent on vortex singularities of the Green's function at the Fermi level. Although such singular behavior might be found in marginal Fermi liquids and Tomonaga-Luttinger liquids, there are many cases beyond these (which we will discuss later in this article) that appear to 
lack a vortex singularity while simultaneously obeying Luttinger's theorem. Only by interpreting the winding number as some phase of the generating functional do Luttinger surface solutions beyond the marginal Fermi liquid and Tomonaga-Luttinger liquid become apparent, as the topologically non-trivial behavior described above is now connected to singularities in $\log(G({\bf k},\,\omega))$ as opposed to singularities in the Green's function itself.

   \section{III. The Fermi surface as an Anomaly}
   
  The versatility of the snark \added{is that it gives us a physical quantity that both Fermi and Luttinger surfaces have in common: namely, the existence of a non-zero topologically-invariant quantity $\mathcal{N}$. The fact that this winding number can be directly interpreted as a topological phase of the quantum field theoretic partition function leads us to conclude that $\mathcal{N}\not=0$ solutions} are the hallmark of an anomaly in the many-body theory. 
  
\added{
Anomalies are defined as a symmetry of the classical Lagrangian which is "lost" in the process of quantization \cite{Bardeen,Bertlmann}. An example of a well-known anomaly can easily be seen by considering the effective action $W[A]$ of a massless Dirac field in the presence of an Abelian gauge field $A$. An infinitesimal chiral transformation on the Dirac field results in a chiral gauge current. The change in the measure of the path integral under such a transformation results in the non-conservation of this current, and hence the system is said to exhibit a chiral or Adler-Bell-Jackiw (ABJ) anomaly \cite{Adler,Bell}. In the presence of a non-Abelian gauge, the real part of $W[A]$ will remain gauge invariant, and thus the spontaneously broken gauge symmetry manifests as a phase contributing to the Dirac determinant $\det(i\hat{D}(A))$ (i.e., the generating functional of the Dirac field). A topological interpretation of the non-Abelian anomaly can be seen by following the result of Alvarez-Gaum\'e and Ginsparg \cite{Ginsparg}. By viewing the gauge transformation $A\rightarrow A^\theta$ as a circle in the gauge connection space surrounding disk ${\it \bf D}$ of a two parameter family of gauge fields, the fermion determinant can be considered a complex function of gauge fields confined to ${\it \bf D}$, and can thus be written as}

 \begin{align}
e^{-W[A^\theta]}= \det (i\hat{D}(A^\theta))=\sqrt{\det (i\gamma^\mu \nabla_\mu (A))}e^{iw(A,\,\theta)}
 \end{align}
 \added{This allows us to consider the phase of the generating functional as a map $S^1\rightarrow U(1)$ (i.e., $\theta \rightarrow e^{iw(A,\,\theta)})$. The presence of an anomaly is therefore analogous to a non-zero winding number of the form }
 \begin{align}
 \mathcal{N}=\frac{1}{2\pi} \int_0^{2\pi }\frac{\partial w(A,\,\theta)}{\partial \theta}d\theta\label{eq18}
 \end{align}

\added{By following a perturbative formulation of the many-body generating functional in terms of the Kadanoff-Baym effective action, we have found a similar anomalous component for $Z[G]$ emerging in the presence of zeroes in the Green's function or inverse Green's function; i.e., $\log(G(\ell))$ plays the role of $w(A,\,\theta)$ in the fermionic many-body system. This leads us to postulate that {\bf the presence of Fermi/Luttinger surfaces in fermionic matter is equivalent to the appearance of an anomaly in the quantized {\color{black} many-body} field theory.} Physically, what this tells us is that the effects of Pauli correlation brought about by anti-symmetrizing the many-body field results in the "loss" of a symmetry once found in the equivalent classical system. This shouldn't be a surprising result; the well-known chiral anomaly is often interpreted as an apparent chiral symmetry breaking in the presence of a Dirac sea; i.e., an "upward" shift of energy levels for particles and a "downward" shift for anti-particles that remains uncompensated at the bottom of the sea in the continuum limit\cite{Ambjorn,Shifman,Casher}. Hence, any instance of many-body fermionic systems that form a Fermi surface trivially experience an anomaly by virtue of Pauli correlation. What is surprising is that, as long as we have a well-defined Luttinger-Ward functional, the explicit form of the anomaly given by Eqn. \eqref{eq9} is seen in fermionic systems with a Luttinger surface as well as those with a Fermi surface. All systems that therefore support a "snark" by definition break some classical symmetry solely by virtue of quantizing the many-body fermionic wavefunction.}  

\added{By interpreting the snark as a many-body anomaly, we can now invoke the Atiyah-Singer index theorem\cite{Atiyah,Ginsparg, Nakahara, Bertlmann,Romer} to better understand the physical implications of Eqn. \eqref{eq9}. In a nutshell, the Atiyah-Singer index theorem states that the topological index is equivalent to the analytical index, the former being defined by a winding number (as given in Eqn. \eqref{eq18}) and the latter being defined as the difference between the dimensions of the kernel and cokernel of some elliptic operator. For the case of the Dirac operator $i\hat{D}(A^\theta)$, the index is given by}
\begin{align}
\textrm{ind}(i\hat{D}(A^\theta))=\nu_+-\nu_-
\end{align}
\added{where $\nu_\pm$ are the number of positive/negative chiral zero modes of $i\hat{D}$. Because the topological and analytical indices of the Dirac operator are equivalent, the difference in the number of chiral modes is given simply by the winding number Eqn. \eqref{eq18}. Consequently, a non-zero winding number about some manifold is a tell-tale sign of an "imbalance" of chiral modes on said manifold. }

{
The above analysis leads us to the following corollary:
\begin{corollary}
The analytical index of the $D$-dimensional {\color{black} generating functional for all two-point Green's functions} cannot distinguish between the presence of a $(D-1)$-dimensional Fermi surface and a $(D-1)$-dimensional Luttinger surface.
\end{corollary}
}

\noindent In other words, the Atiyah-Singer index theorem tells us that {\bf both Luttinger and Fermi surfaces can be mutually defined as lower-dimensional manifolds of gapless chiral excitations.} A non-zero value of $\mathcal{N}$ in a $D$-dimensional fermionic system is synonymous with the existence of a $(D-1)$-dimensional manifold of gapless chiral modes at $\{\omega,\,{\bf k}\}=\{0,\,{\bf k}_F\}$. Chiral symmetry breaking is apparent in a conventional Fermi surface due to the existence of a finite density of states at the Fermi level, where a non-zero condensate of particle-hole pairs with a linearized dispersion results in a violation of helicity and therefore chirality\footnote{Note that this is fundamentally different from the anomalous current seen in Weyl semimetals, where a chiral symmetry is broken due to a negative longitudinal magnetoresistance in the crystal\cite{Nielsen, Jia}}\cite{Banks,Shuryak}. {\color{black} The work of Swingle has similarly explored the possibility that each point on the Fermi surface of (2+1)-D free fermions and Fermi liquids can be considered a $(1+1)$-D fermionic mode with a fixed direction of propagation \cite{Swingle1,Swingle2}, yielding a logarithmic violation of the area law and agreement with the Widom conjecture for fermionic entanglement entropy \cite{Klich}. A lower-dimensional manifold of gapless chiral excitations is therefore a natural way of viewing the sharp Fermi surface inherent to Fermi gases and Fermi liquids.} However, from the form of Eqn. \eqref{eq9}, we can clearly see that such a finite density of states remains in the presence of a Luttinger surface with a well-behaved Luttinger-Ward functional. 
The fact that the snark description holds for both Fermi and Luttinger surfaces makes it a much more robust definition of a generalized Fermi surface than some finite discontinuity in the fermionic distribution function, and is therefore the starting point for our consideration of Luttinger's theorem.

 
Before continuing, it should be noted what the explicit connection is between the topological index $\mathcal{N}$ of the many-body generating function and the topological invariant as introduced by Volovik \cite{Volovik2,Volovik1} (which, for clarity, we will call $\mathcal{\widetilde{N}}$). By invoking standard arguments in algebraic topology, we have shown that {\bf the topological invariant of Volovik is equivalent to the topological index only in the absence of a gap.} One may define a similar winding number $\mathcal{\widetilde{N}}$ in a gapped system, but it can no longer be considered identical to the topological index $\mathcal{N}$ of the {\color{black} functional $Z[G]$} {\color{black} as a non-analytic Luttinger-Ward functional results in a breakdown in the underlying assumptions used in deriving Eqn. \eqref{eq9}.} Indeed, recent studies on topological insulators have considered $\mathcal{\widetilde{N}}$ in the context of "counting" the number of edge states (i.e., poles of $G({\bf k},\,\omega)$) as interactions are turned on \cite{Gurarie1, Gurarie2}, however we cannot attach the presence of such gapless excitations to the existence of a generalized Fermi surface (i.e., a "snark"). In other words, the presence of a finite density of states automatically implies either a manifold of $G^{-1}({\bf k},\,\omega)=0$ or $G({\bf k},\,\omega)=0$, but a manifold of $G^{-1}({\bf k},\,\omega)=0$ or $G({\bf k},\,\omega)=0$ does not automatically imply a finite density of states. 
By the Atiyah-Singer index theorem, it is clear that only for gapless systems can we say with confidence that $\mathcal{\widetilde{N}}=\mathcal{N}$, thereby confirming that both Fermi and Luttinger surfaces may support such a manifold and, hence, obey Luttinger's theorem.

\section{IV. Luttinger's Theorem and $\omega$-dependence of $\Sigma({\bf k},\,\omega)$}
Because the snark solution is applicable to both Fermi liquids and Luttinger liquids \added{(both of which satisfying Eqn. \eqref{eqn1}  \cite{Krastan,Affleck})}, the existence of a manifold of zero modes at \added{the Fermi level} appears to be a promising "hard" requirement for Luttinger's theorem. However, Eqn. \eqref{eq9} tells us that a non-zero value of $\mathcal{N}$ may exist for zeros of $G^{-1}({\bf k},\,\omega)$ or $G({\bf k},\,\omega)$, the latter of which having been noted to contradict the fundamental postulates of Luttinger's theorem \cite{Phillips2,Phillips3,Phillips1,Phillips4,Rosch}. \added{It is therefore worth reviewing the underlying assumptions of Luttinger's theorem, and explicitly seeing what systems (if any) that support Luttinger's theorem contradict the underlying assumptions of the snark.} {Ultimately, we aim to prove the following postulation:
\begin{theorem}
A non-zero value for the topological index of a $D$-dimensional Kadanoff-Baym functional is the sole necessary and sufficient condition for the validity of Luttinger's theorem.
\end{theorem}

From the Atiyah-Singer index theorem, we can restate the above as the following:

\begin{corollary}
The only possible scenario where Luttinger's theorem fails is in the presence of a gap or pseudogap. 
\end{corollary}

}


To begin, recall that for any fermionic system, the applicability of a hard Luttinger's theorem can be boiled down to two main principles:

\begin{subequations}
\begin{align}
&\frac{i}{2\pi}\int \frac{d^d {\bf k}}{(2\pi)^d}\oint_{\mathcal{C}}d\omega \frac{\partial}{\partial \omega}\log(G({\bf k},\,\omega))=\frac{N}{2V}\label{12a}
\end{align}

\begin{align}
&-i\int \frac{d^d {\bf k}}{(2\pi)^d} \oint_{\mathcal{C}}\frac{d\omega}{2\pi} \left\{G({\bf k},\,\omega)\frac{\partial}{\partial \omega}\Sigma({\bf k},\,\omega)
\right\}=0\label{12b}
\end{align}
\end{subequations}
%
\noindent Given the requirements of Eqns. \eqref{12a} and \eqref{12b}, we want to see if they are {\it always} compatible with \added{a non-zero winding number given in} Eqn. \eqref{eq9}. 

We start with the former condition. Recall that we can always write the fermionic Green's function in the K\"allen-Lehmann representation, given as

\begin{align}
&G({\bf k},\,\omega)=(2\pi)^3 \sum_j \left(
\frac{A_j \delta({\bf k}-{\bf k}_j)}{\omega-\epsilon_j^++\mu+i0} 
\pm \frac{B_j \delta({\bf k}+{\bf k}_j)}{\omega-\epsilon_j^-+\mu-i0}
\right)
\end{align}
where the information from the self energy is contained in $A_j$ and $B_j$. 
We can easily see that, under such a representation, $G({\bf k},\,|\omega|\rightarrow \infty)\sim 1/\omega$ regardless of whether or not the system is a Landau-Fermi liquid \cite{Lehmann, Abrikosov}. This makes sense, as the self energy cannot diverge at asymptotically large frequencies \cite{Tremblay2}, and simplifies Eqn. \eqref{12a} to the condition that the low-frequency phase of the retarded Green's function must disappear. This ultimately amounts to the imaginary part of the Green's function (and therefore the imaginary part of $\Sigma({\bf k},\,\omega)$) to converge faster than the real part as $\omega\rightarrow 0$.
%
%
%
%

For some general system, we can relate the real and imaginary parts of the self energy to each other via a simple Kramers-Kronig relation \cite{Tremblay3},
where we assume $\omega$ is small. 
 If we consider some general case $\Im \Sigma(k,\,\omega)\sim \omega^\alpha$, we find
 \\
\begin{align}
P\int\frac{d\omega'}{\pi}\frac{\Im \Sigma(k,\,\omega')}{\omega'-\omega}&\sim P\int\frac{d\omega'}{\pi}\frac{\omega'^{\alpha}}{\omega'-\omega}\notag\\
&\approx P\int \frac{d\omega'}{\pi}\omega'^{\alpha-1} +\omega P\int \frac{d\omega'}{\pi }\omega'^{\alpha-2}
\end{align}

\noindent For 
$\alpha\not =1$, 

\begin{align}
\frac{\partial}{\partial \omega}\Re \Sigma(k,\,\omega)\bigg|_{\omega=0}&\sim P\int \frac{d\omega'}{\pi}\omega'^{\alpha-2}\notag\\
&\sim\frac{1}{\pi(\alpha -1)} \lim_{\omega\rightarrow 0^+}\left\{
\frac{1}{\omega_c^{1-\alpha}}-\frac{1}{\omega^{1-\alpha}}
\right\}\notag\\
\notag\\
\rightarrow \Re\Sigma(k,\,\omega)\bigg|_{\omega=0}&\sim\begin{cases}
\omega,\quad &\alpha>1\\
\omega^\alpha,\quad &\alpha<1
\end{cases}\label{eq11}
\end{align}
\\
\noindent {\color{black} up to a constant independent of $\omega$ and} assuming some UV cutoff $\omega_c$ in the integration limits. {\color{black} Assuming there is no purely momentum-dependent component in the self-energy (which will be discussed shortly), any non-zero $\omega$-independent constant in $\Re \Sigma({\bf k},\,\omega)$ would trivially satisfy Luttinger's theorem by itself, as the self-energy would simply correspond to a shift of the chemical potential \cite{Vilk}. As a consequence we will assume such a constant is absent for simplicity and exclusively focus on the frequency dependence of $\Re \Sigma({\bf k},\,\omega)$ as determined in Eqn. \eqref{eq11}, where we can clearly see that} only the case $\alpha \ge 1$ satisfies Eqn. \eqref{12a} (with the case of $\alpha=1$ being the marginal Fermi liquid), and hence also satisfies Luttinger's theorem. A similar requirement for Luttinger's theorem is observed when we consider Eqn. \eqref{12b}, where the integral will vanish only if we can write the self-energy as an exact differential of the Green's function; i.e., 

\begin{align}
\delta \Phi[G]=\frac{1}{V}\sum_{k\sigma}\int \frac{d\omega}{2\pi i}\Sigma({\bf k},\,\omega)\delta G\label{12c}
\end{align}

\noindent where we recognize $\Phi[G]$ as the Luttinger-Ward functional. For divergent frequency dependence in the self-energy, we are unable to integrate the differential in the neighborhood of the Fermi surface and Luttinger's theorem is, once again, violated. {\color{black} This agrees with recent theoretical work on an SU(N) generalization of the atomic Hubbard model \cite{Phillips3} and ARPES work on the cuprate superconductor Bi$_2$Sr$_2$CaCu$_2$O$_{8+\delta}$ \cite{Phillips5,Reber}, where a well-defined Luttinger-Ward functional is only realized for self-energies with analytic frequency behavior.}
%
%

\begin{figure*}[htbp]
\hspace*{-10mm} 
\begin{subfigure}{.55\columnwidth}
\includegraphics[width=1\columnwidth]{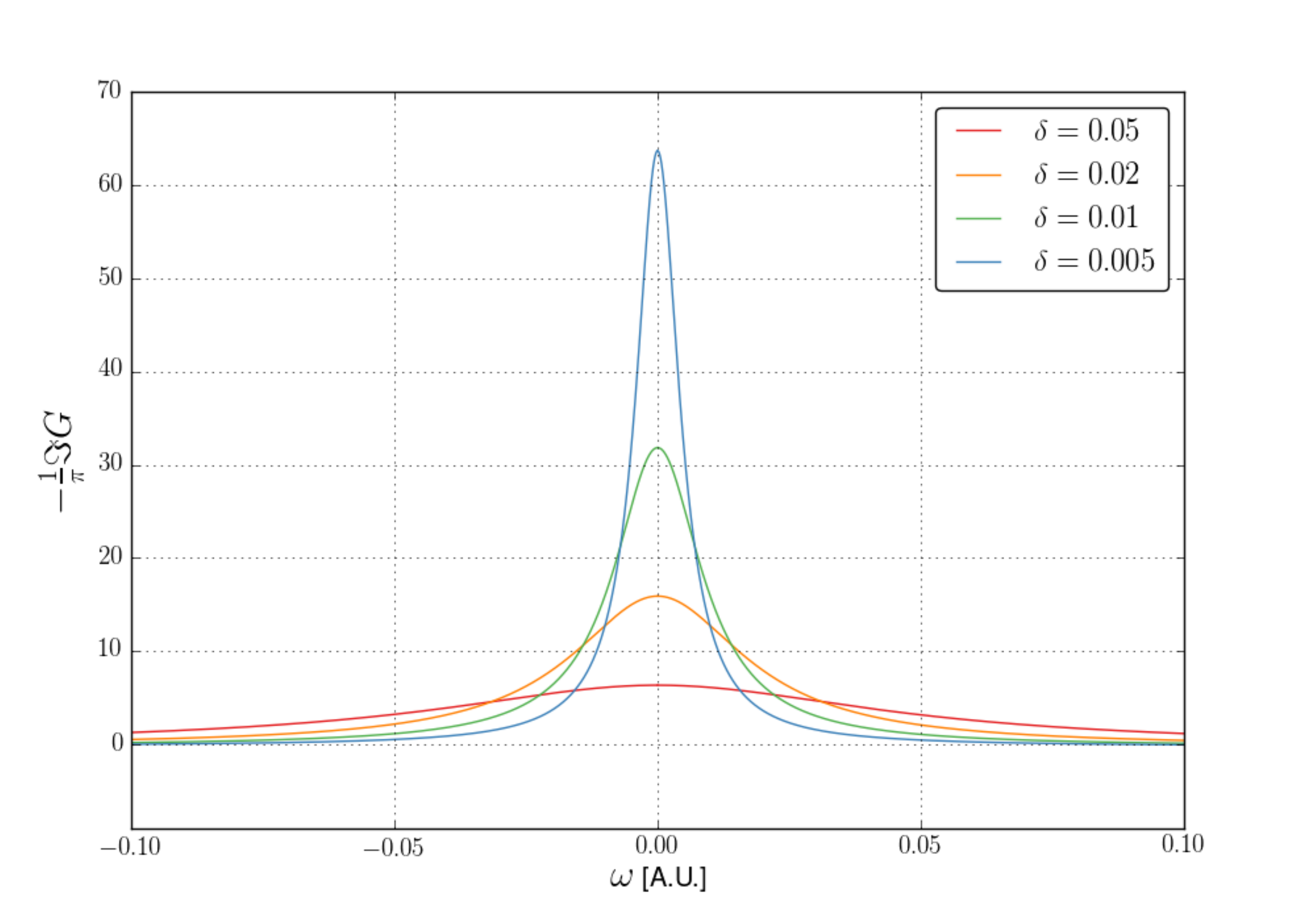}
  \caption{$\alpha=2.25$}\label{fig:2a} 
\end{subfigure}
\begin{subfigure}{.55\columnwidth}
\includegraphics[width=1\columnwidth]{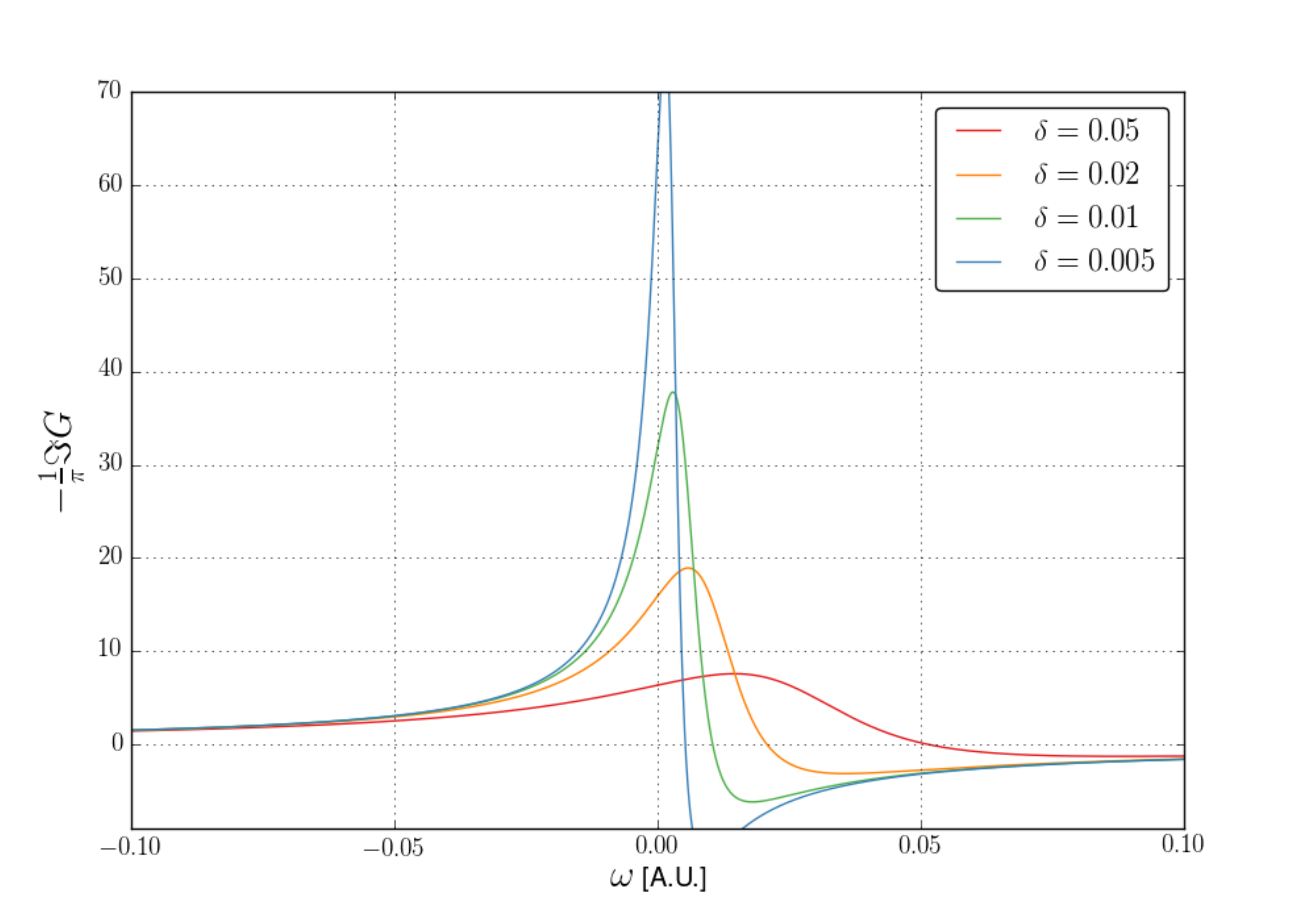} 
  \caption{$\alpha=1.01$}\label{fig:2b} 
\end{subfigure}
\hspace*{-10mm} 
\begin{subfigure}{.55\columnwidth}
 \includegraphics[width=1\columnwidth]{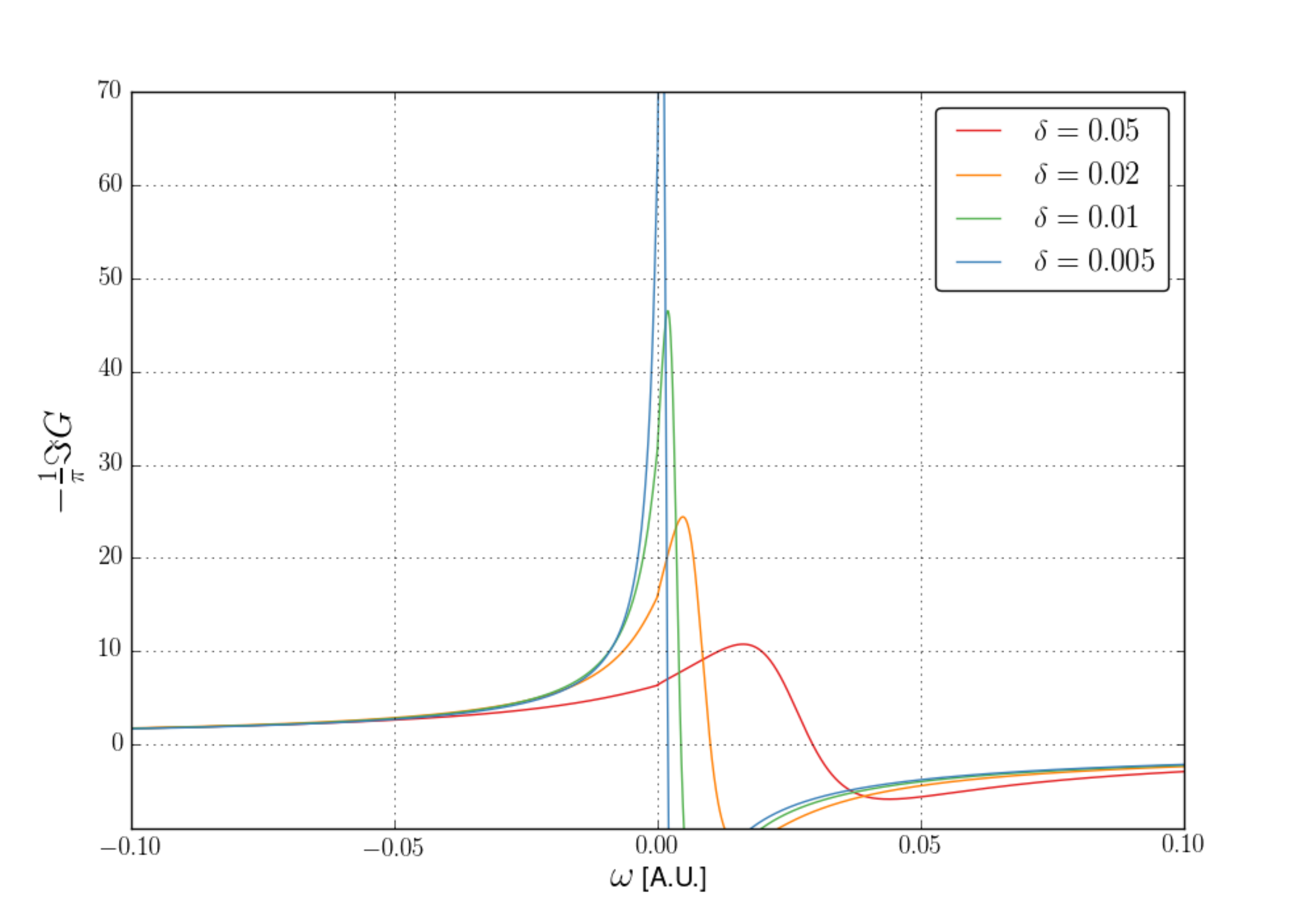} 
   \caption{$\alpha=0.85$}\label{fig:2c} 
\end{subfigure}
\begin{subfigure}{.55\columnwidth}
\includegraphics[width=1\columnwidth]{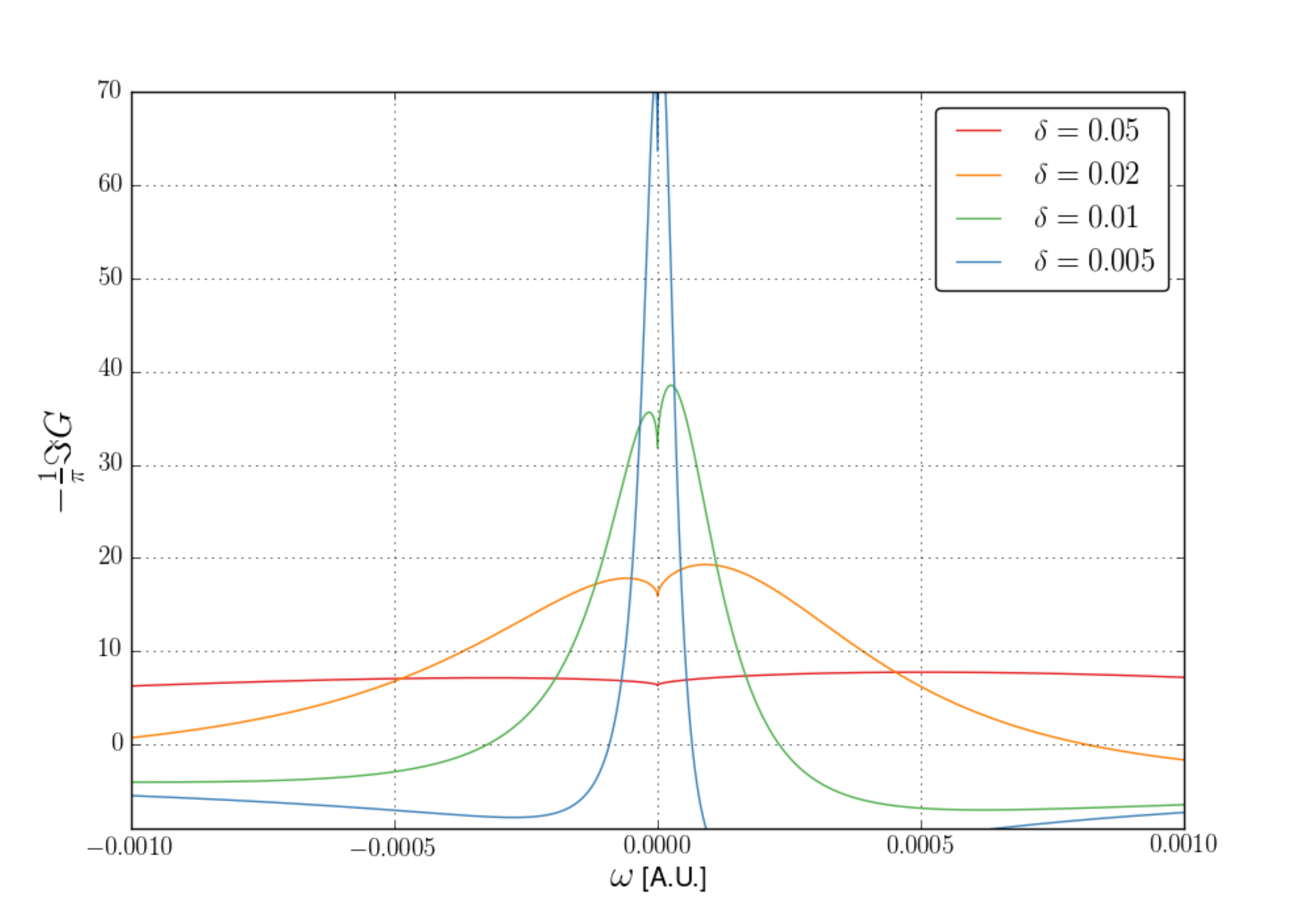} 
   \caption{$\alpha=0.55$}\label{fig:2d} 
\end{subfigure}
\caption{{\color{black} The spectral function $-\frac{1}{\pi}\Im G$ vs. $\omega$ in arbitrary units for various values of the exponent $\alpha$. As we take the limit $\delta\rightarrow 0$, we expect a singularity to emerge at the Fermi surface. Figs \ref{fig:2a} and \ref{fig:2b} show this explicitly, illustrating that all cases with $\alpha>2$ observe near-identical behavior and cases with $1<\alpha<2$ display severe asymmetry across $\omega=0$. In Fig. \ref{fig:2c}, the spectral function begins to show signs of a discontinuity at $\omega=0$ for $0<\alpha<1$, becoming more apparent in Fig. \ref{fig:2d}. Note that the negative density of states for $\alpha<1$ is not necessarily unphysical, as discussed in \cite{Efros}.}} 
\label{Fig2}
\end{figure*}

Now, we will make the connection to gapless excitations and, thus, a non-zero analytical (or topological) index of the {\color{black}quantum field theoretic partition function}. 
%
%
Note that if $\Im \Sigma(k,\,\omega\rightarrow 0)\sim \omega^\alpha$ where $\alpha<0$,
 then both Eqns. \eqref{12a} and \eqref{12b} are violated (with Eqn. \eqref{12a} remaining invalid for $0<\alpha<1$.)
 Physically, this specific $\omega$-dependence is connected to a non-existent or {\color{black}discontinuous} density of states at the Fermi momentum.
 \added{By definition, the density of states $\rho(\omega)$ goes as}
\begin{align}
\rho(\omega) \sim-\frac{1}{\pi}\int\,dr\,\Im G({\bf r},\,{\bf r},\,\omega)
\end{align}
\added{As shown in Appendix C, the regime $\alpha>1$ corresponds to a well-defined density of states at the Fermi level. However, if $\alpha<1$, then either a gap opens (for $\alpha<0$) or the density of states becomes {\color{black} discontinuous} (for $0<\alpha<1$). {\color{black} This is displayed graphically in Figs. \ref{fig:2a}--\ref{fig:2d}, where the spectral function is plotted vs. $\omega$ for several values of $\alpha$. For $0<\alpha<1$, the dip at zero frequency is reminiscent of the minimum in the pseudogap density of states. The identification of the $0<\alpha<1$ regime with the pseudogap phase of the cuprates will be discussed in depth later on.} 

{\color{black} From the arguments given above}, the condition that the low-frequency phase of the retarded Green's function must disappear is equivalent to the condition that the retarded self energy is analytic in the frequency domain in the vicinity of $\omega=0$, which is the same as saying that the condition for a hard Luttinger's theorem is purely based on the existence of a finite density of states at the Fermi level. However, we have already discussed how a finite density of states is identical to our definition of the snark--namely, a lower dimensional manifold of gapless chiral excitations. Because Luttinger surfaces may support such a manifold, materials that exhibit Luttinger surfaces may simultaneously support Luttinger's theorem.
}
As such, we see that {\bf the snark vanishes if and only if Luttinger's theorem fails}, hence proving Theorem 2. 

The above leads to the following corollary:
%
{
\begin{corollary}
Luttinger's theorem may be valid in a system that supports a Luttinger surface as long as the topological index of a $D$-dimensional Kadanoff-Baym functional is non-zero.
\end{corollary}}

%

\added{ The argument in this section based on causality (i.e., the Kramers-Kronig relation) tells us what the conditions for Luttinger's theorem are, but it doesn't tell us if all Luttinger or Fermi surfaces obey those conditions. To answer the latter question, we need to use our more "robust" definition of a Fermi surface in terms of the winding number Eqn. \eqref{eq9}, and utilize the Atiyah-Singer index theorem to connect this number back to the existence of a snark. Because the snark is valid for systems with manifolds of both $G^{-1}({\bf k},\,\omega)=0$ and $G({\bf k},\,\omega)=0$, both solutions can happily coexist with a trivial Luttinger count. The catch, however, is that such solutions must exhibit a self energy that remains analytic in the entirety of the frequency plane.
Therefore, if a generic $G({\bf k},\,\omega)=0$ solution beyond the marginal Fermi liquid is to obey Luttinger's theorem, the singular behavior of the self energy must lie in the momentum-dependence. \added{This is in sharp contrast to 
\cite{Farid2}, where direct application of Volovik's topological argument is used to propose that any Luttinger surface supports Luttinger's theorem. From Phillips' work on the existence of a Luttinger-Ward functional \cite{Phillips3}, we know only a subset of these manifolds fail to introduce/lose the original fermionic DoF, and thus we are motivated to consider the generalized momentum-dependent self energy that supports Luttinger's theorem.} 
}

\section{V. Luttinger's Theorem and ${\bf k}$-dependence of $\Sigma({\bf k},\,\omega)$}
%
{Our goal in this section is to prove the following ansatz:
\begin{theorem}
There exist Luttinger surfaces such that the topological index of the $D$-dimensional Kadanoff-Baym functional is non-zero.
\end{theorem}
This is equivalent to the following statement:
\begin{corollary}
There exists some form of the self-energy such that Luttinger's theorem is implied in the absence of a finite quasiparticle weight in $D\ge1$ dimensions. 
\end{corollary}
}
\noindent When studying such ${\bf k}$-dependent behavior in
strongly correlated matter, a local approximation $\Sigma({\bf k},\,\omega)\sim \Sigma(\omega)$ is {\color{black} often invoked}, as the second-order contribution to the ground-state energy in the lattice vanishes
 as the dimension $D\rightarrow \infty$ \cite{Metzner1, Metzner2,Metzner3}. Although this forms the basis of the highly-successful dynamical mean-field theory (DMFT) \cite{Kotliar1,Kotliar2,Kotliar3}, such an approximation is not always applicable in $D>1$. 
In $D=2$, non-trivial ${\bf k}$-dependence is a core component of GW+DMFT \cite{Tomczak1} and its {\it ab initio} extensions \cite{Galler}, as well as being seen in Monte Carlo simulations of the half-filled Hubbard model on the square lattice \cite{Blumer}. Even in 3D, the local approximation breaks down in the presence of antiferromagnetic fluctuations near second order phase transitions \cite{Tomczak2}.

 \added{To describe some general momentum-dependence, we perform a Laurent expansion of the self energy}:
 
\begin{align}
\Sigma({\bf k},\,\omega)=\sum_{j=0}^\infty A_j (k-k_F)^j+\sum_{j=0}^\infty B_j(k-k_F)^{-j}
\end{align}

\noindent By assuming the self energy is analytic about some annular region near ${\bf k}_F$, it should be clear that solutions of Eqn. \eqref{eq9} correspond to higher-order $m$-derivatives in Eqn. \eqref{eq4}. This allows us to generalize the quasiparticle weight in Eqn. \eqref{eq3} to 

\begin{align}
\frac{\partial^m n(k_F-\delta)}{\partial k^m}-\frac{\partial^m n(k_F+\delta)}{\partial k^m}=Z_k^{(m)}>0 \,\, \forall \,\,m\in \mathbb{N}_0
\end{align}
\vspace{2mm}
\noindent The snark can then be thought of as a "kink" in the bare particle distribution at ${\bf k}_F$. 
These kinks have previously been observed as "critical Fermi surfaces", and indicate non-Fermi liquid behavior in heavy fermion criticality, Mott criticality, and at optimal doping of the cuprates\cite{Senthil,Max}. 
{\bf Much as in the case of a Tomonaga-Luttinger liquid, the existence of a critical Fermi surface coincides with the preservation of Luttinger's theorem.} This agrees with studies of translationally invariant non-Fermi liquids composed of Sachdev-Ye-Kitaev dots, where Luttinger's theorem is shown to coexist happily with the critical Fermi surface\cite{Debanjan}. 

We now introduce the necessary nomenclature to categorize all possible snarks. We call the first $m$th order derivative of the bare particle distribution at $\{0,\,{\bf k}_F\}$ that yields a non-zero $Z_k^{(m)}$ the order of the snark. We include solutions of $m=0$ in the above to account for the local Fermi liquid, which has no ${\bf k}$-dependence\cite{Bedell1,Bedell2,Bedell3}. Generic systems with $Z_k^{(m)}\in \mathbb{R}_{>0}$ for $m>0$ are defined as {\it quasi-local}, and are said to exhibit snarks of the first kind ($n=1$). Physically, quasi-local self energies correspond to some truncation in the Laurent expansion of a general self energy to order $m$ for coefficients $A_j=0\,\,\forall j\in \mathbb{N}_0$.
 Systems where $Z_k^{(m)}\rightarrow \infty$ for $m>0$ are said to be snarks of the second kind ($n=2$). We therefore have the constraint $n\in [1,\,2]$ by definition of the snark's kind.As an example, the snark of a local Fermi liquid would be defined as a $0$th order Fermi surface of the 1st kind, while that of a Tomonaga-Luttinger liquid would be defined as a 1st order Luttinger surface of the 2nd kind (which follows directly from the form of the momentum distribution Eqn. \eqref{eq4}).

\added{To more formally classify all possible snarks, we introduce the shorthand $F_m^{(n)}$ ($L_m^{(n)}$) for an $m$th order Fermi (Luttinger) surface of the $n$th kind. The specific snark classification for the four main behaviors of the self energy (all for $\omega=0$) are given as follows:}
\begin{enumerate}
\begin{subequations}
\item Fermi surface of the 1st kind: positive integer power law 
\begin{align}
\Sigma({\bf k}) \sim A_{m_1}|{\bf k}-{\bf k}_F|^{m_1}+\mathcal{O}(|{\bf k}-{\bf k}_F|^{m_1+1}),\quad A_{m_1}\in \mathbb{R},\,\, m_1 \in \mathbb{N}_1\rightarrow \{{\bf k}_F\} \equiv F_{m_1}^{(1)}\label{60a}
\end{align}
\item Fermi surface of the 2nd kind: positive non-integer power law
\begin{align}
\Sigma({\bf k}) \sim A_{m_2}|{\bf k}-{\bf k}_F|^{m_2}+\mathcal{O}(|{\bf k}-{\bf k}_F|^{m_2+1}),\quad A_{m_2}\in \mathbb{R},\,\,m_2 \in \{ \mathbb{R}_{>0}\backslash \mathbb{N}_1\} \rightarrow \{{\bf k}_F\} \equiv F_{\ceil{m_2}}^{(2)}\label{60b}
\end{align}
\item Luttinger surface of the 1st kind: negative integer power law
\begin{align}
\Sigma({\bf k}) \sim B_{m_3}|{\bf k}-{\bf k}_F|^{m_3}+\mathcal{O}(|{\bf k}-{\bf k}_F|^{m_3+1}),\quad B_{m_3}\in\mathbb{R},\,\,m_3 \in \{ \mathbb{Z}\backslash \mathbb{N}_0\}\rightarrow \{{\bf k}_F\}\equiv L_{m_3}^{(1)}\label{60c}
\end{align}
\item Luttinger surface of the 2nd kind: negative non-integer power law
\begin{align}
\Sigma({\bf k}) \sim B_{m_4}|{\bf k}-{\bf k}_F|^{m_4}+\mathcal{O}(|{\bf k}-{\bf k}_F|^{m_4+1}),\quad B_{m_4}\in \mathbb{R},\,\,m_4 \in \{\mathbb{R}_{<0}\backslash \mathbb{Z}\}\rightarrow \{{\bf k}_F\}\equiv L_{\floor{m_4}}^{(2)}\label{60d}
\end{align}
\end{subequations}
\end{enumerate}

\noindent A table illustrating the behavior of $Z_k^{(0)}$ for snarks of different orders and kinds is given above.

By repeatedly taking $k$-derivatives of the quasiparticle weight $Z_k^{(0)}$, we can devise a taxonomy of all possible self energies that yield a non-zero winding number (Eqn. \eqref{eq9}) and therefore a trivial Luttinger count. 
This exact dependence is derived in detail in Appendix D, and is reproduced below:
\begin{align}
\mathbb{S}^{(n)}_m=\bigg\{{\bf k}_F\iff\,\exists! \,m\in \mathbb{N}_0:&\lim_{{\bf k}\rightarrow {\bf k}_F}\left\{ \frac{\partial^m}{\partial k^m}\left(\frac{\partial \Re\Sigma({\bf k},\,\omega)}{\partial \omega}\right)\bigg|_{\omega=0}\right\}^{-1}
\left(1-\frac{\partial \Re\Sigma({\bf k},\,\omega)}{\partial \omega}\right)^{2}\bigg|_{\omega=0}\notag\\
&=\begin{cases}
\frac{\partial \Re\Sigma(\omega)}{\partial \omega}\bigg|_{\omega=0}\sum_{j=0}^m(-1)^j \frac{(2j-1)!}{(j-1)!}B_j 
,\quad &n=1,\quad m\ge 0\\
0,\quad &n=2,\quad m>0
\end{cases}\label{eq15}
\bigg\}
\end{align}

\begin{table}
\hspace{-16mm}
\scriptsize
\setlength\tabcolsep{2pt}
\begin{tabular}{|c|c|c|c|c|*{5}{c|}}\hline
\slashbox{$F_m^{(2)}$ $\mid$ $L_m^{(2)}$}{$F_m^{(1)}$ $\mid$ $L_m^{(1)}$}
&\makebox[3em]{$\partial^0/\partial k^0$}&\makebox[3em]{$\partial^1/\partial k^1$}&\makebox[3em]{$\partial^2/\partial k^2$}
&\makebox[3em]{$\partial^3/\partial k^3$}&\makebox[3em]{$\partial^4/\partial k^4$}
\\\hline\hline
0th Order $Z_k^{(0)}$ & \slashbox{----- $\mid$ -----}{$\not= 0$ $\mid$ -----}& \slashbox{----- $\mid$ -----}{=0 $\mid$ -----}& \slashbox{----- $\mid$ -----}{=0 $\mid$ -----}& \slashbox{----- $\mid$ -----}{=0 $\mid$ -----}& \slashbox{----- $\mid$  -----}{=0 $\mid$ -----}\\\hline
1st Order $Z_k^{(0)}$ & \slashbox{$\not=0$ $\mid$ $=0$}{$\not=0$ $\mid$ $=0$}&\slashbox{$\infty$ $\mid$ $\infty$}{$\in \mathbb{R}$ $\mid$ $\in \mathbb{R}$} & \slashbox{$\infty$ $\mid$ $\infty$}{$\in \mathbb{R}$ $\mid$ $\in \mathbb{R}$}&\slashbox{$\infty$ $\mid$ $\infty$}{$\in \mathbb{R}$ $\mid$ $\in \mathbb{R}$}&\slashbox{$\infty$ $\mid$ $\infty$}{$\in \mathbb{R}$ $\mid$ $\in \mathbb{R}$}\\\hline
2nd Order $Z_k^{(0)}$ &\slashbox{$\not=0$ $\mid$ $=0$}{$\not=0$ $\mid$ $=0$}&\slashbox{$=0$ $\mid$ $=0$}{$=0$ $\mid$ $=0$}&\slashbox{$\infty$ $\mid$ $\infty$}{$\in \mathbb{R}$ $\mid$ $\in \mathbb{R}$}&\slashbox{$\infty$ $\mid$ $\infty$}{$\in \mathbb{R}$ $\mid$ $\in\mathbb{R}$}&\slashbox{$\infty$ $\mid$ $\infty$}{$\in \mathbb{R}$ $\mid$ $\in \mathbb{R}$}\\\hline
3rd Order $Z_k^{(0)}$ &\slashbox{$\not=0$ $\mid$ $=0$}{$\not=0$ $\mid$ $=0$}& \slashbox{$=0$ $\mid$ $=0$}{$=0$ $\mid$ $=0$}  &\slashbox{$=0$ $\mid$ $=0$}{$=0$ $\mid$ $=0$}&\slashbox{$\infty$ $\mid$ $\infty$}{$\in\mathbb{R}$ $\mid$ $\in \mathbb{R}$}&\slashbox{$\infty$ $\mid$ $\infty$}{$\in \mathbb{R}$ $\mid$ $\in\mathbb{R}$}\\\hline
4th Order $Z_k^{(0)}$ &\slashbox{$\not=0$ $\mid$ $=0$}{$\not=0$ $\mid$ $=0$}& \slashbox{$=0$ $\mid$ $=0$}{$=0$ $\mid$ $=0$}  &\slashbox{$=0$ $\mid$ $=0$}{$=0$ $\mid$ $=0$}&\slashbox{$=0$ $\mid$ $=0$}{$=0$ $\mid$ $=0$}&\slashbox{$\infty$ $\mid$ $\infty$}{$\in\mathbb{R}$ $\mid$ $\in\mathbb{R}$}\\\hline
\end{tabular}
\caption{Behavior of the quasiparticle weight $Z_k^{(0)}$ for several examples of mth order Fermi and Luttinger surfaces of the nth kind (denoted $F_m^{(n)}$ and $L_m^{(n)}$, respectively). \added{The behavior of the quasiparticle weight for any system obeying Luttinger's theorem can then be read off from such a "periodic table" of many-body solutions that support an independent-particle approximation.} }
\end{table}

Note that, as $\Re \Sigma(\omega)\sim \omega$ for all cases that satisfy Luttinger's theorem, $\frac{\partial \Re \Sigma(\omega)}{\partial \omega}\sim$ constant. Therefore, {\bf the set of all $m$th order snarks of the $n$th kind $\mathbb{S}_m^{(n)}$ defines all possible ${\bf k}$-behavior in the self energy that satisfies Luttinger's theorem.} 

\section{VI. The status of Luttinger's theorem in the cuprates and at the Mott transition}

\added{The snark description reveals a deep connection between independent-particle behavior and the absence of an energy gap, as opposed to particle-hole symmetry, the analyticity of the self energy in the entire Fourier space, or the complete absence of a Luttinger surface. The existence of a finite density of states at the Fermi level immediately implies Luttinger's theorem is preserved, and from the discussion above, the former may coexist happily with Luttinger surfaces if the self energy diverges as a function of the momentum ${\bf k}$ as opposed to frequency $\omega$.}

\added{From present studies of the cuprates, we can see clear agreement with our conditions on Luttinger's theorem. Recall that, if $\Im \Sigma(k,\,\omega\rightarrow 0)\sim \omega^\alpha$ where $\alpha<0$, then the system loses a coherent snark and, according to our theory, Luttinger's theorem is violated.} Such behavior is supported experimentally in ARPES data on the cuprate superconductor Bi$_2$Sr$_2$CaCu$_2$O$_{8+\delta}$ in its normal phase, where $\Im\Sigma({\bf k},\,\omega)\sim (\omega^2+T^2)^{\widetilde{\alpha}}$\cite{Reber}. The power $\widetilde{\alpha}$ is a function of doping, with $\widetilde{\alpha}>0.5$ 
($\alpha > 1$) corresponding to the overdoped metallic phase, $\widetilde{\alpha}\approx 0.5$ ($\alpha\approx 1$) to the optimally-doped "strange metal"/marginal Fermi liquid phase, and $\widetilde{\alpha}< 0.5$ ($\alpha<1$) to the underdoped pseudogap phase. Our results confirm the observation that the overdoped phase respects Luttinger's theorem\cite{Dessau,Hussey,Plate,Senthil,Gofron,Georges}, while the underdoped pseudogapped phase violates it\cite{Dessau,Zhang,Schmitt,Phillips3}. \added{A violation of a "hard" Luttinger's theorem in the latter is confirmed in the recent work of A. Tsvelik, where a non-perturbative solution to the Kondo-Heisenberg model yields evidence of a fractionalized Fermi liquid ground state analogous to the pseudogap state\cite{Tsvelik}.} In a similar fashion, the opening of a gap in a antiferromagnetically ordered spin-density-wave state has already been shown to exhibit diverging $\omega$-dependence in $\Sigma({\bf k},\,\omega)$ and subsequently a \added{non-zero} value of Eqn.\eqref{12b}\cite{Morr}. 
%
%
%

%

Whereas previous studies have connected the power-law coefficient in $\Im \Sigma(\bf {k},\,\omega)$ to some anomalous scaling of an unparticle propagator\cite{Phillips5}, the discussion above 
proves that the IPA is always preserved in the normal phase for optimal doping and above, independent of any other internal parameter. Because the cases where Luttinger's theorem fails correspond to the appearance of a (pseudo)gap, Eqn. \eqref{eq9} no longer yields a non-zero winding number as the Luttinger-Ward functional
  is ill-defined and/or the chiral symmetry is at least partially restored at $\{{\bf k}_F\}$. {In a similar vein, our result agrees with self-consistent T-matrix calculations of Fermi systems with large spin population imbalance\cite{Pieri}, where a Luttinger-Ward functional is still appropriately defined and, hence, Luttinger's theorem is shown to be preserved.}

 On the computational side, cellular dynamical mean-field theory (CDMFT)\cite{Kotliar4, Kotliar5,Kotliar6} calculations support the postulate that Luttinger's theorem is violated in the pseudogap phase of the 2D Hubbard model\footnote{We thank Shiro Sakai for bringing this work to our attention.}. Coupled with exact diagonalization techniques, the undoped regime was shown to harbor additional "hidden" fermionic DoF (independent of the cluster-size dependence of the CDMFT) and hence violate Luttinger's theorem\cite{Sakai1, Sakai2}. These additional DoF were later seen to be directly connected to an additional $\omega$-dependent term in Luttinger's spectral representation of the self energy\cite{Luttinger4} which are proportional to $(\omega-\epsilon_{f_1})^{-1}$, where $\epsilon_{f_1}$ is the hidden fermion energy\cite{Sakai3}. As doping increases, this divergent term dies out and Luttinger's theorem is restored, in agreement with the predictions of this article.

\added{Beyond the cuprates}, the snark description resolves the issue of applying Luttinger's theorem at the Mott transition, where the onset of a correlation-induced insulating phase has led to the question of a Fermi gas-like state in these materials\cite{Wigner, Mott_1, Mott_2,Mott_3, Kohn, Essler1,Konik}. For Mott insulators with gapped excitations,
it is well known that Luttinger's theorem is violated\cite{Logan,Essler1}. 
 However, in models such as the large-$U$ limit of the half-filled nearest-neighbor Hubbard model on the triangular lattice\cite{Tremblay} and the weak-tunneling limit of intercoupled 1D Hubbard chains treated in the RPA\cite{Essler1,Essler2}, the gap either remains completely closed (as seen in the former) or negligible compared to the bandwidth (as seen in the latter), supporting Kohn's original premise that the presence of an excitation gap is sufficient but not necessary for insulating behavior\cite{Kohn}. This is similarly supported by the proposal that the Mott transition in 1D and 2D Hubbard models in the $U\rightarrow \infty$ limit is a Pokrovsky-Talapov (commensurate-incommensurate) transition, and are thus integrable\cite{Bedell4}. Because Luttinger's theorem remains in the presence of a gapless Luttinger surface, we predict that the IPA remains applicable to this special class of insulators.

 The divergent behavior \added{in the ${\bf k}$-dependence of the self-energy required for the existence of a Luttinger's theorem-obeying system with a Luttinger surface has similarly been} hinted at in numerical studies of the Mott-Hubbard metal-insulator transition in the unfrustrated 2D Hubbard model\cite{Toschi1} as well as in a functional renormalization group extension of DMFT applied to the 2D Hubbard model at half filling\cite{Toschi2}. A more rigorous proposal of quasi-local behavior in 2D materials is seen in \cite{Luther,Anderson1,Anderson4}, where the applicability of the Bethe Ansatz in $D>1$ allows us to describe excitations near the Fermi surface in terms of phase-shift variables.
The presence of a unrenormalizable Fermi surface phase-shift results in the sudden collapse of the quasiparticle weight with the addition of even a single external particle; a phenomenon known as the "orthogonal catastrophe"\cite{Anderson2,Anderson3,Louis}. A direct consequence of this is that the Landau parameter for this 2D system goes as $f_{kk'}\sim 1/|{\bf k}-{\bf k}'|$, and is therefore divergent for forward scattering. This interaction then leads to marginal Fermi liquid behavior in $\Im \Sigma({\bf k},\,\omega)$ with the addition of a term $\sim \log(q_c v_F)$, where $q_c$ is an upper momentum cutoff\cite{Stamp1,Stamp2}. Because we can always take a different branch cut in the low-$\omega$ integral of the logarithm, the Luttinger-Ward functional is still well-defined in any case of marginal Fermi liquid behavior of $\Sigma(\omega)$ (as expected\cite{Zimanyi}).
Therefore, although the 2D Landau-Fermi liquid formalism might break down in the presence of forward-scattering near the Fermi surface, a 1st order Luttinger surface of the 2nd kind is present, and thus Luttinger's theorem and the IPA remains. This is in agreement with the work of Haldane, where the bosonized $D\ge 1$ fermionic system is shown to obey Luttinger's theorem even when there is no Landau quasiparticle\cite{Haldane}. Our general result is similarly in agreement with experimental studies on dilute 2D materials (such as the low-disordered silicon metal-oxide semiconductor field-effect transistors), where evidence is found for a strongly-correlated metallic ground state despite the absence of a Landau-like quasiparticle\cite{Kravchenko,Finkel1,Finkel2,Castel1,Castel2,Valles}. 

{
As a result of the above discussion, we can see that the coexistence of Luttinger surfaces with a trivial Luttinger count 
is most likely in dimensions $D\le 2$, where quasilocal ${\bf k}$-dependence in the self energy is most probable. The existence of un-conventional, scale-invariant physics that breaks the IPA in the absence of a spectral gap would then be confined to noncompact dimensions much larger than our own, as already hinted in the work of Randall and Sundrum\cite{Randall}.}

 \section{VII. The role of particle-hole symmetry and limiting behaviors on Luttinger's theorem}
\added{As apparent in the above discussion}, two main models have been considered \added{in the study of Luttinger's theorem}: the Hubbard model\cite{Phillips2} and the Tomonaga-Luttinger liquid\cite{Phillips4}. \added{In the current literature,} the requirements of Luttinger's theorem in the former has been reduced to the disappearance of $\Im G({\bf k},\,\omega)$ at $\omega=0$ and $\omega\rightarrow -\infty$ and the existence of particle-hole symmetry, while the requirements of Luttinger's theorem in the latter has been boiled down to a constraint on the scaling dimension of the many-particle Green's function; namely, $G\sim 1/(\omega-\epsilon_{\bf k})^{\alpha}$, $1<\alpha<2$. Given the clear overlap of our work with these studies, we \added{will now} address how our general prescription fits into these model-based analyses.

First, we concern requirements of Luttinger's theorem in the Hubbard model. The condition of a disappearing imaginary Green's function at $\omega\rightarrow 0,\,-\infty$ is obviously important for Luttinger's theorem in any generic system, as already addressed. Whether or not a fermionic system obeys Luttinger's theorem, we expect that the phase of the retarded Green's function will approach $\pi$ as $\omega\rightarrow -\infty$. The more significant limit is when $\omega\rightarrow 0$. This is directly dependent on the behavior of the imaginary part of the self energy near the Fermi surface, from which the discussion above follows. Interestingly, we have shown that a more crucial condition for Luttinger's theorem is not the low-frequency behavior of the imaginary part of the self energy {\it per se}, but instead the low-frequency behavior of the imaginary part of the self energy {\it relative to} the real part. For $0<\alpha<1$, we have clearly shown that, despite $\Im G(\omega=0)\rightarrow 0$ and the existence of a well-defined Luttinger-Ward function, Luttinger's theorem breaks down. This is to be expected, as the regime of $0<\alpha<1$ corresponds to the pseudogapped phase, where (as derived in Appendix C) the density of states becomes {\color{black}discontinuous}. This regime of parameters $0<\alpha<1$ was explored numerically in \cite{Phillips4}, where it was confirmed that Luttinger's theorem breaks down for $0<\alpha<0.7$. Whereas the numerical integration techniques for $\alpha>0.7$ are unstable, the analytical derivation above based on Kramers-Kronig relations illustrates the importance of $\Im \Sigma$ relative to $\Re \Sigma$ as opposed to the behavior of $\Im \Sigma$ itself. 

As for particle-hole symmetry, it is worth noting that, by isolating the power law behavior of $\Sigma(\omega)$ from that of the total Green's function, it is not necessary to invoke particle-hole symmetry to verify Luttinger's theorem in a generic many-body system \cite{Rosch}. Indeed, the lack of particle-hole symmetry simply means an asymmetric density of states, and in many cases a Luttinger's theorem (and even a Landau-Fermi liquid prescription \cite{Quader,Carlos}) remains appropriate \cite{Krastan,Affleck,Phillips4}. Considering that particle-hole symmetry is present in a superconducting state (where Luttinger's theorem clearly fails) yet is absent in certain Landau-Fermi liquids (where Luttinger's theorem clearly succeeds), we interpret the behavior of the imaginary component of the self energy relative to the real component (i.e., the existence of a snark) as a much more robust condition on an independent particle approximation in strongly correlated matter.  

Of course, \cite{Phillips4} has indicated that particle-hole symmetry is not a necessary condition of Luttinger's theorem in the case of Luttinger liquids, which \cite{Krastan, Affleck, Oshikawa} have illustrated to exhibit a trivial Luttinger count. In \cite{Phillips4}, these limits are "special cases" when the scaling dimension of the Green's function itself is constrained to be between unity and two. This agrees with our work, where the power of the self energy for some general fermionic system must not pass under one for the snark to remain well-defined. Nevertheless, by noticing that the Luttinger's theorem constraint depends specifically on the frequency-dependence of the self energy as opposed to the general scaling behavior of the total Green's function, we can say with confidence that any scaling parameter $\alpha>1$ will lead to a well-defined single-particle approximation of the many-body system. Moreover, our calculations have shown that zeroes of the Green's function do not necessarily indicate a gap, as diverging ${\bf k}$-dependence in a "quasi-local" system (as already indicated in \cite{Toschi1,Toschi2,Luther,Anderson1,Anderson4}) is perfectly compatible with Luttinger's theorem.

\section{VIII. Conclusion}
Many condensed matter physicists study the properties of strongly interacting electron systems; how they interact with each other, themselves, and their environment. The presence of coherence might force the interacting regime to exhibit emergent phenomenon unlike anything seen in the non-interacting limit, but at the end of the day an independent-particle picture is always reduced to an extreme inconvenience rather than an absolute impossibility \cite{Anderson_book}.

Luttinger's theorem is a powerful tool that tells us when an independent particle approximation is salvageable. Previous studies have suggested that such cases are rare, and instead an "un-particle" approximation must be used for the great majority of models where the IR limit loses any resemblance to the UV. {The work given above implies that Luttinger's theorem is, instead, possible in systems with disappearing quasiparticle weight beyond the well-known Tomonaga-Luttinger liquid case. Our main argument is summarized as follows:
\begin{enumerate}
\item By defining a generalized Fermi surface in terms of a non-zero topological index of the many-body generating functional (i.e., Eqn. \eqref{eq9}), we show that it is physically possible for a gapless fermionic system to support a Luttinger surface in some arbitrary dimension.
\item The existence of a generalized Fermi surface as defined in (i) is the sole requirement for a fermionic system to uphold a hard Luttinger's theorem.
\item From (i) and (ii), it is implied that materials which exhibit a Luttinger surface may simultaneously satisfy a hard Luttinger's theorem as long as the self energy satisfies Eqn. \eqref{eq15}.
\end{enumerate}}
\noindent { \added{It is worth noting that} the mutual coexistence of gapless excitations with stable Fermi surfaces has previously been suggested via an Atiyah-Bott-Shapiro construction of the K-theory group $K({\bf R}^k)=\pi_{k-1}(GL(N,\,{\bf C}))$ \cite{Horava} as well as via conformal field theory arguments in the IR \cite{Swingle1,Swingle2}. In a similar vein, the topological nature of the Luttinger count and Luttinger's theorem itself have previously been suggested in \cite{Seki} and \cite{Oshikawa}, respectively. However, unlike these studies, the snark description \added{and its anomalous interpretation} explicitly connects the microscopic details of the many-body propagator to the existence of a topologically robust manifold of chiral gapless excitations in Fourier space. 
In other words, we have shown what properties the propagator in Eqn. \eqref{eqn1} must have to ensure the Fermi volume remains invariant as interactions become arbitrarily large, a feat which no one to our knowledge has previously accomplished.
}


%

In light of recent experiments, numerics, and theoretical models regarding strongly correlated matter, we believe that our study brings Luttinger's theorem out of the "folklore" of recent years, and opens new avenues to solving the many-body problem with common-sense IR physics.
\section{Acknowledgements}
We thank Matthew Gochan and Tong Yang for useful discussions on the ideas presented in this paper; \added{Thomas Gasenzer, Philip Phillips, Shiro Sakai, Giancarlo Strinati, and Alexei Tsvelik for their comments and suggestions}; \added{Tyler Dodge for a thorough review of the paper}; and Krastan Blagoev for his guidance, encouragement, and input throughout this paper's development. J.T.H. would also like to thank the organizers of the 2018 International Summer School on Computational Quantum Materials at the Jouvence resort in Quebec, Canada. The excellent lectures at this workshop greatly contributed to a deep understanding of the preliminary many-body techniques required for the calculations in this paper. This work was partially supported by the John H. Rourke endowment fund at Boston College.  
\newpage 
\vspace{5mm}
\section{Supplemental Material}
\subsection{Proof of Luttinger's Theorem in a Landau-Fermi Liquid}
We now re-derive the well-known proof of Luttinger's theorem  in a Landau-Fermi liquid. We hope that this will fill in certain gaps not appropriately addressed in the main body of the text. 

We begin by recalling the form of the Green's function for a bare particle in the interacting system:
\begin{align}
G({\bf k},\,\omega)=\frac{1}{\omega-\xi_k-\Sigma({\bf k},\,\omega)}
\end{align}
where $\xi_k$ is measured with the respect to the chemical potential and an infinitesimal value of $i\delta$ is implied. The total density can be written as\cite{Abrikosov}
\begin{align}
\frac{N}{2V}&=\frac{1}{2}\langle \psi_{\alpha}^\dagger(x) \psi_\alpha (x)\rangle\notag\\
&=-i\lim_{{\bf r}\rightarrow {\bf r}'}G(x-x')\notag\\
&=-i\int \frac{ d^D {\bf k}}{(2\pi)^D}\oint_{\mathcal{C}} \frac{d \omega}{2\pi}G({\bf k},\,\omega)
\end{align}
To solve the above integral, we take the frequency derivative of the log of the Green's function, which yields
\begin{align}
\frac{\partial}{\partial \omega}\log G({\bf k},\,\omega)
&=-\frac{1}{\omega-\xi_k-\Sigma({\bf k},\,\omega)}\left(
1-\frac{\partial}{\partial \omega}\Sigma({\bf k},\,\omega)\right)\notag\\
&=-G({\bf k},\,\omega)+G({\bf k},\,\omega) \frac{\partial}{\partial \omega}\Sigma({\bf k},\,\omega)
\end{align}
This yields the following form of the Green's function:
\begin{align}
G({\bf k},\,\omega)=-\frac{\partial}{\partial \omega}\log G({\bf k},\,\omega)+G({\bf k},\,\omega)\frac{\partial}{\partial \omega}\Sigma({\bf k},\,\omega)
\end{align}
The particle density then becomes

\begin{align}
\frac{N}{2V}
&=-i\int \frac{d^D {\bf k}}{(2\pi)^D} \oint_{\mathcal{C}}\frac{d\omega}{2\pi} \left\{ -\frac{\partial}{\partial \omega}\log G({\bf k},\,\omega)+G(k,\,\omega)\frac{\partial}{\partial \omega}\Sigma({\bf k},\,\omega)
\right\}
\end{align}
Which yields the integrals Eqns. \eqref{12a} and \eqref{12b}, respectively. 

We'll start with the general solution of the first integral, which we'll solve by introducing the retarded Green's function $G^R({\bf k},\,\omega)$. Because there is only a pole in the upper half plane, any closed contour will then yield zero for $G^R$, as we can shift the contour to the regime where $\Im(\omega)$ is infinite. Therefore, the above integral becomes

\begin{align}
&\frac{i}{2\pi}\int \frac{d^D {\bf k}}{(2\pi)^D}\oint_{\mathcal{C}}d\omega \frac{\partial}{\partial \omega}\log(G({\bf k},\,\omega))\notag\\
=&\frac{i}{2\pi}\int \frac{d^D {\bf k}}{(2\pi)^D}\oint_{\mathcal{C}}d\omega \frac{\partial}{\partial \omega}\log\left(\frac{G({\bf k},\,\omega)}{G^R({\bf k},\,\omega)}\right)
\end{align}
Note that, if $\omega>0$, then $G^R({\bf k},\,\omega)=G({\bf k},\,\omega)$, while $G^R({\bf k},\,\omega)=G^*({\bf k},\,\omega)$ for $\omega<0$. Therefore, 

\begin{align}
&\frac{i}{2\pi}\int \frac{d^D {\bf k}}{(2\pi)^D}\oint_{\mathcal{C}}d\omega \frac{\partial}{\partial \omega}\log\left(\frac{G({\bf k},\,\omega)}{G^R({\bf k},\,\omega)}\right)\notag\\
=&\frac{i}{2\pi}\int \frac{d^D {\bf k}}{(2\pi)^D}\oint_0^\infty d\omega \frac{\partial}{\partial \omega}\log\left(\frac{G({\bf k},\,\omega)}{G({\bf k},\,\omega)}\right)+\frac{i}{2\pi}\int \frac{d^d {\bf k}}{(2\pi)^d}\oint_{-\infty}^0 d\omega \frac{\partial}{\partial \omega}\log\left(\frac{G({\bf k},\,\omega)}{G^*({\bf k},\,\omega)}\right)\notag\\
=&\frac{i}{2\pi} \int \frac{d^D {\bf k}}{(2\pi)^D}\log\left(\frac{G({\bf k},\,\omega)}{G^*({\bf k},\,\omega)}\right)\bigg|_{-\infty}^0
\end{align}
If we write $G({\bf k},\,\omega)=e^{i \phi(\omega)}|G({\bf k},\,\omega)|$, we find that

\begin{align}
\frac{i}{2\pi} \int \frac{d^D {\bf k}}{(2\pi)^D}\log\left(\frac{G({\bf k},\,\omega)}{G^*({\bf k},\,\omega)}\right)\bigg|_{-\infty}^0
&=\frac{i}{2\pi} \int \frac{d^D {\bf k}}{(2\pi)^D}\log\left(e^{2i\phi(\omega)}\right)\bigg|_{-\infty}^0\notag\\
&=-\frac{1}{\pi}\int \frac{d^D {\bf k}}{(2\pi)^D } \left\{
\phi(0)-\phi(-\infty)
\right\}
\end{align}
We therefore find that, as discussed in the main body of the text, that a key component of Luttinger's theorem is dependent upon the phase $\phi(\omega)$ of the Green's function. Note that $\Im(G({\bf k},\,\omega))>0$ when $\omega<0$, with $\Im(G({\bf k},\,\omega))=0$ for $\omega=0$. $\Im(G({\bf k},\,\omega))$ does not change sign, so we are confined in the upper half plane. Similarly, as $\omega\rightarrow -\infty$, $\Im(G({\bf k},\,\omega))$ falls off more rapidly then $\Re(G({\bf k},\,\omega))$, because $G({\bf k},\,\omega)\sim 1/\omega$ for $\omega\rightarrow \pm \infty$. This directly follows from the K\"allen-Lehmann representation,
\begin{align}
G({\bf k},\,\omega)=(2\pi)^3 \sum_s \left(
\frac{A_s \delta({\bf k}-{\bf k}_s)}{\omega-\epsilon_s^++\mu+i0} 
\pm \frac{B_s \delta({\bf k}+{\bf k}_s)}{\omega-\epsilon_s^-+\mu-i0}
\right)
\end{align}
where the information from the self energy is contained in $A_s$ and $B_s$. Therefore, the ratio of imaginary to real parts of the Green's function goes to $0$. Now, from our definition of the phase above, we can easily see that $\Re(G({\bf k},\,\omega))=\cos(\phi(\omega))|G(k,\,\omega)|$, while $\Im(G(k,\,\omega))=\sin(\phi(\omega))|G(k,\,\omega)|$. Hence,
\begin{align}
\frac{\Im(G({\bf k},\,\omega))}{\Re(G({\bf k},\,\omega))}=\tan(\phi(\omega))
\end{align}
For this to go to zero as $\omega\rightarrow -\infty$, $\phi(-\infty)=\pi$. This modifies Eqn. \eqref{12a} to the simpler form 

\begin{align}
\frac{i}{2\pi}\int \frac{d^D {\bf k}}{(2\pi)^D}\oint_{\mathcal{C}}d\omega \frac{\partial}{\partial \omega}\log(G({\bf k},\,\omega))=\int \frac{d^D k}{(2\pi)^D}\left\{
1-\frac{\phi(0)}{\pi}
\right\}
\end{align}
We are now left to solve for the phase of the Green's function at low frequency. Note that the above yields the well-known solution if we assume that the imaginary part of the Green's function "disappears" faster than the real component in the limit of $\omega\rightarrow 0$, which is equivalent to saying that the imaginary part of the self energy disappears faster than the real part in this limit. If this occurs, then $\tan(\phi(0))=0$, which occurs when $\phi(0)=0$ or $\phi(0)=\pi$. The former case corresponds to $G({\bf k},\,0)>0$, or when we are below the Fermi surface, while the latter case corresponds to $G({\bf k},\,0)<0$ or when we are below the Fermi surface. Therefore,

\begin{align}
-\frac{1}{\pi}\int \frac{d^D k}{(2\pi)^D } \left\{
1-\frac{\phi(0)}{\pi}
\right\}
&=\int \frac{d^D k}{(2\pi)^D } 
\Theta(k_F-k)
\notag\\
&=\frac{1}{(2\pi)^D}\int_{G({\bf k},\,\omega=0)>0}d^D {\bf k}\label{eq27}
\end{align}
We then left with proving that the imaginary part of the self energy disappears faster than the real part at small frequency. This can be seen in the case of Landau-Fermi liquid by finding the explicit frequency-dependence of $\Im \Sigma({\bf k},\,\omega)$, which can be done by first finding the lifetime of a quasiparticle near the Fermi surface. If we considered a free particle, the Green's function would be given by

\begin{align}
G^{(0)}({\bf k},\,t)=-i\Theta(t)e^{-i\xi_k t}
\end{align}
It is well known that the spectral function of the above is a perfect delta function. When considering a Landau quasiparticle, we include an additional component proportional to the quasiparticle lifetime $\tau$:

\begin{align}
G({\bf k},\,t)=-i\Theta(t)e^{-i\xi_k t}e^{-t/\tau}
\end{align}
The spectral function of the above is a "widened" delta function with width $1/\tau$:
\begin{align}
A({\bf k},\,\omega)=\frac{1/\tau}{(\omega-\xi_k)^2+(1/\tau)^2}
\end{align}
Compare this with the form of the spectral function from the full K\"allen-Lehmann representation:
\begin{align}
A({\bf k},\,\omega)=-\frac{1}{\pi}\frac{\Im \Sigma({\bf k},\,\omega)}{(\omega-\xi_k)^2+\Im \Sigma({\bf k},\,\omega)^2}
\end{align}
Hence, we can easily see that $\Im \Sigma({\bf k},\,\omega)\sim 1/\tau$. Therefore, by calculating the lifetime, we can find the dependence of the self-energy on the frequency $\omega$. This can be done by writing down Fermi's golden rule to find the decay rate towards $n$ particle-hole pairs and subsequently replacing the scattering amplitude with a Fermi surface average\footnote{See \cite{Coleman_book} for a complete discussion of this derivation from Fermi's golden rule.}:

\begin{align}
W_{2n+1}(\omega_1)&\sim 2\pi\sum_{2,\,3,\,...,\,2n+2} |a|^2 \delta\left(
\omega_1-(\omega_2'+\omega_3+\omega_4'+...+\omega_{2n+2})
\right)\notag\\
&=2\pi \sum_{2,\,3,\,...,\,2n+2} \prod_{i=1}^{2n+1} \int_{-\infty}^\infty d\omega_i \delta(\omega_i)|a|^2 \delta\left(
\omega_1-(\omega_2'+\omega_3+\omega_4'+...+\omega_{2n+2})
\right)\notag\\
&=2\pi \langle |a_{2n+1}|^2\rangle\int_0^{\infty} d\omega_2' d\omega_3...d\omega_{2n+1} \delta\left(
\omega_1-(\omega_2'+\omega_3+\omega_4'+...+\omega_{2n+2})
\right)\notag\\
&\sim \frac{\omega^{2n}}{(2n)!}
\end{align} 
where the primed terms denote the quasihole energies and we inserted the identity into the second line. Therefore, because the inverse of the lifetime $1/\tau$ is given by the sum of all possibly allowed decay possibilities, $\Im \Sigma(k,\,\omega)\sim \omega^2$ for the special case of a Landau-Fermi liquid (where this quasiparticle picture makes sense).  From the Kramers-Kronig relation given in Eqn \eqref{eq11}, it is then obvious that $\Re \Sigma({\bf k},\,\omega)\sim \omega$, and thus the imaginary part of the self energy disappears faster than the real component whenever a Landau quasiparticle picture is applicable, and thus Eqn. \eqref{eq27} remains valid. 

We now move onto the second integral, which is easier to solve:

\begin{align}
\int d^D k \oint_{\mathcal{C}'}G \frac{\partial}{\partial \omega}\Sigma({\bf k},\,\omega)&=\int d^d k \left(G({\bf k},\,\omega)\Sigma({\bf k},\,\omega)\bigg|_{-\infty}^\infty-\oint_{\mathcal{C}'}\Sigma({\bf k},\,\omega) \frac{\partial}{\partial \omega}G({\bf k},\,\omega) \right)\notag\\
&=-\oint_{\mathcal{C}'}\Sigma({\bf k},\,\omega) \frac{\partial}{\partial \omega} G({\bf k},\,\omega)
\end{align}
If we can write the self energy as an exact differential (as in Eqn. \eqref{12c}), then the above integral disappears. To ensure this, we have to make sure that the self energy $\Sigma({\bf k},\,\omega)$ doesn't exhibit any divergent frequency dependence. However, we have already proven this when solving for the previous integral! Therefore, we automatically have a well-defined Luttinger-Ward functional for the Landau-Fermi liquid, and we are left with

\begin{align}
\int d^D k \oint_{\mathcal{C}'}G \frac{\partial}{\partial \omega}\Sigma({\bf k},\,\omega)&=0
\end{align}
Therefore, we come to the final form of Luttinger's theorem:
\begin{align}
\frac{N}{2V}&=-i\int \frac{ d^D {\bf k}}{(2\pi)^D}\oint_{\mathcal{C}} \frac{d \omega}{2\pi}G({\bf k},\,\omega)\notag\\
&=-i\int \frac{d^D {\bf k}}{(2\pi)^D} \oint_{\mathcal{C}}\frac{d\omega}{2\pi} \left\{ -\frac{\partial}{\partial \omega}\log G({\bf k},\,\omega)+G(k,\,\omega)\frac{\partial}{\partial \omega}\Sigma({\bf k},\,\omega)
\right\}\notag\\
&=\frac{1}{(2\pi)^D}\int_{G({\bf k},\,\omega=0)>0}d^D {\bf k}
\end{align}
The proof of Luttinger's theorem in the text is built form of the above derivation. From the calculation of Fermi's golden rule, we see that all non-Fermi liquid behavior is contained within the frequency dependence of the imaginary part of the self energy, where a deviation from the $\omega^2$ behavior can be interpreted as a breakdown of the quasiparticle paradigm. Analogously, also note that only in the quasiparticle approximation of Fermi's golden rule did we assume any perturbative approximation in our derivation. In this way, we can apply the above to any system with a non-analytic self energy as long as we take the diverging form {\it after} we perform the calculation and, as explained in the text, only if the non-analyticity is purely in the momentum-dependence of $\Sigma({\bf k},\,\omega)$.

\subsection{Derivation of the Kadanoff-Baym functional}

We will now illustrate how to obtain the form of the snark by solving for the general form of the Kadanoff-Baym functional. We will primarily follow the derivation given in \cite{Tremblay2}. For this reason, we will take the same notation and define arguments $(n)=({\bf x}_n,\,\tau_n;\,\sigma_n)$, while an overbar means integrals over the space-time coordinates and spin sums

We define the quantum action $W[J]$ in terms of the partition function: 
\begin{align}
Z[K]=\exp \left(W[K]\right)\rightarrow W[L]=\log Z[K]
\end{align}
where $K$ is some source field and for simplicity we omit $J$.\footnote{Note that in Baym's original 1962 paper, he uses $U$ instead of $K$. We will keep using $J$ here to be consistent with contemporary notation.} Therefore, the Green function $G(2,\,1)$ is given by the functional derivative of $W$ with respect to the external field:
\begin{align}
\frac{\delta W[K]}{\delta K(1,\,2)}=G(2,\,1)
\end{align}
The effective action $\Gamma[G]$ is then the Legendre transform of the above:

\begin{align}
\Gamma[G]=W[K]-\textrm{Tr}[K G]
\end{align}
The functional $\Gamma[G]$ is the Kadanoff-Baym functional. The existence is dependent on the existence of a renormalization group picture and an appropriate cutoff $\Lambda$, such that $\Gamma_{k\rightarrow \Lambda}\approx S$ and $\Gamma_{k\rightarrow 0}=\Gamma$, where $S$ is the un-quantized action. 
We can similarly find that

\begin{align}
\frac{\delta \Gamma[G]}{\delta G(1,\,2)}=-K(2,\,1)
\end{align}
To proceed, we need to simplify the r.h.s of the above. For this, we utilize the equation of motion for the Green's function:
\begin{align}
\left(\frac{\partial}{\partial \tau_1}-\frac{\nabla_1^2}{2m}-\mu\right)G(1,\,2)=-\delta(1-2)
+\langle
T[
\psi^\dagger (\bar{2^+})V(1-\bar{2})\psi(\bar{2})\psi(1)\psi^\dagger (2)
]
\rangle-K(1,\,\bar{2})G(\bar{2},\,2)
\end{align}
Now, let us define the inverse of the non-interacting Green function to be

\begin{align}
G_0^{-1}(1,\bar{2})=-\left(
\frac{\partial}{\partial \tau_1}-\frac{\nabla_1^2}{2m}-\mu
\right)\delta(1-\bar{2})
\end{align}
%
Plugging this into the equation of motion and simplifying, we have 

\begin{align}
\left(G_0^{-1}(1,\,\bar{2})-K(1,\,\bar{2})\right)G(\bar{2},\,2)=\delta(1-2)-\langle
T[
\psi^\dagger (\bar{2^+})V(1-\bar{2})\psi(\bar{2})\psi(1)\psi^\dagger (2)
]
\rangle
\end{align}
If we take the limit of $K=0$, then we see that the above corresponds to the Dyson equation if we take 

\begin{align}
&\Sigma(1,\,\bar{2})G(\bar{2},\,2)=-V(1-\bar{2})\langle 
T[
\phi^2(\bar{2^+})\psi(\bar{2})\psi(1)\psi^\dagger (2)
]
\rangle\notag\\
&\rightarrow
\left(G_0^{-1}(1,\,\bar{2})-K(1,\,\bar{2})-\Sigma(1,\,\bar{2})\right)G(\bar{2},\,2)=\delta(1-2)
\end{align}
Which can be re-written to give

\begin{align}
G^{-1}(1,\,2)=G_0^{-1}(1,\,2)-K(1,\,2)-\Sigma(1,\,2)
\end{align}
This simplifies the above equation for $\Gamma[G]$ by solving the above to represent $J(1,\,2)$ in terms of Green's functions and the self energy:

\begin{align}
\frac{\delta \Gamma[G]}{\delta G(1,\,2)}&=G^{-1}(2,\,1)-G_0^{-1}(2,\,1)+\Sigma(2,\,1)
\end{align}
which yields zero in the limit of $K=0$, as in that case Dyson's equation is exact. 

We can now solve the above for the Baym-Kadanoff functional. It is easy to see that the above becomes

\begin{align}
\delta \Gamma[G]=\textrm{Tr}[G^{-1}\delta G]-\textrm{Tr}[
G_0^{-1}\delta G
]
+
\textrm{Tr}[\Sigma \delta G]
\end{align}
where we have dropped the argument $(2,\,1)$ for conciseness.
For the first term, we perform the simplification

\begin{align}
\textrm{Tr}[G^{-1}\delta G]&=-\textrm{Tr}[G\delta G^{-1}]\notag\\
&=\textrm{Tr}[\delta \log (-G)]
\end{align}
which can be seen with simple calculus. For the second term, we simplify it the following way:

\begin{align}
\textrm{Tr}[G_0^{-1}\delta G]&=\textrm{Tr}[\delta(
G_0^{-1}G-1)
]\notag\\
&=\textrm{Tr}[
\delta \left\{(G_0^{-1}-G^{-1})G\right\}
]
\end{align}
For the final term, recall that we can write the change of the Luttinger-Ward functional $\Phi$ as
\begin{align}
\delta \Phi&=\int d1\,d1'\frac{\delta \Phi}{\delta G(1,\,1')}\delta G(1',\,1^+)\notag\\
&=\int d1\,d1'\Sigma(1,\,1')\delta G(1',\,1^+)
\end{align}
\\
which is $\textrm{Tr}[\Sigma \delta G]$. Hence, the final term is just $\delta \Phi$. This tells us that the differential of the Kadanoff-Baym functional is just

\begin{align}
\delta \Gamma[G]=\delta \Phi-\textrm{Tr}[\delta \left\{
(G_0^{-1}-G^{-1})G
\right\}   ]+\textrm{Tr}[\delta \log(-G)]
\end{align}
The solution for the Kadanoff-Baym functional is now trivial:

\begin{align}
\Gamma[G]\approx \Phi-\textrm{Tr}[(G_0^{-1}-G^{-1})G]+\textrm{Tr}[\log (-G)]
\end{align}
\\
From which Eqn. \eqref{eq7} directly follows.
%
Interestingly, because we have included the Luttinger-Ward functional and we assume that it is well-behaved, the above result for the effective action is {\it exact} as long as the Luttinger-Ward functional exists. 

\subsection{Derivation of self energy dependence on density of states}

The main result of this paper concerns when the eigenstates of a generic many-body system can be approximated as the collective behavior of independent, interacting particles. This ultimately boils down to studying the regime of validity of a "hard" version of Luttinger's theorem, which we have shown is restricted to models with $\Im \Sigma({\bf k},\,\omega)\sim \omega^\alpha$ where $\alpha\ge 1$. Experimentally, ARPES data tells us that this corresponds to either the overdoped phase (for $\alpha>1$) or the optimally-doped "strange-metal" phase ($\alpha=1$), both of which respects Luttinger's theorem, while the cases of $0<\alpha<1$ (the pseudogap phase) and $\alpha<0$ (the insulating phase) violate Luttinger's theorem. In this appendix we will briefly show that the condition $\alpha\ge 1$ is analogous to the existence of a well-defined, non-zero density of states at ${\bf k}={\bf k}_F$ and $\omega=0$. Because this is also the frequency regime where the density of states is always well-defined, this further confirms that the snark definition given in Eqn. \eqref{eq9} is a necessary and sufficient condition for the validity of Luttinger's theorem by the Atiyah-Singer index theorem. 

The general definition of density of states $\rho(\omega)$ for a many-body fermionic system is given by\cite{Abrikosov}

\begin{align}
\rho(\omega) \sim-\frac{1}{\pi}\int\,dr\,\Im G({\bf r},\,{\bf r},\,\omega)
\end{align}

\noindent When we're at the Fermi energy, we can write the Green's function as
\begin{align}
G({\bf k}_F,\,\omega)=\frac{1}{\omega-\Sigma({\bf k},\,\omega)+i\delta}
\end{align}
where we restrict ourselves to the surface defined by the non-interacting Fermi momentum ${\bf k}_F$.

As discussed in the text, a Kramers-Kronig relation connects the imaginary and real parts of the self energy. ARPES data and the requirements of Luttinger's theorem indicates two specific cases (excluding the marginal Fermi liquid), as outlined in Eqn. \eqref{eq11}. The first case we will consider is $\Im \Sigma({\bf k},\,\omega)\sim \omega^\alpha$, where $\alpha>1$. In this case, $\Re \Sigma({\bf k},\,\omega)\sim \omega$. The imaginary part of the Green's function is then given by

\begin{align}
\Im G({\bf k}_F,\,\omega)&=\begin{cases}
\dfrac{\omega^\alpha-\delta}{\delta^2+\omega^2-2 \omega^\alpha \delta +\omega^{2\alpha}},\quad &\omega>0\notag\\
\notag\\
\dfrac{|\omega|^\alpha\cos(\pi \alpha)-\delta}{
\delta^2+\omega^2-2 |\omega|^\alpha \cos(\pi \alpha)\delta+2\textrm{sgn}(\omega)|\omega|^{\alpha+1} \sin(\pi \alpha)+|\omega|^{2\alpha}
},\quad &\omega<0
\end{cases}
\end{align}
\\
Because we are considering the limit $\omega\rightarrow 0$, $\delta\rightarrow 0$, we can ignore terms higher than $\mathcal{O}(\omega^2)$, $\mathcal{O}(\delta^2)$, or $\mathcal{O}(\omega \delta)$. Singularities then occur when $\omega\rightarrow \pm i\delta$. Note that both cases have zeroes when $\omega\sim \delta^{1/\alpha}$, however they are of little concern because $\alpha>1$, and therefore the singularity will always be "closer" to $\omega=0$ then the zero $\forall\, \delta<1$. We can then conclude that the limit approaching the singularity is well-defined, and hence there exists a well-defined density of states (and, hence, gapless chiral excitations) for $\Im \Sigma({\bf k},\,\omega)\sim \omega^\alpha$ if $\alpha>1$, as already suggested by ARPES data.

The more interesting case occurs when $\alpha<1$. This corresponds to $\Im \Sigma({\bf k},\,\omega)\sim\Re \Sigma({\bf k},\,\omega)\sim \omega^\alpha$. The imaginary component of the Green's function then becomes

\begin{align}
&\Im G({\bf k}_F,\,\omega)\notag\\
=&\begin{cases}
&\dfrac{\omega^\alpha-\delta}{\delta^2+\omega^2-2 \omega^{2\alpha}\delta-2\omega^{1+\alpha}+2\omega^{2\alpha}},\quad \omega>0\notag\\
&\notag\\
&\dfrac{|\omega|^\alpha \{\cos(\pi \alpha)+\sin(\pi \alpha)\}-\delta}{
\delta^2 +\omega^2 -2 |\omega|^\alpha \delta\{\cos(\pi \alpha)+\sin(\pi \alpha)\}+2|\omega|^{\alpha+1}\{\cos (\pi \alpha)-\sin(\pi \alpha)\}+2|\omega|^{2\alpha}},\quad \omega<0
\end{cases}
\end{align}
\\
We first deal with the regime of $0<\alpha<1$. Under such circumstances, we can ignore terms that go larger than $\mathcal{O}(\omega^\alpha)$. We can easily see that, for both $\omega>0$ and $\omega<0$, singularities and zeroes of the imaginary part of the Green's function occur when $\omega\sim \pm \delta^{1/\alpha}$. Therefore, taking the limit $\omega\rightarrow 0$, $\delta\rightarrow 0$ no longer {\color{black} guarantees a sharp singularity}, and a {\color{black} finite} density of states {\color{black} at $\omega=0$} is not {\color{black} universally observed as it was} for $\alpha>1$. Once again, this agrees with ARPES data, as $0<\alpha<1$ corresponds to the pseudogap state where a partial energy gap occurs. {\color{black} This is similarly seen in a plot of the spectral function given in Figs. \ref{fig:2a}--\ref{fig:2d} in the text, where said function is seen to exhibit a discontinuous dip at $\omega=0$ for $0<\alpha<1$.}

For the case of $\alpha<0$, it is clear to see that $\Im G({\bf k}_F,\,\omega)\sim 1/|\omega|^{\alpha}$ in the limit under consideration. Because $\alpha<0$, it vanishes as we approach $\omega=0$, and thus the density of states (and, hence, gapless chiral excitations) disappears for these self energies.

\subsection{Classification of self energy momentum dependencies that yield snark solutions}

The goal of this appendix is to derive Eqn. \eqref{eq15}. Before we can do this, let's recall the classification theme we have already introduced for snarks. The {\bf order} $m$ of the snark is the lowest ${\bf k}$-derivative of the Landau quasiparticle weight $Z_k^{(0)}$ that either yields a singularity or some real number. In principle, the order could be any natural number. The {\bf kind} $n$ of the snark tells us if the $m$th order derivative diverges or not. If it diverges, then it's a snark of the second kind. If the $m$th order derivative is a real number, then it's said to be of the first kind or quasi-local. Therefore, we have the constraint $m\in [1,\,2]$ by definition of the snark's kind. If the quasiparticle weight itself is non-zero, then it's said to be a {\bf Fermi surface}. If the quasiparticle weight is vanishing, then it's said to be a {\bf Luttinger surface}.

We are now in a position to derive Eqn. \eqref{eq15}. We start by looking at the 1st-order ${\bf k}$-derivative of the quasiparticle weight of a 1st order snark of the 2nd kind:
\begin{align}
Z_k^{(1)}&=\frac{\partial Z_k^{(0)}}{\partial k}\notag\\
&=\frac{\partial^2 \Re\Sigma({\bf k},\,\omega)}{\partial k\partial \omega}\left(1-\frac{\partial \Re\Sigma({\bf k},\,\omega)}{\partial \omega}\bigg|_{\genfrac{}{}{0pt}{}{{\bf k}={\bf k}_F}{\hspace{-2mm}\omega=0}} \right)^{-2}\notag\\
&=\left(Z^{(0)}\right)^2 \frac{\partial}{\partial k}\left(\frac{\partial \Re\Sigma({\bf k},\,\omega)}{\partial \omega}\bigg|_{\genfrac{}{}{0pt}{}{{\bf k}={\bf k}_F}{\hspace{-2mm}\omega=0}} \right)
\end{align}

\noindent Remember that the first order snark has a singularity for $Z_k^{(1)}$. Because $Z_k^{(0)}$ is always bounded by one, the divergent term must be the derivative of the self energy at the Fermi energy:
\\
\begin{align}
\frac{\partial}{\partial k}\left(\frac{\partial \Re\Sigma({\bf k},\,\omega)}{\partial \omega}\bigg|_{\genfrac{}{}{0pt}{}{{\bf k}={\bf k}_F}{\hspace{-2mm}\omega=0}}\right) \rightarrow \infty
\end{align}
\\
We have assumed that the self-energy is analytic in frequency space, otherwise there will not exist a well-defined Luttinger-Ward functional. Therefore, $\partial \Re \Sigma({\bf k},\,\omega)/\partial \omega$ is some finite value, and the divergent term must be the momentum derivative. For some general non-Fermi liquid system, however, $Z_k^{(0)}\rightarrow 0$, meaning $\partial \Re \Sigma({\bf k},\,\omega)/\partial \omega \rightarrow -\infty$ as the self energy approaches the Fermi energy. If the system is a non-Fermi liquid and has such divergent behavior in the frequency derivative of the self energy, then the condition for a first order Fermi boundary is that the term
\begin{align}
\frac{\partial}{\partial k}\left(\frac{\partial \Re\Sigma({\bf k},\,\omega)}{\partial \omega}\bigg|_{\genfrac{}{}{0pt}{}{{\bf k}={\bf k}_F}{\hspace{-2mm}\omega=0}}\right)\quad \textrm{diverges faster than}\quad \left(1-\frac{\partial \Re\Sigma({\bf k},\,\omega)}{\partial \omega}\bigg|_{\genfrac{}{}{0pt}{}{{\bf k}={\bf k}_F}{\hspace{-2mm}\omega=0}}\right)^2
\end{align}
This condition will give us $\lim_{k\rightarrow k_F} Z_k^{(1)}\rightarrow \infty$, or, in other words, $\lim_{k\rightarrow k_F}(1/Z_k^{(1)})\rightarrow 0$. For our perturbative Green's function approach to make sense, it's not the frequency derivative that diverges; rather, it's the momentum dependence and momentum derivative. In other words, if the frequency derivative diverges, then the above expansion of the self energy is invalid. Instead, we are saying that the momentum derivative of the self energy must diverge faster than the self energy itself at the Fermi energy; i.e., 

\begin{align}
\lim_{k\rightarrow k_F}\left\{ \frac{\partial}{\partial k}\left(\frac{\partial \Re\Sigma({\bf k},\,\omega)}{\partial \omega}\right)\right\}^{-1}
\left(1-\frac{\partial \Re \Sigma({\bf k},\,\omega)}{\partial \omega}\right)^{2}
=0
\end{align}

\noindent Note that we have to be careful how we take the limit in the above. The residue is only well-defined as $k\rightarrow k_F$. Because $Z_k^{(m)}$ diverges for the $m$th derivative, the limit and the derivative might not commute. It is therefore implied that the above limit is taken {\it after} we take the derivative. If this is ensured, then the above defines the self-energy dependence for a 1st order Fermi surface of the 2nd kind.

 We can extend this idea to the 2nd order snark of the 2nd kind:
\begin{align}
Z_k^{(2)}&=\frac{\partial Z_k^{(1)}}{\partial k}\notag\\
&=\frac{\partial}{\partial k}\left\{
\left(Z_k^{(0)}\right)^2 \frac{\partial}{\partial k}\left(\frac{\partial \Re\Sigma({\bf k},\,\omega)}{\partial \omega}\right)\bigg|_{\genfrac{}{}{0pt}{}{{\bf k}={\bf k}_F}{\hspace{-2mm}\omega=0}}
\right\}\notag\\
&=2Z^{(0)}\frac{\partial Z_k^{(0)}}{\partial k}\frac{\partial}{\partial k}\left(\frac{\partial \Re \Sigma({\bf k},\,\omega)}{\partial \omega}\bigg|_{\genfrac{}{}{0pt}{}{{\bf k}={\bf k}_F}{\hspace{-2mm}\omega=0}}\right)+\left(Z_k^{(0)}\right)^2 \frac{\partial^2}{\partial k^2}\left(\frac{\partial \Re\Sigma({\bf k},\,\omega)}{\partial \omega}\bigg|_{\genfrac{}{}{0pt}{}{{\bf k}={\bf k}_F}{\hspace{-2mm}\omega=0}}\right)\notag\\
&=2Z_k^{(0)}Z_k^{(1)}\frac{\partial}{\partial k}\left(\frac{\partial \Re\Sigma({\bf k},\,\omega)}{\partial \omega}\bigg|_{\genfrac{}{}{0pt}{}{{\bf k}={\bf k}_F}{\hspace{-2mm}\omega=0}}\right)+\left(Z_k^{(0)}\right)^2 \frac{\partial^2}{\partial k^2}\left(\frac{\partial \Re \Sigma({\bf k},\,\omega)}{\partial \omega}\bigg|_{\genfrac{}{}{0pt}{}{{\bf k}={\bf k}_F}{\hspace{-2mm}\omega=0}}\right)\notag\\
&=2\left(Z_k^{(0)}\right)^3\left(\frac{\partial}{\partial k}\left(\frac{\partial \Re\Sigma({\bf k},\,\omega)}{\partial \omega}\bigg|_{\genfrac{}{}{0pt}{}{{\bf k}={\bf k}_F}{\hspace{-2mm}\omega=0}}\right)\right)^2+\left(Z_k^{(0)}\right)^2 \frac{\partial^2}{\partial k^2}\left(\frac{\partial \Re\Sigma({\bf k},\,\omega)}{\partial \omega}\bigg|_{\genfrac{}{}{0pt}{}{{\bf k}={\bf k}_F}{\hspace{-2mm}\omega=0}}\right)
\end{align}
This gives us two possibilities for a singular value of $Z_k^{(2)}$ as we approach the Fermi energy: either

\begin{align}
\frac{\partial}{\partial k}\left(\frac{\partial \Re\Sigma({\bf k},\,\omega)}{\partial \omega}\bigg|_{\genfrac{}{}{0pt}{}{{\bf k}={\bf k}_F}{\hspace{-2mm}\omega=0}}\right)\quad \textrm{diverges faster than} \quad \left(1-\frac{\partial \Re \Sigma({\bf k},\,\omega)}{\partial \omega}\bigg|_{\genfrac{}{}{0pt}{}{{\bf k}={\bf k}_F}{\hspace{-2mm}\omega=0}}\right)^{3/2}
\end{align}
\\
\noindent or
\\
\begin{align}
\frac{\partial^2}{\partial k^2}\left(\frac{\partial \Re \Sigma({\bf k},\,\omega)}{\partial \omega}\bigg|_{\genfrac{}{}{0pt}{}{{\bf k}={\bf k}_F}{\hspace{-2mm}\omega=0}}\right)\quad \textrm{diverges faster than} \quad \left(1-\frac{\partial \Re \Sigma({\bf k},\,\omega)}{\partial \omega}\bigg|_{\genfrac{}{}{0pt}{}{{\bf k}={\bf k}_F}{\hspace{-2mm}\omega=0}}\right)^{2}
\end{align}

\noindent One or both of these conditions is necessary for $\lim_{k\rightarrow k_F}(1/Z_k^{(2)})\rightarrow 0$. The former is a weaker condition than the case of the 1st order snark; namely, if the first order derivative diverges faster than the zeroth order derivative to the power of $2$, then it will obviously diverge faster than the zeroth order derivative to the power $3/2$. In other words, the first term tells us that a $1$st order snark is automatically a second order snark. The first expression in the above is not the defining characteristic of the $2$nd order snark. Instead, the unique condition for the $2$nd order snark is given by
\begin{align}
\lim_{\genfrac{}{}{0pt}{}{\hspace{2mm}{\bf k}\rightarrow {\bf k}_F}{\hspace{-0mm}\omega\rightarrow 0}}\left\{ \frac{\partial^2}{\partial k^2}\left(\frac{\partial \Re \Sigma({\bf k},\,\omega)}{\partial \omega}\right)\right\}^{-1}
\left(1-\frac{\partial \Re \Sigma({\bf k},\,\omega)}{\partial \omega}\right)^{2}
=0
\end{align}

We quote the next order derivative:

\begin{align}
Z_k^{(3)}&=6(Z_k^{(0)})^4 \left(\frac{\partial}{\partial k}\left(\frac{\partial \Re \Sigma({\bf k},\,\omega)}{\partial \omega}\bigg|_{\genfrac{}{}{0pt}{}{{\bf k}={\bf k}_F}{\hspace{-2mm}\omega=0}}\right)\right)^3+6(Z_k^{(0)})^3 \frac{\partial}{\partial k}\left(\frac{\partial \Re \Sigma({\bf k},\,\omega)}{\partial \omega}\bigg|_{\genfrac{}{}{0pt}{}{{\bf k}={\bf k}_F}{\hspace{-2mm}\omega=0}}\right)\frac{\partial^2}{\partial k^2}\left(\frac{\partial \Re \Sigma({\bf k},\,\omega)}{\partial \omega}\bigg|_{\genfrac{}{}{0pt}{}{{\bf k}={\bf k}_F}{\hspace{-2mm}\omega=0}}\right)\notag\\
&\phantom{=}+(Z_k^{(0)})^2\frac{\partial^3}{\partial k^3}\left(\frac{\partial \Re \Sigma({\bf k},\,\omega)}{\partial \omega}\bigg|_{\genfrac{}{}{0pt}{}{{\bf k}={\bf k}_F}{\hspace{-2mm}\omega=0}}\right)
\end{align}

This gives three conditions for divergence. Either
\\
\begin{align}
\frac{\partial}{\partial k}\left(\frac{\partial \Re \Sigma({\bf k},\,\omega)}{\partial \omega}\bigg|_{\genfrac{}{}{0pt}{}{{\bf k}={\bf k}_F}{\hspace{-2mm}\omega=0}}\right)\quad \textrm{diverges faster than} \quad \left(1-\frac{\partial \Re \Sigma({\bf k},\,\omega)}{\partial \omega}\bigg|_{\genfrac{}{}{0pt}{}{{\bf k}={\bf k}_F}{\hspace{-2mm}\omega=0}}\right)^{4/3}
\end{align}
or
\begin{align}
\frac{\partial}{\partial k}\left(\frac{\partial \Re \Sigma({\bf k},\,\omega)}{\partial \omega}\bigg|_{\genfrac{}{}{0pt}{}{{\bf k}={\bf k}_F}{\hspace{-2mm}\omega=0}}\right)\frac{\partial^2}{\partial k^2}\left(\frac{\partial \Re \Sigma({\bf k},\,\omega)}{\partial \omega}\bigg|_{\genfrac{}{}{0pt}{}{{\bf k}={\bf k}_F}{\hspace{-2mm}\omega=0}}\right)\quad \textrm{diverges faster than} \quad \left(1-\frac{\partial \Re \Sigma({\bf k},\,\omega)}{\partial \omega}\bigg|_{\genfrac{}{}{0pt}{}{{\bf k}={\bf k}_F}{\hspace{-2mm}\omega=0}}\right)^{3}
\end{align}
or
\begin{align}
\frac{\partial^3}{\partial k^3}\left(\frac{\partial \Re \Sigma({\bf k},\,\omega)}{\partial \omega}\bigg|_{\genfrac{}{}{0pt}{}{{\bf k}={\bf k}_F}{\hspace{-2mm}\omega=0}}\right)\quad \textrm{diverges faster than} \quad \left(1-\frac{\partial \Re \Sigma({\bf k},\,\omega)}{\partial \omega}\bigg|_{\genfrac{}{}{0pt}{}{{\bf k}={\bf k}_F}{\hspace{-2mm}\omega=0}}\right)^{2}
\end{align}
\\
If the system obeys the first condition, then it could also be a $1$st or $2$nd order snark, so the first condition is not unique for the $3$rd order snark. Furthermore, if some $2$nd order snark has the first order momentum derivative diverge faster than $(1-\frac{\partial \Sigma}{\partial \omega})$, then the second term is not unique for the $3$rd order Fermi boundary. Therefore, the only unique condition for the $3$rd order snark is that 
\begin{align}
\lim_{\genfrac{}{}{0pt}{}{\hspace{2mm}{\bf k}\rightarrow {\bf k}_F}{\hspace{-0mm}\omega\rightarrow 0}}\left\{ \frac{\partial^2}{\partial k^2}\left(\frac{\partial \Re \Sigma(k,\,\omega)}{\partial \omega}\right)\right\}^{-1}
\left(1-\frac{\partial \Re \Sigma(k,\,\omega)}{\partial \omega}\right)^{3}
=0
\end{align}

\noindent Without loss of generality, we can write the $m$th order derivative of $Z_k$ 

\begin{align}
Z^{(m)}_k&=m! (Z_k^{(0)})^{m+1}\left(\frac{\partial}{\partial k}\left(\frac{\partial \Re \Sigma({\bf k},\,\omega)}{\partial \omega}\bigg|_{\genfrac{}{}{0pt}{}{{\bf k}={\bf k}_F}{\hspace{-2mm}\omega=0}}\right)\right)^m+...+(Z^{(0)})^2\frac{\partial^m}{\partial k^m}\left(\frac{\partial \Re \Sigma({\bf k},\,\omega)}{\partial \omega}\bigg|_{\genfrac{}{}{0pt}{}{{\bf k}={\bf k}_F}{\hspace{-2mm}\omega=0}}\right)\notag\\
&=Z^{(m)}_k\sum_{j=1}^{m} A_m (Z_k^{(0)})^{j+1} \left(\frac{\partial}{\partial k}\left(\frac{\partial \Re \Sigma({\bf k},\,\omega)}{\partial \omega}\bigg|_{\genfrac{}{}{0pt}{}{{\bf k}={\bf k}_F}{\hspace{-2mm}\omega=0}}\right)\right)^j  \frac{\partial^{m-j}}{\partial k^{m-j}}\left(\frac{\partial \Re \Sigma({\bf k},\,\omega)}{\partial \omega}\bigg|_{\genfrac{}{}{0pt}{}{{\bf k}={\bf k}_F}{\hspace{-2mm}\omega=0}}\right)\notag\\
&=\sum_{j=1}^{m} A_m \left(1-\frac{\partial \Re \Sigma({\bf k},\,\omega)}{\partial \omega}\bigg|_{\genfrac{}{}{0pt}{}{{\bf k}={\bf k}_F}{\hspace{-2mm}\omega=0}}\right)^{-(j+1)} \left(\frac{\partial}{\partial k}\left(\frac{\partial  \Re \Sigma({\bf k},\,\omega)}{\partial \omega}\right)\right)^j  \frac{\partial^{m-j}}{\partial k^{m-j}}\left(\frac{\partial \Re \Sigma({\bf k},\,\omega)}{\partial \omega}\bigg|_{\genfrac{}{}{0pt}{}{{\bf k}={\bf k}_F}{\hspace{-2mm}\omega=0}}\right)
\end{align}
\\
\noindent where $A_m$ is some constant. The only unique constraint for some general $m$th order snark is the $j=1$ term. Therefore, the general condition for some $m$th order snark of the second kind is  
\begin{align}
\lim_{\genfrac{}{}{0pt}{}{\hspace{2mm}{\bf k}\rightarrow {\bf k}_F}{\hspace{-0mm}\omega\rightarrow 0}}\left\{ \frac{\partial^m}{\partial k^m}\left(\frac{\partial \Re \Sigma({\bf k},\,\omega)}{\partial \omega}\right)\right\}^{-1}
\left(1-\frac{\partial \Re \Sigma({\bf k},\,\omega)}{\partial \omega}\right)^{2}
=0
\end{align}
for some integer $m$. 

Of course, the above argument only makes sense for $m$th order Fermi and Luttinger surfaces of the second kind, as we have assumed that the $m$th order derivative diverges. From the form of Eqns. \eqref{60a} and \eqref{60c}, we see that $m$th order snarks of the first kind are more complicated, as their $m$th order derivative is a constant. This can trivially be quantified for the $0$th order Fermi surface of the first kind, and thus we begin with a $1$st order snark of the first kind:

\begin{align}
\Re \Sigma({\bf k},\,\omega)=\Re \Sigma(\omega)\frac{B_1}{|{\bf k}-{\bf k}_F|}
\end{align}
\\
\noindent where $B_1$ is some constant. Using this form of the self energy, we can find what constant the previously-derived equation yields:
\begin{align}
&\phantom{=}\lim_{\genfrac{}{}{0pt}{}{\hspace{2mm}{\bf k}\rightarrow {\bf k}_F}{\hspace{-0mm}\omega\rightarrow 0}}\left\{ \frac{\partial}{\partial k}\left(\frac{\partial \Re \Sigma({\bf k},\,\omega)}{\partial \omega}\right)\right\}^{-1}
\left(1-\frac{\partial \Re \Sigma({\bf k},\,\omega)}{\partial \omega}\right)^{2}\notag\\
&=\lim_{\genfrac{}{}{0pt}{}{\hspace{2mm}{\bf k}\rightarrow {\bf k}_F}{\hspace{-0mm}\omega\rightarrow 0}}\left\{ \frac{\partial}{\partial k}\left(\frac{B_1}{|{\bf k}-{\bf k}_F|}\frac{\partial \Re \Sigma(\omega)}{\partial \omega}\right)\right\}^{-1}
\left(1-\frac{B_1}{|{\bf k}-{\bf k}_F|}\frac{\partial \Sigma(\omega)}{\partial \omega}\right)^{2}\notag\\
&=\lim_{\genfrac{}{}{0pt}{}{\hspace{2mm}{\bf k}\rightarrow {\bf k}_F}{\hspace{-0mm}\omega\rightarrow 0}} \left\{
-\frac{B_1}{|{\bf k}-{\bf k}_F|^2}\frac{\partial \Re \Sigma(\omega)}{\partial \omega}
\right\}^{-1}\left(1-\frac{B_1}{|{\bf k}-{\bf k}_F|}\frac{\partial \Re \Sigma(\omega)}{\partial \omega}\right)^{2}\notag\\
&=\lim_{\genfrac{}{}{0pt}{}{\hspace{2mm}{\bf k}\rightarrow {\bf k}_F}{\hspace{-0mm}\omega\rightarrow 0}} \left(
-\frac{|{\bf k}-{\bf k}_F|^2}{B_1} \frac{1}{\partial \Sigma(\omega)/\partial \omega}
\right)\left(
1-\frac{2B_1}{|{\bf k}-{\bf k}_F|}\frac{\partial \Re \Sigma(\omega)}{\partial \omega}
+\frac{B_1^2}{|{\bf k}-{\bf k}_F|^2}\left(\frac{\partial \Re \Sigma(\omega)}{\partial \omega}\right)^2
\right)\notag\\
&=\lim_{\genfrac{}{}{0pt}{}{\hspace{2mm}{\bf k}\rightarrow {\bf k}_F}{\hspace{-0mm}\omega\rightarrow 0}}\left\{
-\frac{|{\bf k}-{\bf k}_F|^2}{B_1}\frac{1}{\partial \Sigma(\omega)/\partial \omega}
+2|{\bf k}-{\bf k}_F|-B_1 \frac{\partial \Re \Sigma(\omega)}{\partial \omega}
\right\}\notag\\
&=-B_1 \frac{\partial \Re \Sigma(\omega)}{\partial \omega}
\end{align}

All higher derivatives are clearly zero. However, we have to be careful here because we have singular behavior in $Z_k$ (or $1/Z_k$). In the above calculation, the limit and derivative are interchangeable, as $Z_k$ is just some constant. This is easily seen from a back-of-the envelope calculation where we take the limit $\lim_{k\rightarrow k_F}Z_k$ first:

\begin{align}
\lim_{k\rightarrow k_F} Z_k^{(1)}&=\lim_{\genfrac{}{}{0pt}{}{\hspace{2mm}{\bf k}\rightarrow {\bf k}_F}{\hspace{-0mm}\omega\rightarrow 0}} \left(1-\frac{\partial \Re \Sigma(\omega)}{\partial \omega}\frac{B_1}{|{\bf k}-{\bf k}_F|}\right)^{-1}\notag\\
&\sim -\lim_{\genfrac{}{}{0pt}{}{\hspace{2mm}{\bf k}\rightarrow {\bf k}_F}{\hspace{-0mm}\omega\rightarrow 0}}\frac{1}{B_1 \partial \Re \Sigma(\omega)/\partial \omega} |{\bf k}-{\bf k}_F|
\end{align}
Thus, the first derivative is a constant. However, for higher derivatives, we see that the above is zero. Because $Z_k^{(1)}$ is only defined near $k_F$, we take the above solution for higher derivatives, and hence $Z_k^{(1)}=0$ for higher derivatives, rather than $1/Z_k^{(1)}=0$ as implied when we took the derivative first. 

Following Eqn. \eqref{60a} and \eqref{60c}, we can now suggest a form of the self energy for 2nd order snarks of the 2nd kind:
\begin{align}
\Re \Sigma(k,\,\omega)=\Re \Sigma(\omega)\frac{B_2}{|{\bf k}-{\bf k}_F|^2}
\end{align}
The quasilocal case of the snark condition is therefore given by
\begin{align}
&\phantom{=}\lim_{\genfrac{}{}{0pt}{}{\hspace{2mm}{\bf k}\rightarrow {\bf k}_F}{\hspace{-0mm}\omega\rightarrow 0}}\left\{ \frac{\partial^2}{\partial k^2}\left(\frac{\partial \Re \Sigma({\bf k},\,\omega)}{\partial \omega}\right)\right\}^{-1}
\left(1-\frac{\partial \Re \Sigma({\bf k},\,\omega)}{\partial \omega}\right)^{2}\notag\\
&=\left\{ \frac{\partial^2}{\partial k^2}\left(\frac{B_2}{|{\bf k}-{\bf k}_F|^2}\frac{\partial \Re \Sigma(\omega)}{\partial \omega}\right)\right\}^{-1}
\left(1-\frac{B_2}{|k-k_F|^2}\frac{\partial \Re \Sigma(\omega)}{\partial \omega}\right)^{2}\notag\\
&=\lim_{\genfrac{}{}{0pt}{}{\hspace{2mm}{\bf k}\rightarrow {\bf k}_F}{\hspace{-0mm}\omega\rightarrow 0}} \left\{
\frac{6B_2}{|{\bf k}-{\bf k}_F|^4}\frac{\partial \Re \Sigma(\omega)}{\partial \omega}
\right\}^{-1}\left(1-\frac{B_2}{|{\bf k}-{\bf k}_F|^2}\frac{\partial \Re \Sigma(\omega)}{\partial \omega}\right)^{2}\notag\\
&=\lim_{\genfrac{}{}{0pt}{}{\hspace{2mm}{\bf k}\rightarrow {\bf k}_F}{\hspace{-0mm}\omega\rightarrow 0}} \left(
\frac{6|{\bf k}-{\bf k}_F|^4}{B_2} \frac{1}{\partial \Re \Sigma(\omega)/\partial \omega}
\right)\left(
1-\frac{2B_2}{|{\bf k}-{\bf k}_F|^2}\frac{\partial \Re \Sigma(\omega)}{\partial \omega}
+\frac{B_2^2}{|k-k_F|^4}\left(\frac{\partial\Re \Sigma(\omega)}{\partial \omega}\right)^2
\right)\notag\\
&=\lim_{\genfrac{}{}{0pt}{}{\hspace{2mm}{\bf k}\rightarrow {\bf k}_F}{\hspace{-0mm}\omega\rightarrow 0}}\left\{
\frac{6|{\bf k}-{\bf k}_F|^4}{B_2}\frac{1}{\partial \Re \Sigma(\omega)/\partial \omega}
-12|{\bf k}-{\bf k}_F|^2+6B_2 \frac{\partial \Re \Sigma(\omega)}{\partial \omega}
\right\}\notag\\
&=6B_2 \frac{\partial \Re \Sigma(\omega)}{\partial \omega}
\end{align}

\noindent Note that the first derivative diverges, while all higher derivatives are zero. However, from the previous discussion, we know that all higher derivatives of $Z_k^{(2)}$ are, in fact, zero, from the subtle issue of interchanging derivatives and limits. The first derivative of the above also goes to zero.
%
%
In general, the condition for the $m$th order Fermi/Luttinger surface of the first kind becomes 
\begin{align}
\lim_{\genfrac{}{}{0pt}{}{\hspace{2mm}{\bf k}\rightarrow {\bf k}_F}{\hspace{-0mm}\omega\rightarrow 0}}\left\{ \frac{\partial^m}{\partial k^m}\left(\frac{\partial \Re \Sigma({\bf k},\,\omega)}{\partial \omega}\right)\right\}^{-1}
\left(1-\frac{\partial \Re \Sigma({\bf k},\,\omega)}{\partial \omega}\right)^{2}&=m! {-m\choose m}B_m \frac{\partial \Re\Sigma(\omega)}{\partial \omega}
\notag\\
&=(-1)^m \frac{(2m-1)!}{(m-1)!}B_m \frac{\partial \Re \Sigma(\omega)}{\partial \omega}
\end{align}
The form of Eqn. \eqref{eq15} follows by noting that terms in the above expression with parameters $j<m$ are allowed for $m$th order snarks of the 1st kind. As such, we see that Eqn. \eqref{eq15} is the sole behavior the self energy $\Sigma({\bf k},\,\omega)$ must observe if a snark is to be present and, hence, Luttinger's theorem preserved.

\newpage 
\newpage

\bibliographystyle{iopart-num}
\bibliography{main}{}

\providecommand{\newblock}{}
\begin{thebibliography}{100}
\expandafter\ifx\csname url\endcsname\relax
  \def\url#1{{\tt #1}}\fi
\expandafter\ifx\csname urlprefix\endcsname\relax\def\urlprefix{URL }\fi
\providecommand{\eprint}[2][]{\url{#2}}

\bibitem{Schrieffer}
Schrieffer J~R 1970 {\em Journal of Research of the National Bureau of
  Standards-A. Physics and Chemistry\/} {\bf 74A}(4) 537--541 invited paper
  presented at the 3d Materials Research Symposium, Electronic Density of
  States , November 3-6,1969, Gaithersburg, Md.
  \urlprefix\url{https://nvlpubs.nist.gov/nistpubs/jres/74A/jresv74An4p537_A1b.pdf}

\bibitem{Pauli}
Pauli W 1961 "zur {\"a}lteren und neueren geschichte des neutrinos" {\em
  Aufs{\"a}dtze und Vortr{\"a}ge {\"u}ber Physik und Erkenntnistheorie\/} ed
  Pauli W (Zurich: Fr: Vieweg \& Sohn) pp 156--180

\bibitem{Glashow}
Glashow S~L 1961 {\em Nuclear Physics\/} {\bf 22} 579 -- 588 ISSN 0029-5582
  \urlprefix\url{http://www.sciencedirect.com/science/article/pii/0029558261904692}

\bibitem{Guralnik}
Guralnik G~S, Hagen C~R and Kibble T~W~B 1964 {\em Phys. Rev. Lett.\/} {\bf
  13}(20) 585--587
  \urlprefix\url{https://link.aps.org/doi/10.1103/PhysRevLett.13.585}

\bibitem{Weinberg}
Weinberg S 1967 {\em Phys. Rev. Lett.\/} {\bf 19}(21) 1264--1266
  \urlprefix\url{https://link.aps.org/doi/10.1103/PhysRevLett.19.1264}

\bibitem{Landau1}
Landau L 1956 {\em JETP\/} {\bf 3}(6) 1058
  \urlprefix\url{http://www.jetp.ac.ru/cgi-bin/e/index/e/3/6/p920?a=list}

\bibitem{Benfatto}
Benfatto G and Gallavotti G 1990 {\em Phys. Rev. B\/} {\bf 42}(16) 9967--9972
  \urlprefix\url{https://link.aps.org/doi/10.1103/PhysRevB.42.9967}

\bibitem{Polchinski}
Polchinski J 1992 {\em Effective Field Theory and the Fermi Surface\/} vol~4
  (Proceedings, "Recent directions in particle theory", Boulder TASI)
  \urlprefix\url{https://arxiv.org/abs/hep-th/9210046}

\bibitem{Shankar1}
Shankar R 1994 {\em Rev. Mod. Phys.\/} {\bf 66}(1) 129--192
  \urlprefix\url{https://link.aps.org/doi/10.1103/RevModPhys.66.129}

\bibitem{Chitov}
Chitov G~Y and S\'en\'echal D 1998 {\em Phys. Rev. B\/} {\bf 57}(3) 1444--1456
  \urlprefix\url{https://link.aps.org/doi/10.1103/PhysRevB.57.1444}

\bibitem{Shankar2}
Shankar R 2011 {\em Phil. Trans. R. Soc. A\/} {\bf 369} 2612--2624
  \urlprefix\url{https://royalsocietypublishing.org/doi/pdf/10.1098/rsta.2010.0385}

\bibitem{Georgi}
Georgi H 2007 {\em Phys. Rev. Lett.\/} {\bf 98}(22) 221601
  \urlprefix\url{https://link.aps.org/doi/10.1103/PhysRevLett.98.221601}

\bibitem{Georgi2}
Georgi H 2007 {\em Physics Letters B\/} {\bf 650} 275 -- 278 ISSN 0370-2693
  \urlprefix\url{http://www.sciencedirect.com/science/article/pii/S0370269307006296}

\bibitem{Nikolic}
Nikolic H 2008 {\em Mod. Phys. Lett. A\/} {\bf 23} 2645--2649
  \urlprefix\url{https://www.worldscientific.com/doi/abs/10.1142/S021773230802820X#}

\bibitem{Varma}
Varma C~M, Littlewood P~B, Schmitt-Rink S, Abrahams E and Ruckenstein A~E 1989
  {\em Phys. Rev. Lett.\/} {\bf 63}(18) 1996--1999
  \urlprefix\url{https://link.aps.org/doi/10.1103/PhysRevLett.63.1996}

\bibitem{LeBlanc}
LeBlanc J~P~F and Grushin A~G 2015 {\em New Journal of Physics\/} {\bf 17}
  033039 \urlprefix\url{https://doi.org/10.1088%2F1367-2630%2F17%2F3%2F033039}

\bibitem{Phillips5}
Leong Z, Setty C, Limtragool K and Phillips P~W 2017 {\em Phys. Rev. B\/} {\bf
  96}(20) 205101
  \urlprefix\url{https://link.aps.org/doi/10.1103/PhysRevB.96.205101}

\bibitem{Phillips1}
Phillips P~W, Langley B~W and Hutasoit J~A 2013 {\em Phys. Rev. B\/} {\bf
  88}(11) 115129
  \urlprefix\url{https://link.aps.org/doi/10.1103/PhysRevB.88.115129}

\bibitem{Kingman}
Cheung K, Keung W~Y and Yuan T~C 2007 {\em Phys. Rev. D\/} {\bf 76}(5) 055003
  \urlprefix\url{https://link.aps.org/doi/10.1103/PhysRevD.76.055003}

\bibitem{Luttinger2}
Kohn W and Luttinger J~M 1960 {\em Phys. Rev.\/} {\bf 118}(1) 41--45
  \urlprefix\url{https://link.aps.org/doi/10.1103/PhysRev.118.41}

\bibitem{Luttinger3}
Luttinger J~M and Ward J~C 1960 {\em Phys. Rev.\/} {\bf 118}(5) 1417--1427
  \urlprefix\url{https://link.aps.org/doi/10.1103/PhysRev.118.1417}

\bibitem{Luttinger1}
Luttinger J~M 1960 {\em Phys. Rev.\/} {\bf 119}(4) 1153--1163
  \urlprefix\url{https://link.aps.org/doi/10.1103/PhysRev.119.1153}

\bibitem{Farid}
Farid B 1999 {\em Philosophical Magazine B\/} {\bf 79} 1097--1143
  \urlprefix\url{https://www.tandfonline.com/doi/abs/10.1080/13642819908218309}

\bibitem{Dzyal}
Dzyaloshinskii I 2003 {\em Phys. Rev. B\/} {\bf 68}(8) 085113
  \urlprefix\url{https://link.aps.org/doi/10.1103/PhysRevB.68.085113}

\bibitem{Eisaki}
Yoshida T, Zhou X~J, Tanaka K, Yang W~L, Hussain Z, Shen Z~X, Fujimori A,
  Sahrakorpi S, Lindroos M, Markiewicz R~S, Bansil A, Komiya S, Ando Y, Eisaki
  H, Kakeshita T and Uchida S 2006 {\em Phys. Rev. B\/} {\bf 74}(22) 224510
  \urlprefix\url{https://link.aps.org/doi/10.1103/PhysRevB.74.224510}

\bibitem{Kidd}
Yang H~B, Rameau J~D, Pan Z~H, Gu G~D, Johnson P~D, Claus H, Hinks D~G and Kidd
  T~E 2011 {\em Phys. Rev. Lett.\/} {\bf 107}(4) 047003
  \urlprefix\url{https://link.aps.org/doi/10.1103/PhysRevLett.107.047003}

\bibitem{Phillips_book}
Phillips P  Chapter of the book "Quantum Criticality in Condensed Matter", pp.
  133-158, World Scientific (editor: Janusz J\c{e}drzejewski) (2015)
  \urlprefix\url{https://www.worldscientific.com/doi/abs/10.1142/9789814704090_0005}

\bibitem{Karch}
Karch A, Limtragool K and Phillips P~W 2016 {\em Journal of High Energy
  Physics\/} {\bf 2016}
  \urlprefix\url{https://link.springer.com/article/10.1007/JHEP03(2016)175}

\bibitem{Phillips4}
Limtragool K, Leong Z and Phillips P~W 2018 {\em SciPost Phys.\/} {\bf 5}(5) 49
  \urlprefix\url{https://scipost.org/10.21468/SciPostPhys.5.5.049}

\bibitem{Heath}
Heath J 2020  (Under review) (\textit{Preprint} \eprint{arXiv:2001.08230})
  \urlprefix\url{https://arxiv.org/abs/2001.08230}

\bibitem{Krastan}
Blagoev K~B and Bedell K~S 1997 {\em Phys. Rev. Lett.\/} {\bf 79}(6) 1106--1109
  \urlprefix\url{https://link.aps.org/doi/10.1103/PhysRevLett.79.1106}

\bibitem{Affleck}
Yamanaka M, Oshikawa M and Affleck I 1997 {\em Phys. Rev. Lett.\/} {\bf 79}(6)
  1110--1113
  \urlprefix\url{https://link.aps.org/doi/10.1103/PhysRevLett.79.1110}

\bibitem{Oshikawa}
Oshikawa M 2000 {\em Phys. Rev. Lett.\/} {\bf 84}(15) 3370--3373
  \urlprefix\url{https://link.aps.org/doi/10.1103/PhysRevLett.84.3370}

\bibitem{Farid3}
Farid B 2007  (\textit{Preprint} \eprint{arXiv:0711.0952v1})
  \urlprefix\url{https://arxiv.org/abs/0711.0952}

\bibitem{Rosch2}
Rosch A 2007  (\textit{Preprint} \eprint{arXiv:0711.3093v1})
  \urlprefix\url{https://arxiv.org/abs/0711.3093}

\bibitem{Farid4}
Farid B 2007  (\textit{Preprint} \eprint{arXiv:0711.3195v1})
  \urlprefix\url{https://arxiv.org/abs/0711.3195}

\bibitem{Phillips2}
Stanescu T~D, Phillips P and Choy T~P 2007 {\em Phys. Rev. B\/} {\bf 75}(10)
  104503 \urlprefix\url{https://link.aps.org/doi/10.1103/PhysRevB.75.104503}

\bibitem{Phillips3}
Dave K~B, Phillips P~W and Kane C~L 2013 {\em Phys. Rev. Lett.\/} {\bf 110}(9)
  090403
  \urlprefix\url{https://link.aps.org/doi/10.1103/PhysRevLett.110.090403}

\bibitem{Essler2}
Essler F~H~L and Tsvelik A~M 2005 {\em Phys. Rev. B\/} {\bf 71}(19) 195116
  \urlprefix\url{https://link.aps.org/doi/10.1103/PhysRevB.71.195116}

\bibitem{Rosch}
Rosch A 2007 {\em The European Physical Journal B\/} {\bf 59} 495--502 ISSN
  1434-6036 \urlprefix\url{https://doi.org/10.1140/epjb/e2007-00312-3}

\bibitem{Migdal}
Migdal A 1957 {\em JETP\/} {\bf 5}(2) 399
  \urlprefix\url{http://www.jetp.ac.ru/cgi-bin/e/index/e/5/2/p333?a=list}

\bibitem{Sachdev}
Senthil T, Sachdev S and Vojta M 2003 {\em Phys. Rev. Lett.\/} {\bf 90}(21)
  216403 \urlprefix\url{https://link.aps.org/doi/10.1103/PhysRevLett.90.216403}

\bibitem{Haldane}
Haldane F Luttinger's theorem and bosonization of the fermi surface Proceedings
  of the International School of Physics "Enrico Fermi", Course CXXI:
  "Perspectives in Many-Particle Physics",eds. R. Broglia and J. R. Schrieffer
  (North Holland, Amsterdam 1994), pp 5-30
  \urlprefix\url{https://arxiv.org/abs/cond-mat/0505529}

\bibitem{Tomonaga}
Tomonaga S~i 1950 {\em Progress of Theoretical Physics\/} {\bf 5} 544--569 ISSN
  0033-068X \urlprefix\url{https://dx.doi.org/10.1143/ptp/5.4.544}

\bibitem{Luttinger_liquid}
Luttinger J~M 1963 {\em Journal of Mathematical Physics\/} {\bf 4} 1154--1162
  \urlprefix\url{https://doi.org/10.1063/1.1704046}

\bibitem{Mattis}
Mattis D~C and Lieb E~H 1965 {\em Journal of Mathematical Physics\/} {\bf 6}
  304--312 \urlprefix\url{https://doi.org/10.1063/1.1704281}

\bibitem{Haldane_Luttinger_liquid}
Haldane F~D~M 1981 {\em Journal of Physics C: Solid State Physics\/} {\bf 14}
  2585--2609
  \urlprefix\url{https://iopscience.iop.org/article/10.1088/0022-3719/14/19/010/meta}

\bibitem{Matt}
Gochan M~P and Bedell K~S 2018 {\em Journal of Physics: Condensed Matter\/}
  {\bf 30} 445603
  \urlprefix\url{https://iopscience.iop.org/article/10.1088/1361-648X/aae3cb/meta}

\bibitem{Cornwall}
Cornwall J~M, Jackiw R and Tomboulis E 1974 {\em Phys. Rev. D\/} {\bf 10}(8)
  2428--2445 \urlprefix\url{https://link.aps.org/doi/10.1103/PhysRevD.10.2428}

\bibitem{Baym1}
Baym G 2000 {\em Conservation Laws and the Quantum Theory of Transport:. the
  Early Days\/} in "Progress in Nonequilibrium Green's Functions". Proceedings
  of the Conference "Kadanoff-Baym Equations: Progress and Perspectives for
  Many-body Physics". Held 20-24 September 1999 in Rostock, Germany. Edited by
  Michael Bonitz (Universit\"at Rostock, German). Published by World Scientific
  Publishing Co. Pte. Ltd., 2000. ISBN \#9789812793812, pp. 17-32
  \urlprefix\url{https://jfi.uchicago.edu/~leop/Kb.pdf}

\bibitem{Baym2}
Baym G and Kadanoff L~P 1961 {\em Phys. Rev.\/} {\bf 124}(2) 287--299
  \urlprefix\url{https://link.aps.org/doi/10.1103/PhysRev.124.287}

\bibitem{Tremblay2}
Tremblay A~M~T 2008 {\em Lecture notes for Cifar-PiTP International Summer
  School on Numerical Methods for Correlated Systems in Condensed Matter\/}
  \urlprefix\url{https://pitp.phas.ubc.ca/confs/sherbrooke/archives.html}

\bibitem{Polonyi}
Polonyi J and Schwenk A~e 2012 {\em Renormalization Group and Effective Field
  Theory Approaches to Many-Body Systems\/} (Heidelberg: Springer-Verlag) ISBN
  978-3-642-27319-3

\bibitem{Hagen}
Hagen K 2016 {\em Particles and Quantum Fields\/} (5 Toh Tuck Link, Singapore
  596224: World Scientific) ISBN 978-9814740906

\bibitem{Rammer}
Rammer J 2007 {\em Quantum Field Theory of Non-Equilibrium States\/} (The
  Edinburgh Building, Cambridge CB2 8RU, UK: Cambridge University Press) ISBN
  978-0-521-87499-1

\bibitem{Gasenzer}
Gasenzer T {\it Ultracool Gases far from Equilibrium}, Lecture at the BEC10
  Summer School, MPIPKS Dresden, Aug 10, 2010 (Lecture 2)
  \urlprefix\url{https://www.pks.mpg.de/~bec10/Gasenzer_Lecture2.pdf}

\bibitem{Gasenzer2}
Gasenzer T 2009 {\em The European Physical Journal Special Topics\/} {\bf 168}
  89--148
  \urlprefix\url{https://link.springer.com/article/10.1140/epjst/e2009-00960-5}

\bibitem{Mudry}
Mudry C 2014 {\em Lecture Notes on Field Theory in Condensed Matter Physics\/}
  (5 Toh Tuck Link, Singapore 596224: World Scientific) ISBN 978-981-4449-09-0

\bibitem{Volovik2}
Volovik G~E 1991 {\em JETP Letters\/} {\bf 53}(4) 222
  \urlprefix\url{http://www.jetpletters.ac.ru/ps/1149/article_17392.shtml}

\bibitem{Volovik1}
Volovik G~E 2003 {\em The Universe in a Helium Droplet\/} (Great Clarendon
  Street, Oxford OX2 6DP: Oxford University Press) ISBN 0-19-850782-8

\bibitem{Ruckenstein}
Ruckenstein A and Varma C 1991 {\em Physica C: Superconductivity\/} {\bf
  185-189} 134 -- 140 ISSN 0921-4534
  \urlprefix\url{http://www.sciencedirect.com/science/article/pii/0921453491919624}

\bibitem{Mermin1}
Mermin N 1977 {\em Physica B+C\/} {\bf 90} 1 -- 10 ISSN 0378-4363
  \urlprefix\url{http://www.sciencedirect.com/science/article/pii/0378436377900031}

\bibitem{Mermin2}
Mermin N 1981 {\em Physics Today\/}  46--53
  \urlprefix\url{http://www.economics.soton.ac.uk/staff/aldrich/boojum.pdf}

\bibitem{Bardeen}
Bardeen W~A 1969 {\em Phys. Rev.\/} {\bf 184}(5) 1848--1859
  \urlprefix\url{https://link.aps.org/doi/10.1103/PhysRev.184.1848}

\bibitem{Bertlmann}
Bertlmann R~A 1996 {\em Anomalies in Quantum Field Theory\/} (Walton Street,
  Oxford OX2 6DP, UK: Oxford University Press) ISBN 0-19-852047-6

\bibitem{Adler}
Adler S~L 1969 {\em Phys. Rev.\/} {\bf 177}(5) 2426--2438
  \urlprefix\url{https://link.aps.org/doi/10.1103/PhysRev.177.2426}

\bibitem{Bell}
Bell J~S and Jackiw R 1969 {\em Il Nuovo Cimento A (1965-1970)\/} {\bf 60}
  47--61 ISSN 1826-9869 \urlprefix\url{https://doi.org/10.1007/BF02823296}

\bibitem{Ginsparg}
Alvarez-Gaum\'e L and Ginsparg P 1984 {\em Nuclear Physics B\/} {\bf 243} 449
  -- 474 ISSN 0550-3213
  \urlprefix\url{http://www.sciencedirect.com/science/article/pii/0550321384904875}

\bibitem{Ambjorn}
Ambj{\o}rn J, Greensite J and Peterson C 1983 {\em Nuclear Physics B\/} {\bf
  221} 381 -- 408 ISSN 0550-3213
  \urlprefix\url{http://www.sciencedirect.com/science/article/pii/0550321383905850}

\bibitem{Shifman}
Shifman M 1991 {\em Physics Reports\/} {\bf 209} 341 -- 378 ISSN 0370-1573
  \urlprefix\url{http://www.sciencedirect.com/science/article/pii/037015739190020M}

\bibitem{Casher}
Casher A 1979 {\em Physics Letters B\/} {\bf 83} 395 -- 398 ISSN 0370-2693
  \urlprefix\url{http://www.sciencedirect.com/science/article/pii/0370269379911377}

\bibitem{Atiyah}
Atiyah M~F and Singer I~M 1968 {\em Annals of Mathematics\/}  484--530
  \urlprefix\url{https://www.jstor.org/stable/1970715}

\bibitem{Nakahara}
Nakahara M 2003 {\em Geometry, Topology and Physics\/} (6000 Broken Sound
  Parkway NW, Suite 300, Boca Raton, FL, USA: Taylor \& Francis) ISBN
  0-7503-0606-8

\bibitem{Romer}
R{\"o}mer H 1981 Atiyah-singer index theorem and quantum field theory {\em
  Differential Geometric Methods in Mathematical Physics\/} ed Doebner H~D
  (Berlin, Heidelberg: Springer Berlin Heidelberg) pp 167--211 ISBN
  978-3-540-38573-8

\bibitem{Nielsen}
Nielsen H and Ninomiya M 1983 {\em Physics Letters B\/} {\bf 130} 389 -- 396
  ISSN 0370-2693
  \urlprefix\url{http://www.sciencedirect.com/science/article/pii/0370269383915290}

\bibitem{Jia}
Jia S, Xu S~Y and Hasan M~Z 2016 {\em Nature Materials\/} {\bf 15} 1140--1144
  \urlprefix\url{https://www.nature.com/articles/nmat4787}

\bibitem{Banks}
Banks T and Casher A 1980 {\em Nuclear Physics B\/} {\bf 169} 103 -- 125 ISSN
  0550-3213
  \urlprefix\url{http://www.sciencedirect.com/science/article/pii/0550321380902552}

\bibitem{Shuryak}
Shuryak E 2014 {\em Nuclear Physics A\/} {\bf 928} 138 -- 143 ISSN 0375-9474
  special Issue Dedicated to the Memory of Gerald E Brown (1926-2013)
  \urlprefix\url{http://www.sciencedirect.com/science/article/pii/S0375947414000645}

\bibitem{Swingle1}
Swingle B 2010 {\em Phys. Rev. Lett.\/} {\bf 105}(5) 050502
  \urlprefix\url{https://link.aps.org/doi/10.1103/PhysRevLett.105.050502}

\bibitem{Swingle2}
Swingle B 2012 {\em Phys. Rev. B\/} {\bf 86}(3) 035116
  \urlprefix\url{https://link.aps.org/doi/10.1103/PhysRevB.86.035116}

\bibitem{Klich}
Gioev D and Klich I 2006 {\em Phys. Rev. Lett.\/} {\bf 96}(10) 100503
  \urlprefix\url{https://link.aps.org/doi/10.1103/PhysRevLett.96.100503}

\bibitem{Gurarie1}
Gurarie V 2011 {\em Phys. Rev. B\/} {\bf 83}(8) 085426
  \urlprefix\url{https://link.aps.org/doi/10.1103/PhysRevB.83.085426}

\bibitem{Gurarie2}
Essin A~M and Gurarie V 2011 {\em Phys. Rev. B\/} {\bf 84}(12) 125132
  \urlprefix\url{https://link.aps.org/doi/10.1103/PhysRevB.84.125132}

\bibitem{Lehmann}
Lehmann H 1954 {\em Il Nuovo Cimento\/} {\bf 11} 417--434
  \urlprefix\url{http://users.physik.fu-berlin.de/~kamecke/ps/Lehmann.pdf}

\bibitem{Abrikosov}
Abrikosov A, Gorkov L and Dzyaloshinski I 1975 {\em Methods of Quantum Field
  Theory in Statistical Physics\/} (New York, USA: Dover Publications, Inc.)
  ISBN 978-0-486-63228-5

\bibitem{Tremblay3}
Arsenault L~F~m~c, S\'emon P and Tremblay A~M~S 2012 {\em Phys. Rev. B\/} {\bf
  86}(8) 085133
  \urlprefix\url{https://link.aps.org/doi/10.1103/PhysRevB.86.085133}

\bibitem{Vilk}
Vilk Y and Tremblay A~M 1997 {\em J. Phys. I France\/} {\bf 7}(11) 1309--1368
  \urlprefix\url{https://jp1.journaldephysique.org/articles/jp1/abs/1997/11/jp1v7p1309/jp1v7p1309.html}

\bibitem{Reber}
Reber T {\em et~al.\/}  (\textit{Preprint} \eprint{arXiv:1509.01611v1})
  \urlprefix\url{https://arxiv.org/abs/1509.01611}

\bibitem{Efros}
Efros A~L 2008 {\em Phys. Rev. B\/} {\bf 78}(15) 155130
  \urlprefix\url{https://link.aps.org/doi/10.1103/PhysRevB.78.155130}

\bibitem{Farid2}
Farid B and Tsvelik A~M  (\textit{Preprint} \eprint{arXiv:0909.2886v1})
  \urlprefix\url{https://arxiv.org/abs/0909.2886}

\bibitem{Metzner1}
Metzner W and Vollhardt D 1989 {\em Phys. Rev. Lett.\/} {\bf 62}(3) 324--327
  \urlprefix\url{https://link.aps.org/doi/10.1103/PhysRevLett.62.324}

\bibitem{Metzner2}
Metzner W and Vollhardt D 1989 {\em Phys. Rev. Lett.\/} {\bf 62}(9) 1066--1066
  \urlprefix\url{https://link.aps.org/doi/10.1103/PhysRevLett.62.1066.2}

\bibitem{Metzner3}
Metzner W and Vollhardt D 1987 {\em Phys. Rev. Lett.\/} {\bf 59}(1) 121--124
  \urlprefix\url{https://link.aps.org/doi/10.1103/PhysRevLett.59.121}

\bibitem{Kotliar1}
Georges A and Kotliar G 1992 {\em Phys. Rev. B\/} {\bf 45}(12) 6479--6483
  \urlprefix\url{https://link.aps.org/doi/10.1103/PhysRevB.45.6479}

\bibitem{Kotliar2}
Georges A, Kotliar G, Krauth W and Rozenberg M~J 1996 {\em Rev. Mod. Phys.\/}
  {\bf 68}(1) 13--125
  \urlprefix\url{https://link.aps.org/doi/10.1103/RevModPhys.68.13}

\bibitem{Kotliar3}
Kotliar G and Vollhardt D 2004 {\em Physics Today\/} {\bf 57}(3) 53
  \urlprefix\url{https://physicstoday.scitation.org/doi/10.1063/1.1712502}

\bibitem{Tomczak1}
Tomczak J~M, Liu P, Toschi A, Kresse G and Held K 2017 {\em The European
  Physical Journal Special Topics\/} {\bf 226} 2565--2590 ISSN 1951-6401
  \urlprefix\url{https://doi.org/10.1140/epjst/e2017-70053-1}

\bibitem{Galler}
Galler A {\em et~al.\/}  (\textit{Preprint} \eprint{arXiv:1710.06651v1})
  \urlprefix\url{https://arxiv.org/abs/1710.06651}

\bibitem{Blumer}
Pudleiner P, Sch\"afer T, Rost D, Li G, Held K and Bl\"umer N 2016 {\em Phys.
  Rev. B\/} {\bf 93}(19) 195134
  \urlprefix\url{https://link.aps.org/doi/10.1103/PhysRevB.93.195134}

\bibitem{Tomczak2}
Sch\"afer T, Toschi A and Tomczak J~M 2015 {\em Phys. Rev. B\/} {\bf 91}(12)
  121107 \urlprefix\url{https://link.aps.org/doi/10.1103/PhysRevB.91.121107}

\bibitem{Senthil}
Senthil T 2008 {\em Phys. Rev. B\/} {\bf 78}(3) 035103
  \urlprefix\url{https://link.aps.org/doi/10.1103/PhysRevB.78.035103}

\bibitem{Max}
Nandkishore R, Metlitski M~A and Senthil T 2012 {\em Phys. Rev. B\/} {\bf
  86}(4) 045128
  \urlprefix\url{https://link.aps.org/doi/10.1103/PhysRevB.86.045128}

\bibitem{Debanjan}
Chowdhury D, Werman Y, Berg E and Senthil T 2018 {\em Phys. Rev. X\/} {\bf
  8}(3) 031024
  \urlprefix\url{https://link.aps.org/doi/10.1103/PhysRevX.8.031024}

\bibitem{Bedell1}
Engelbrecht J~R and Bedell K~S 1995 {\em Phys. Rev. Lett.\/} {\bf 74}(21)
  4265--4268
  \urlprefix\url{https://link.aps.org/doi/10.1103/PhysRevLett.74.4265}

\bibitem{Bedell2}
Bedell K~S and Engelbrecht J 1996 {\em Philosophical Magazine B\/} {\bf 74}
  633--640 \urlprefix\url{https://doi.org/10.1080/01418639608240363}

\bibitem{Bedell3}
Bedell K, Engelbrecht J and Blagoev K  (\textit{Preprint}
  \eprint{arXiv:cond-mat/9808203})
  \urlprefix\url{https://arxiv.org/abs/cond-mat/9808203}

\bibitem{Dessau}
Dessau D~S, Shen Z~X, King D~M, Marshall D~S, Lombardo L~W, Dickinson P~H,
  Loeser A~G, DiCarlo J, Park C~H, Kapitulnik A and Spicer W~E 1993 {\em Phys.
  Rev. Lett.\/} {\bf 71}(17) 2781--2784
  \urlprefix\url{https://link.aps.org/doi/10.1103/PhysRevLett.71.2781}

\bibitem{Hussey}
Hussey N {\em et~al.\/} 2003 {\em Nature\/} {\bf 425} 814--817
  \urlprefix\url{https://www.nature.com/articles/nature01981}

\bibitem{Plate}
Plat\'e M, Mottershead J~D~F, Elfimov I~S, Peets D~C, Liang R, Bonn D~A, Hardy
  W~N, Chiuzbaian S, Falub M, Shi M, Patthey L and Damascelli A 2005 {\em Phys.
  Rev. Lett.\/} {\bf 95}(7) 077001
  \urlprefix\url{https://link.aps.org/doi/10.1103/PhysRevLett.95.077001}

\bibitem{Gofron}
Gofron K, Campuzano J~C, Abrikosov A~A, Lindroos M, Bansil A, Ding H, Koelling
  D and Dabrowski B 1994 {\em Phys. Rev. Lett.\/} {\bf 73}(24) 3302--3305
  \urlprefix\url{https://link.aps.org/doi/10.1103/PhysRevLett.73.3302}

\bibitem{Georges}
Berthod C, Giamarchi T, Biermann S and Georges A 2006 {\em Phys. Rev. Lett.\/}
  {\bf 97}(13) 136401
  \urlprefix\url{https://link.aps.org/doi/10.1103/PhysRevLett.97.136401}

\bibitem{Zhang}
Yang K~Y, Rice T~M and Zhang F~C 2006 {\em Phys. Rev. B\/} {\bf 73}(17) 174501
  \urlprefix\url{https://link.aps.org/doi/10.1103/PhysRevB.73.174501}

\bibitem{Schmitt}
Schmitt S 2010 {\em Phys. Rev. B\/} {\bf 82}(15) 155126
  \urlprefix\url{https://link.aps.org/doi/10.1103/PhysRevB.82.155126}

\bibitem{Tsvelik}
Tsvelik A~M 2016 {\em Phys. Rev. B\/} {\bf 94}(16) 165114
  \urlprefix\url{https://link.aps.org/doi/10.1103/PhysRevB.94.165114}

\bibitem{Morr}
Altshuler B~L, Chubukov A~V, Dashevskii A, Finkel'stein A~M and Morr D~K 1998
  {\em Europhysics Letters ({EPL})\/} {\bf 41} 401--406
  \urlprefix\url{https://iopscience.iop.org/article/10.1209/epl/i1998-00164-y/meta}

\bibitem{Pieri}
Pieri P and Strinati G~C 2017 {\em The European Physical Journal B\/} {\bf 90}
  68 ISSN 1434-6036 \urlprefix\url{https://doi.org/10.1140/epjb/e2017-80071-2}

\bibitem{Kotliar4}
Stanescu T~D and Kotliar G 2006 {\em Phys. Rev. B\/} {\bf 74}(12) 125110
  \urlprefix\url{https://link.aps.org/doi/10.1103/PhysRevB.74.125110}

\bibitem{Kotliar5}
Stanescu T~D, Civelli M, Haule K and Kotliar G 2006 {\em Annals of Physics\/}
  {\bf 321} 1682 -- 1715 ISSN 0003-4916 july 2006 Special Issue
  \urlprefix\url{http://www.sciencedirect.com/science/article/pii/S0003491606000972}

\bibitem{Kotliar6}
Kotliar G, Savrasov S~Y, P\'alsson G and Biroli G 2001 {\em Phys. Rev. Lett.\/}
  {\bf 87}(18) 186401
  \urlprefix\url{https://link.aps.org/doi/10.1103/PhysRevLett.87.186401}

\bibitem{Sakai1}
Sakai S, Motome Y and Imada M 2009 {\em Phys. Rev. Lett.\/} {\bf 102}(5) 056404
  \urlprefix\url{https://link.aps.org/doi/10.1103/PhysRevLett.102.056404}

\bibitem{Sakai2}
Sakai S, Sangiovanni G, Civelli M, Motome Y, Held K and Imada M 2012 {\em Phys.
  Rev. B\/} {\bf 85}(3) 035102
  \urlprefix\url{https://link.aps.org/doi/10.1103/PhysRevB.85.035102}

\bibitem{Luttinger4}
Luttinger J~M 1961 {\em Phys. Rev.\/} {\bf 121}(4) 942--949
  \urlprefix\url{https://link.aps.org/doi/10.1103/PhysRev.121.942}

\bibitem{Sakai3}
Sakai S, Civelli M and Imada M 2016 {\em Phys. Rev. B\/} {\bf 94}(11) 115130
  \urlprefix\url{https://link.aps.org/doi/10.1103/PhysRevB.94.115130}

\bibitem{Wigner}
Wigner E 1938 {\em Trans. Faraday Soc.\/} {\bf 34}(0) 678--685
  \urlprefix\url{http://dx.doi.org/10.1039/TF9383400678}

\bibitem{Mott_1}
Mott N~F 1949 {\em Proceedings of the Physical Society. Section A\/} {\bf 62}
  416--422
  \urlprefix\url{https://iopscience.iop.org/article/10.1088/0370-1298/62/7/303/meta}

\bibitem{Mott_2}
Mott N 1952 {\em Progress in Metal Physics\/} {\bf 3} 76 -- 114 ISSN 0502-8205
  \urlprefix\url{http://www.sciencedirect.com/science/article/pii/0502820552900051}

\bibitem{Mott_3}
Mott N~F and Zinamon Z 1970 {\em Reports on Progress in Physics\/} {\bf 33}
  881--940
  \urlprefix\url{https://iopscience.iop.org/article/10.1088/0034-4885/33/3/302/meta}

\bibitem{Kohn}
Kohn W 1964 {\em Phys. Rev.\/} {\bf 133}(1A) A171--A181
  \urlprefix\url{https://link.aps.org/doi/10.1103/PhysRev.133.A171}

\bibitem{Essler1}
Essler F~H~L and Tsvelik A~M 2002 {\em Phys. Rev. B\/} {\bf 65}(11) 115117
  \urlprefix\url{https://link.aps.org/doi/10.1103/PhysRevB.65.115117}

\bibitem{Konik}
Konik R~M, Rice T~M and Tsvelik A~M 2006 {\em Phys. Rev. Lett.\/} {\bf 96}(8)
  086407 \urlprefix\url{https://link.aps.org/doi/10.1103/PhysRevLett.96.086407}

\bibitem{Logan}
Logan D~E and Galpin M~R 2015 {\em Journal of Physics: Condensed Matter\/} {\bf
  28} 025601
  \urlprefix\url{https://iopscience.iop.org/article/10.1088/0953-8984/28/2/025601/meta}

\bibitem{Tremblay}
C{\^{o}}t{\'{e}} R and Tremblay A~M~S 1995 {\em Europhysics Letters ({EPL})\/}
  {\bf 29} 37--42
  \urlprefix\url{https://iopscience.iop.org/article/10.1209/0295-5075/29/1/007}

\bibitem{Bedell4}
Gulacsi M and Bedell K~S 1995 {\em Integrated Ferroelectrics\/} {\bf 6}
  265--279
  \urlprefix\url{https://www.tandfonline.com/doi/abs/10.1080/10584589508019370}

\bibitem{Toschi1}
Sch\"afer T, Geles F, Rost D, Rohringer G, Arrigoni E, Held K, Bl\"umer N,
  Aichhorn M and Toschi A 2015 {\em Phys. Rev. B\/} {\bf 91}(12) 125109
  \urlprefix\url{https://link.aps.org/doi/10.1103/PhysRevB.91.125109}

\bibitem{Toschi2}
Taranto C, Andergassen S, Bauer J, Held K, Katanin A, Metzner W, Rohringer G
  and Toschi A 2014 {\em Phys. Rev. Lett.\/} {\bf 112}(19) 196402
  \urlprefix\url{https://link.aps.org/doi/10.1103/PhysRevLett.112.196402}

\bibitem{Luther}
Luther A 1979 {\em Phys. Rev. B\/} {\bf 19}(1) 320--330
  \urlprefix\url{https://link.aps.org/doi/10.1103/PhysRevB.19.320}

\bibitem{Anderson1}
Anderson P~W 1990 {\em Phys. Rev. Lett.\/} {\bf 64}(15) 1839--1841
  \urlprefix\url{https://link.aps.org/doi/10.1103/PhysRevLett.64.1839}

\bibitem{Anderson4}
Anderson P~W 1990 {\em Phys. Rev. Lett.\/} {\bf 65}(18) 2306--2308
  \urlprefix\url{https://link.aps.org/doi/10.1103/PhysRevLett.65.2306}

\bibitem{Anderson2}
Anderson P~W 1967 {\em Phys. Rev. Lett.\/} {\bf 18}(24) 1049--1051
  \urlprefix\url{https://link.aps.org/doi/10.1103/PhysRevLett.18.1049}

\bibitem{Anderson3}
Anderson P~W 1967 {\em Phys. Rev.\/} {\bf 164}(2) 352--359
  \urlprefix\url{https://link.aps.org/doi/10.1103/PhysRev.164.352}

\bibitem{Louis}
Baeriswyl D and others (eds) 1995 {\em The Hubbard Model: Its Physics and
  Mathematical Physics\/} (223 Spring Street, New York, N.Y. 10013: Plenum
  Press) ISBN 0-306-45003-8 in the Proceedings of a NATO Advanced Research
  Workshop on the Physics and Mathematical Physics of the Hubbard Model, held
  October 3-8, 1993, in San Sebastian, Spain

\bibitem{Stamp1}
Stamp P~C~E 1992 {\em Phys. Rev. Lett.\/} {\bf 68}(14) 2180--2183
  \urlprefix\url{https://link.aps.org/doi/10.1103/PhysRevLett.68.2180}

\bibitem{Stamp2}
Stamp P~C~E 1993 {\em J. Phys. I France\/} {\bf 3}(2) 2180--2183
  \urlprefix\url{https://jp1.journaldephysique.org/articles/jp1/abs/1993/02/jp1v3p625/jp1v3p625.html}

\bibitem{Zimanyi}
Zimanyi G~T and Bedell K~S 1991 {\em Phys. Rev. Lett.\/} {\bf 66}(2) 228--231
  \urlprefix\url{https://link.aps.org/doi/10.1103/PhysRevLett.66.228}

\bibitem{Kravchenko}
Kravchenko S~V and Sarachik M~P 2003 {\em Reports on Progress in Physics\/}
  {\bf 67} 1--44
  \urlprefix\url{https://iopscience.iop.org/article/10.1088/0034-4885/67/1/R01/meta}

\bibitem{Finkel1}
Finkel'shtein 1983 {\em JETP\/} {\bf 57}(1) 168
  \urlprefix\url{http://www.jetp.ac.ru/cgi-bin/e/index/e/57/1/p97?a=list}

\bibitem{Finkel2}
Finkel'stein A~M 1984 {\em Zeitschrift f{\"u}r Physik B Condensed Matter\/}
  {\bf 56} 189--196 ISSN 1431-584X
  \urlprefix\url{https://doi.org/10.1007/BF01304171}

\bibitem{Castel1}
Castellani C, Di~Castro C, Lee P~A and Ma M 1984 {\em Phys. Rev. B\/} {\bf
  30}(2) 527--543
  \urlprefix\url{https://link.aps.org/doi/10.1103/PhysRevB.30.527}

\bibitem{Castel2}
Castellani C, Di~Castro C and Lee P~A 1998 {\em Phys. Rev. B\/} {\bf 57}(16)
  R9381--R9384
  \urlprefix\url{https://link.aps.org/doi/10.1103/PhysRevB.57.R9381}

\bibitem{Valles}
Dobrosavljevi\'c V, Trivedi N and Valles J~M~J 2012 {\em Conductor-Insulator
  Quantum Phase Transitions\/} (Great Clarendon Street, Oxford, OX2 6DP: Oxford
  University Press) ISBN 978-0-19-959259-3

\bibitem{Randall}
Randall L and Sundrum R 1999 {\em Phys. Rev. Lett.\/} {\bf 83}(23) 4690--4693
  \urlprefix\url{https://link.aps.org/doi/10.1103/PhysRevLett.83.4690}

\bibitem{Quader}
Quader K~F and Bedell K~S 1985 {\em Journal of Low Temperature Physics\/} {\bf
  58} 89--125 ISSN 1573-7357 \urlprefix\url{https://doi.org/10.1007/BF00682568}

\bibitem{Carlos}
Sanchez-Castro C~R, Bedell K~S and Wiegers S~A~J 1989 {\em Phys. Rev. B\/} {\bf
  40}(1) 437--453
  \urlprefix\url{https://link.aps.org/doi/10.1103/PhysRevB.40.437}

\bibitem{Anderson_book}
Anderson P 1984 {\em Basic Notions of Condensed Matter Physics\/} (6000 Broken
  Sound Parkway NW, Suite 300, Boca Raton, FL, USA: Taylor \& Francis) ISBN
  978-0-201-32830-1

\bibitem{Horava}
Ho\ifmmode~\check{r}\else \v{r}\fi{}ava P 2005 {\em Phys. Rev. Lett.\/} {\bf
  95}(1) 016405
  \urlprefix\url{https://link.aps.org/doi/10.1103/PhysRevLett.95.016405}

\bibitem{Seki}
Seki K and Yunoki S 2017 {\em Phys. Rev. B\/} {\bf 96}(8) 085124
  \urlprefix\url{https://link.aps.org/doi/10.1103/PhysRevB.96.085124}

\bibitem{Coleman_book}
Coleman P 2015 {\em Introduction to Many-Body Physics\/} (University Preiting
  House, Cambridge CB2 8BS, United Kingdom: Cambridge University Press) ISBN
  978-0-521-86488-6

\end{thebibliography}


\begin{thebibliography}{109}%
\makeatletter
\providecommand \@ifxundefined [1]{%
 \@ifx{#1\undefined}
}%
\providecommand \@ifnum [1]{%
 \ifnum #1\expandafter \@firstoftwo
 \else \expandafter \@secondoftwo
 \fi
}%
\providecommand \@ifx [1]{%
 \ifx #1\expandafter \@firstoftwo
 \else \expandafter \@secondoftwo
 \fi
}%
\providecommand \natexlab [1]{#1}%
\providecommand \enquote  [1]{``#1''}%
\providecommand \bibnamefont  [1]{#1}%
\providecommand \bibfnamefont [1]{#1}%
\providecommand \citenamefont [1]{#1}%
\providecommand \href@noop [0]{\@secondoftwo}%
\providecommand \href [0]{\begingroup \@sanitize@url \@href}%
\providecommand \@href[1]{\@@startlink{#1}\@@href}%
\providecommand \@@href[1]{\endgroup#1\@@endlink}%
\providecommand \@sanitize@url [0]{\catcode `\\12\catcode `\$12\catcode
  `\&12\catcode `\#12\catcode `\^12\catcode `\_12\catcode `\%12\relax}%
\providecommand \@@startlink[1]{}%
\providecommand \@@endlink[0]{}%
\providecommand \url  [0]{\begingroup\@sanitize@url \@url }%
\providecommand \@url [1]{\endgroup\@href {#1}{\urlprefix }}%
\providecommand \urlprefix  [0]{URL }%
\providecommand \Eprint [0]{\href }%
\providecommand \doibase [0]{http://dx.doi.org/}%
\providecommand \selectlanguage [0]{\@gobble}%
\providecommand \bibinfo  [0]{\@secondoftwo}%
\providecommand \bibfield  [0]{\@secondoftwo}%
\providecommand \translation [1]{[#1]}%
\providecommand \BibitemOpen [0]{}%
\providecommand \bibitemStop [0]{}%
\providecommand \bibitemNoStop [0]{.\EOS\space}%
\providecommand \EOS [0]{\spacefactor3000\relax}%
\providecommand \BibitemShut  [1]{\csname bibitem#1\endcsname}%
\let\auto@bib@innerbib\@empty
\bibitem [{\citenamefont {Dirac}(1928)}]{Dirac1}%
  \BibitemOpen
  \bibfield  {author} {\bibinfo {author} {\bibfnamefont {P.~A.~M.}\
  \bibnamefont {Dirac}},\ }\href {\doibase 10.1098/rspa.1928.0023} {\bibfield
  {journal} {\bibinfo  {journal} {Proceedings of the Royal Society of London A:
  Mathematical, Physical and Engineering Sciences}\ }\textbf {\bibinfo {volume}
  {117}},\ \bibinfo {pages} {610} (\bibinfo {year} {1928})}\BibitemShut
  {NoStop}%
\bibitem [{\citenamefont {Darwin}(1928)}]{Darwin}%
  \BibitemOpen
  \bibfield  {author} {\bibinfo {author} {\bibfnamefont {C.~G.}\ \bibnamefont
  {Darwin}},\ }\href {\doibase 10.1098/rspa.1928.0076} {\bibfield  {journal}
  {\bibinfo  {journal} {Proceedings of the Royal Society of London A:
  Mathematical, Physical and Engineering Sciences}\ }\textbf {\bibinfo {volume}
  {118}},\ \bibinfo {pages} {654} (\bibinfo {year} {1928})}\BibitemShut
  {NoStop}%
\bibitem [{\citenamefont {Dirac}(1930)}]{Dirac2}%
  \BibitemOpen
  \bibfield  {author} {\bibinfo {author} {\bibfnamefont {P.~A.~M.}\
  \bibnamefont {Dirac}},\ }\href {\doibase 10.1098/rspa.1930.0013} {\bibfield
  {journal} {\bibinfo  {journal} {Proceedings of the Royal Society of London A:
  Mathematical, Physical and Engineering Sciences}\ }\textbf {\bibinfo {volume}
  {126}},\ \bibinfo {pages} {360} (\bibinfo {year} {1930})}\BibitemShut
  {NoStop}%
\bibitem [{\citenamefont {Eddington}(1918)}]{Eddington}%
  \BibitemOpen
  \bibfield  {author} {\bibinfo {author} {\bibfnamefont {A.}~\bibnamefont
  {Eddington}},\ }\href
  {https://www.jstor.org/stable/95117?seq=1#page_scan_tab_contents} {\bibfield
  {journal} {\bibinfo  {journal} {Proceedings of the Royal Society A,
  Containing Papers of a Mathematical and Physical Character}\ }\textbf
  {\bibinfo {volume} {121}},\ \bibinfo {pages} {524} (\bibinfo {year}
  {1918})}\BibitemShut {NoStop}%
\bibitem [{\citenamefont {Kilmister}(1994)}]{Kilmister}%
  \BibitemOpen
  \bibfield  {author} {\bibinfo {author} {\bibfnamefont {C.}~\bibnamefont
  {Kilmister}},\ }\href@noop {} {\emph {\bibinfo {title} {Eddington's search
  for a fundamental theory: A key to the universe}}}\ (\bibinfo  {publisher}
  {Cambridge University Press},\ \bibinfo {address} {United Kingdom},\ \bibinfo
  {year} {1994})\BibitemShut {NoStop}%
\bibitem [{\citenamefont {Majorana}(1937)}]{Majorana}%
  \BibitemOpen
  \bibfield  {author} {\bibinfo {author} {\bibfnamefont {E.}~\bibnamefont
  {Majorana}},\ }\href {https://link.springer.com/article/10.1007/BF02961314}
  {\bibfield  {journal} {\bibinfo  {journal} {Il Nuovo Cimento}\ }\textbf
  {\bibinfo {volume} {14}},\ \bibinfo {pages} {171} (\bibinfo {year}
  {1937})}\BibitemShut {NoStop}%
\bibitem [{\citenamefont {Park}(2010)}]{Park}%
  \BibitemOpen
  \bibfield  {author} {\bibinfo {author} {\bibfnamefont {J.-H.}\ \bibnamefont
  {Park}},\ }\href
  {http://home.sogang.ac.kr/sites/lab7616/3/Lists/b10/Attachments/92/gamma[0].pdf}
  {\emph {\bibinfo {title} {Lecture note on Clifford algebra}}},\ \bibinfo
  {number} {Supplement for lecture at Modave Summer School in Mathematical
  Physics, Belgium, June, 2005}\ (\bibinfo {year} {2010})\BibitemShut {NoStop}%
\bibitem [{\citenamefont {Anderson}(1933)}]{Anderson}%
  \BibitemOpen
  \bibfield  {author} {\bibinfo {author} {\bibfnamefont {C.~D.}\ \bibnamefont
  {Anderson}},\ }\href {\doibase 10.1103/P1hysRev.43.491} {\bibfield  {journal}
  {\bibinfo  {journal} {Phys. Rev.}\ }\textbf {\bibinfo {volume} {43}},\
  \bibinfo {pages} {491} (\bibinfo {year} {1933})}\BibitemShut {NoStop}%
\bibitem [{\citenamefont {Sivaguru}(2012)}]{Sivaguru}%
  \BibitemOpen
  \bibfield  {author} {\bibinfo {author} {\bibfnamefont {A.}~\bibnamefont
  {Sivaguru}},\ }\emph {\bibinfo {title} {Majorana Fermions}},\ \href
  {http://www.imperial.ac.uk/media/imperial-college/research-centres-and-groups/theoretical-physics/msc/dissertations/2012/Aran-Sivaguru-Dissertation.pdf}
  {Master's thesis},\ \bibinfo  {school} {Imperial College London} (\bibinfo
  {year} {2012})\BibitemShut {NoStop}%
\bibitem [{\citenamefont {Cork}\ \emph {et~al.}(1956)\citenamefont {Cork},
  \citenamefont {Lambertson}, \citenamefont {Piccioni},\ and\ \citenamefont
  {Wenzel}}]{Cork}%
  \BibitemOpen
  \bibfield  {author} {\bibinfo {author} {\bibfnamefont {B.}~\bibnamefont
  {Cork}}, \bibinfo {author} {\bibfnamefont {G.~R.}\ \bibnamefont
  {Lambertson}}, \bibinfo {author} {\bibfnamefont {O.}~\bibnamefont
  {Piccioni}}, \ and\ \bibinfo {author} {\bibfnamefont {W.~A.}\ \bibnamefont
  {Wenzel}},\ }\href {\doibase 10.1103/PhysRev.104.1193} {\bibfield  {journal}
  {\bibinfo  {journal} {Phys. Rev.}\ }\textbf {\bibinfo {volume} {104}},\
  \bibinfo {pages} {1193} (\bibinfo {year} {1956})}\BibitemShut {NoStop}%
\bibitem [{\citenamefont {Furry}(1938)}]{Furry1}%
  \BibitemOpen
  \bibfield  {author} {\bibinfo {author} {\bibfnamefont {W.~H.}\ \bibnamefont
  {Furry}},\ }\href {\doibase 10.1103/PhysRev.54.56} {\bibfield  {journal}
  {\bibinfo  {journal} {Phys. Rev.}\ }\textbf {\bibinfo {volume} {54}},\
  \bibinfo {pages} {56} (\bibinfo {year} {1938})}\BibitemShut {NoStop}%
\bibitem [{\citenamefont {Furry}(1939)}]{Furry2}%
  \BibitemOpen
  \bibfield  {author} {\bibinfo {author} {\bibfnamefont {W.~H.}\ \bibnamefont
  {Furry}},\ }\href {\doibase 10.1103/PhysRev.56.1184} {\bibfield  {journal}
  {\bibinfo  {journal} {Phys. Rev.}\ }\textbf {\bibinfo {volume} {56}},\
  \bibinfo {pages} {1184} (\bibinfo {year} {1939})}\BibitemShut {NoStop}%
\bibitem [{\citenamefont {Schechter}\ and\ \citenamefont
  {Valle}(1982)}]{Valle}%
  \BibitemOpen
  \bibfield  {author} {\bibinfo {author} {\bibfnamefont {J.}~\bibnamefont
  {Schechter}}\ and\ \bibinfo {author} {\bibfnamefont {J.~W.~F.}\ \bibnamefont
  {Valle}},\ }\href {\doibase 10.1103/PhysRevD.25.2951} {\bibfield  {journal}
  {\bibinfo  {journal} {Phys. Rev. D}\ }\textbf {\bibinfo {volume} {25}},\
  \bibinfo {pages} {2951} (\bibinfo {year} {1982})}\BibitemShut {NoStop}%
\bibitem [{\citenamefont {Agostini}\ \emph {et~al.}(2013)\citenamefont
  {Agostini} \emph {et~al.}}]{Agostini}%
  \BibitemOpen
  \bibfield  {author} {\bibinfo {author} {\bibfnamefont {M.}~\bibnamefont
  {Agostini}} \emph {et~al.},\ }\href {\doibase 10.1103/PhysRevLett.111.122503}
  {\bibfield  {journal} {\bibinfo  {journal} {Phys. Rev. Lett.}\ }\textbf
  {\bibinfo {volume} {111}},\ \bibinfo {pages} {122503} (\bibinfo {year}
  {2013})},\ \bibinfo {note} {(with GERDA Collaboration)}\BibitemShut {NoStop}%
\bibitem [{\citenamefont {Agostini}\ \emph {et~al.}(2017)\citenamefont
  {Agostini} \emph {et~al.}}]{GERDA}%
  \BibitemOpen
  \bibfield  {author} {\bibinfo {author} {\bibfnamefont {M.}~\bibnamefont
  {Agostini}} \emph {et~al.},\ }\href {http://dx.doi.org/10.1038/nature21717}
  {\bibfield  {journal} {\bibinfo  {journal} {Nature}\ }\textbf {\bibinfo
  {volume} {544}},\ \bibinfo {pages} {47} (\bibinfo {year} {2017})},\ \bibinfo
  {note} {(with GERDA Collaboration)}\BibitemShut {NoStop}%
\bibitem [{\citenamefont {Fukuda}\ \emph {et~al.}(1998)\citenamefont {Fukuda}
  \emph {et~al.}}]{Fukuda}%
  \BibitemOpen
  \bibfield  {author} {\bibinfo {author} {\bibfnamefont {Y.}~\bibnamefont
  {Fukuda}} \emph {et~al.},\ }\href {\doibase 10.1103/PhysRevLett.81.1562}
  {\bibfield  {journal} {\bibinfo  {journal} {Phys. Rev. Lett.}\ }\textbf
  {\bibinfo {volume} {81}},\ \bibinfo {pages} {1562} (\bibinfo {year}
  {1998})},\ \bibinfo {note} {(with Super-Kamiokande
  Collaboration)}\BibitemShut {NoStop}%
\bibitem [{\citenamefont {Ahmad}\ \emph {et~al.}(2001)\citenamefont {Ahmad}
  \emph {et~al.}}]{Ahmad1}%
  \BibitemOpen
  \bibfield  {author} {\bibinfo {author} {\bibfnamefont {Q.~R.}\ \bibnamefont
  {Ahmad}} \emph {et~al.},\ }\href {\doibase 10.1103/PhysRevLett.87.071301}
  {\bibfield  {journal} {\bibinfo  {journal} {Phys. Rev. Lett.}\ }\textbf
  {\bibinfo {volume} {87}},\ \bibinfo {pages} {071301} (\bibinfo {year}
  {2001})},\ \bibinfo {note} {(with SNO Collaboration)}\BibitemShut {NoStop}%
\bibitem [{\citenamefont {Ahmad}\ \emph {et~al.}(2002)\citenamefont {Ahmad}
  \emph {et~al.}}]{Ahmad2}%
  \BibitemOpen
  \bibfield  {author} {\bibinfo {author} {\bibfnamefont {Q.~R.}\ \bibnamefont
  {Ahmad}} \emph {et~al.},\ }\href {\doibase 10.1103/PhysRevLett.89.011301}
  {\bibfield  {journal} {\bibinfo  {journal} {Phys. Rev. Lett.}\ }\textbf
  {\bibinfo {volume} {89}},\ \bibinfo {pages} {011301} (\bibinfo {year}
  {2002})},\ \bibinfo {note} {(with SNO Collaboration)}\BibitemShut {NoStop}%
\bibitem [{\citenamefont {Mohapatra}\ and\ \citenamefont
  {Senjanovi\ifmmode~\acute{c}\else \'{c}\fi{}}(1980)}]{Mohapatra}%
  \BibitemOpen
  \bibfield  {author} {\bibinfo {author} {\bibfnamefont {R.~N.}\ \bibnamefont
  {Mohapatra}}\ and\ \bibinfo {author} {\bibfnamefont {G.}~\bibnamefont
  {Senjanovi\ifmmode~\acute{c}\else \'{c}\fi{}}},\ }\href {\doibase
  10.1103/PhysRevLett.44.912} {\bibfield  {journal} {\bibinfo  {journal} {Phys.
  Rev. Lett.}\ }\textbf {\bibinfo {volume} {44}},\ \bibinfo {pages} {912}
  (\bibinfo {year} {1980})}\BibitemShut {NoStop}%
\bibitem [{\citenamefont {Yanagida}(1980)}]{Tsutomu}%
  \BibitemOpen
  \bibfield  {author} {\bibinfo {author} {\bibfnamefont {T.}~\bibnamefont
  {Yanagida}},\ }\href {\doibase 10.1143/PTP.64.1103} {\bibfield  {journal}
  {\bibinfo  {journal} {Progress of Theoretical Physics}\ }\textbf {\bibinfo
  {volume} {64}},\ \bibinfo {pages} {1103} (\bibinfo {year}
  {1980})}\BibitemShut {NoStop}%
\bibitem [{\citenamefont {Lavoura}(1994)}]{Lavoura}%
  \BibitemOpen
  \bibfield  {author} {\bibinfo {author} {\bibfnamefont {L.}~\bibnamefont
  {Lavoura}},\ }\href {\doibase 10.1103/PhysRevD.50.523} {\bibfield  {journal}
  {\bibinfo  {journal} {Phys. Rev. D}\ }\textbf {\bibinfo {volume} {50}},\
  \bibinfo {pages} {523} (\bibinfo {year} {1994})}\BibitemShut {NoStop}%
\bibitem [{\citenamefont {Wilczek}(2009)}]{Wilczek}%
  \BibitemOpen
  \bibfield  {author} {\bibinfo {author} {\bibfnamefont {F.}~\bibnamefont
  {Wilczek}},\ }\href
  {http://www.nature.com/nphys/journal/v5/n9/full/nphys1380.html} {\bibfield
  {journal} {\bibinfo  {journal} {Nat. Phys.}\ }\textbf {\bibinfo {volume}
  {5}},\ \bibinfo {pages} {614} (\bibinfo {year} {2009})}\BibitemShut {NoStop}%
\bibitem [{\citenamefont {Beenakker}(2013)}]{Beenakker3}%
  \BibitemOpen
  \bibfield  {author} {\bibinfo {author} {\bibfnamefont {C.}~\bibnamefont
  {Beenakker}},\ }\href {\doibase 10.1146/annurev-conmatphys-030212-184337}
  {\bibfield  {journal} {\bibinfo  {journal} {Annual Review of Condensed Matter
  Physics}\ }\textbf {\bibinfo {volume} {4}},\ \bibinfo {pages} {113} (\bibinfo
  {year} {2013})}\BibitemShut {NoStop}%
\bibitem [{\citenamefont {Wilczek}\ and\ \citenamefont
  {Esposito}(2014)}]{Wilczek_book}%
  \BibitemOpen
  \bibfield  {author} {\bibinfo {author} {\bibfnamefont {F.}~\bibnamefont
  {Wilczek}}\ and\ \bibinfo {author} {\bibfnamefont {S.}~\bibnamefont
  {Esposito}},\ }\enquote {\bibinfo {title} {Majorana and condensed matter
  physics},}\ in\ \href {\doibase 10.1017/CBO9781107358362.014} {\emph
  {\bibinfo {booktitle} {The Physics of Ettore Majorana: Theoretical,
  Mathematical, and Phenomenological}}}\ (\bibinfo  {publisher} {Cambridge
  University Press},\ \bibinfo {year} {2014})\ pp.\ \bibinfo {pages}
  {279--302}\BibitemShut {NoStop}%
\bibitem [{\citenamefont {Leggett}(2016)}]{Leggett}%
  \BibitemOpen
  \bibfield  {author} {\bibinfo {author} {\bibfnamefont {A.~J.}\ \bibnamefont
  {Leggett}},\ }\href {\doibase 10.1142/S0217979216300127} {\bibfield
  {journal} {\bibinfo  {journal} {International Journal of Modern Physics B}\
  }\textbf {\bibinfo {volume} {30}},\ \bibinfo {pages} {1630012} (\bibinfo
  {year} {2016})}\BibitemShut {NoStop}%
\bibitem [{\citenamefont {Elliott}\ and\ \citenamefont
  {Franz}(2015)}]{Elliott}%
  \BibitemOpen
  \bibfield  {author} {\bibinfo {author} {\bibfnamefont {S.~R.}\ \bibnamefont
  {Elliott}}\ and\ \bibinfo {author} {\bibfnamefont {M.}~\bibnamefont
  {Franz}},\ }\href {\doibase 10.1103/RevModPhys.87.137} {\bibfield  {journal}
  {\bibinfo  {journal} {Rev. Mod. Phys.}\ }\textbf {\bibinfo {volume} {87}},\
  \bibinfo {pages} {137} (\bibinfo {year} {2015})}\BibitemShut {NoStop}%
\bibitem [{\citenamefont {Leijnse}\ and\ \citenamefont
  {Flensberg}(2012)}]{Flensberg}%
  \BibitemOpen
  \bibfield  {author} {\bibinfo {author} {\bibfnamefont {M.}~\bibnamefont
  {Leijnse}}\ and\ \bibinfo {author} {\bibfnamefont {K.}~\bibnamefont
  {Flensberg}},\ }\href {http://stacks.iop.org/0268-1242/27/i=12/a=124003}
  {\bibfield  {journal} {\bibinfo  {journal} {Semiconductor Science and
  Technology}\ }\textbf {\bibinfo {volume} {27}},\ \bibinfo {pages} {124003}
  (\bibinfo {year} {2012})}\BibitemShut {NoStop}%
\bibitem [{\citenamefont {Volovik}(1999)}]{Volovik}%
  \BibitemOpen
  \bibfield  {author} {\bibinfo {author} {\bibfnamefont {G.~E.}\ \bibnamefont
  {Volovik}},\ }\href {\doibase 10.1134/1.568223} {\bibfield  {journal}
  {\bibinfo  {journal} {Journal of Experimental and Theoretical Physics
  Letters}\ }\textbf {\bibinfo {volume} {70}},\ \bibinfo {pages} {609}
  (\bibinfo {year} {1999})}\BibitemShut {NoStop}%
\bibitem [{\citenamefont {Read}\ and\ \citenamefont {Green}(2000)}]{Read}%
  \BibitemOpen
  \bibfield  {author} {\bibinfo {author} {\bibfnamefont {N.}~\bibnamefont
  {Read}}\ and\ \bibinfo {author} {\bibfnamefont {D.}~\bibnamefont {Green}},\
  }\href {\doibase 10.1103/PhysRevB.61.10267} {\bibfield  {journal} {\bibinfo
  {journal} {Phys. Rev. B}\ }\textbf {\bibinfo {volume} {61}},\ \bibinfo
  {pages} {10267} (\bibinfo {year} {2000})}\BibitemShut {NoStop}%
\bibitem [{\citenamefont {Kitaev}(2001)}]{Kitaev}%
  \BibitemOpen
  \bibfield  {author} {\bibinfo {author} {\bibfnamefont {A.}~\bibnamefont
  {Kitaev}},\ }\href {http://stacks.iop.org/1063-7869/44/i=10S/a=S29}
  {\bibfield  {journal} {\bibinfo  {journal} {Physics-Uspekhi}\ }\textbf
  {\bibinfo {volume} {44}},\ \bibinfo {pages} {131} (\bibinfo {year}
  {2001})}\BibitemShut {NoStop}%
\bibitem [{\citenamefont {Fu}\ and\ \citenamefont {Kane}(2008)}]{FuKane}%
  \BibitemOpen
  \bibfield  {author} {\bibinfo {author} {\bibfnamefont {L.}~\bibnamefont
  {Fu}}\ and\ \bibinfo {author} {\bibfnamefont {C.~L.}\ \bibnamefont {Kane}},\
  }\href {\doibase 10.1103/PhysRevLett.100.096407} {\bibfield  {journal}
  {\bibinfo  {journal} {Phys. Rev. Lett.}\ }\textbf {\bibinfo {volume} {100}},\
  \bibinfo {pages} {096407} (\bibinfo {year} {2008})}\BibitemShut {NoStop}%
\bibitem [{\citenamefont {Bravyi}\ and\ \citenamefont {Kitaev}(2002)}]{Bravyi}%
  \BibitemOpen
  \bibfield  {author} {\bibinfo {author} {\bibfnamefont {S.~B.}\ \bibnamefont
  {Bravyi}}\ and\ \bibinfo {author} {\bibfnamefont {A.~Y.}\ \bibnamefont
  {Kitaev}},\ }\href {\doibase http://dx.doi.org/10.1006/aphy.2002.6254}
  {\bibfield  {journal} {\bibinfo  {journal} {Annals of Physics}\ }\textbf
  {\bibinfo {volume} {298}},\ \bibinfo {pages} {210 } (\bibinfo {year}
  {2002})}\BibitemShut {NoStop}%
\bibitem [{\citenamefont {Wu}\ \emph {et~al.}(2014)\citenamefont {Wu},
  \citenamefont {He},\ and\ \citenamefont {Kou}}]{SuPeng}%
  \BibitemOpen
  \bibfield  {author} {\bibinfo {author} {\bibfnamefont {Y.-J.}\ \bibnamefont
  {Wu}}, \bibinfo {author} {\bibfnamefont {J.}~\bibnamefont {He}}, \ and\
  \bibinfo {author} {\bibfnamefont {S.-P.}\ \bibnamefont {Kou}},\ }\href
  {\doibase 10.1103/PhysRevA.90.022324} {\bibfield  {journal} {\bibinfo
  {journal} {Phys. Rev. A}\ }\textbf {\bibinfo {volume} {90}},\ \bibinfo
  {pages} {022324} (\bibinfo {year} {2014})}\BibitemShut {NoStop}%
\bibitem [{\citenamefont {Das~Sarma}\ \emph {et~al.}(2015)\citenamefont
  {Das~Sarma}, \citenamefont {Freedman},\ and\ \citenamefont
  {Nayak}}]{DasSarma}%
  \BibitemOpen
  \bibfield  {author} {\bibinfo {author} {\bibfnamefont {S.}~\bibnamefont
  {Das~Sarma}}, \bibinfo {author} {\bibfnamefont {M.}~\bibnamefont {Freedman}},
  \ and\ \bibinfo {author} {\bibfnamefont {C.}~\bibnamefont {Nayak}},\ }\href
  {https://www.nature.com/articles/npjqi20151} {\bibfield  {journal} {\bibinfo
  {journal} {npj Quantum Information}\ }\textbf {\bibinfo {volume} {1}},\
  \bibinfo {pages} {15001} (\bibinfo {year} {2015})}\BibitemShut {NoStop}%
\bibitem [{\citenamefont {Pachos}(2012)}]{Pachos}%
  \BibitemOpen
  \bibfield  {author} {\bibinfo {author} {\bibfnamefont {K.~J.}\ \bibnamefont
  {Pachos}},\ }\href@noop {} {\emph {\bibinfo {title} {Introduction to
  Topological Quantum Computation}}}\ (\bibinfo  {publisher} {Cambridge
  University Press},\ \bibinfo {address} {United Kingdom},\ \bibinfo {year}
  {2012})\BibitemShut {NoStop}%
\bibitem [{\citenamefont {Nadj-Perge}\ \emph {et~al.}(2014)\citenamefont
  {Nadj-Perge}, \citenamefont {Drozdov}, \citenamefont {Li}, \citenamefont
  {Chen}, \citenamefont {Jeon}, \citenamefont {Seo}, \citenamefont {MacDonald},
  \citenamefont {Bernevig},\ and\ \citenamefont {Yazdani}}]{Yazdani}%
  \BibitemOpen
  \bibfield  {author} {\bibinfo {author} {\bibfnamefont {S.}~\bibnamefont
  {Nadj-Perge}}, \bibinfo {author} {\bibfnamefont {I.~K.}\ \bibnamefont
  {Drozdov}}, \bibinfo {author} {\bibfnamefont {J.}~\bibnamefont {Li}},
  \bibinfo {author} {\bibfnamefont {H.}~\bibnamefont {Chen}}, \bibinfo {author}
  {\bibfnamefont {S.}~\bibnamefont {Jeon}}, \bibinfo {author} {\bibfnamefont
  {J.}~\bibnamefont {Seo}}, \bibinfo {author} {\bibfnamefont {A.~H.}\
  \bibnamefont {MacDonald}}, \bibinfo {author} {\bibfnamefont {B.~A.}\
  \bibnamefont {Bernevig}}, \ and\ \bibinfo {author} {\bibfnamefont
  {A.}~\bibnamefont {Yazdani}},\ }\href {\doibase 10.1126/science.1259327}
  {\bibfield  {journal} {\bibinfo  {journal} {Science}\ }\textbf {\bibinfo
  {volume} {346}},\ \bibinfo {pages} {602} (\bibinfo {year}
  {2014})}\BibitemShut {NoStop}%
\bibitem [{\citenamefont {He}\ \emph {et~al.}(2017)\citenamefont {He} \emph
  {et~al.}}]{He}%
  \BibitemOpen
  \bibfield  {author} {\bibinfo {author} {\bibfnamefont {Q.~L.}\ \bibnamefont
  {He}} \emph {et~al.},\ }\href {\doibase 10.1126/science.aag2792} {\bibfield
  {journal} {\bibinfo  {journal} {Science}\ }\textbf {\bibinfo {volume}
  {357}},\ \bibinfo {pages} {294} (\bibinfo {year} {2017})}\BibitemShut
  {NoStop}%
\bibitem [{\citenamefont {Wen}(2004)}]{Wen}%
  \BibitemOpen
  \bibfield  {author} {\bibinfo {author} {\bibfnamefont {X.-G.}\ \bibnamefont
  {Wen}},\ }\href@noop {} {\emph {\bibinfo {title} {Quantum Field Theory of
  Many-Body Systems}}}\ (\bibinfo  {publisher} {Oxford University Press},\
  \bibinfo {address} {United Kingdom},\ \bibinfo {year} {2004})\BibitemShut
  {NoStop}%
\bibitem [{\citenamefont {Ivanov}(2001)}]{Ivanov}%
  \BibitemOpen
  \bibfield  {author} {\bibinfo {author} {\bibfnamefont {D.~A.}\ \bibnamefont
  {Ivanov}},\ }\href {\doibase 10.1103/PhysRevLett.86.268} {\bibfield
  {journal} {\bibinfo  {journal} {Phys. Rev. Lett.}\ }\textbf {\bibinfo
  {volume} {86}},\ \bibinfo {pages} {268} (\bibinfo {year} {2001})}\BibitemShut
  {NoStop}%
\bibitem [{\citenamefont {Teo}\ and\ \citenamefont {Kane}(2010)}]{Teo}%
  \BibitemOpen
  \bibfield  {author} {\bibinfo {author} {\bibfnamefont {J.~C.~Y.}\
  \bibnamefont {Teo}}\ and\ \bibinfo {author} {\bibfnamefont {C.~L.}\
  \bibnamefont {Kane}},\ }\href {\doibase 10.1103/PhysRevLett.104.046401}
  {\bibfield  {journal} {\bibinfo  {journal} {Phys. Rev. Lett.}\ }\textbf
  {\bibinfo {volume} {104}},\ \bibinfo {pages} {046401} (\bibinfo {year}
  {2010})}\BibitemShut {NoStop}%
\bibitem [{\citenamefont {Zheng}\ \emph {et~al.}(2015)\citenamefont {Zheng},
  \citenamefont {Dua},\ and\ \citenamefont {Jiang}}]{Zheng}%
  \BibitemOpen
  \bibfield  {author} {\bibinfo {author} {\bibfnamefont {H.}~\bibnamefont
  {Zheng}}, \bibinfo {author} {\bibfnamefont {A.}~\bibnamefont {Dua}}, \ and\
  \bibinfo {author} {\bibfnamefont {L.}~\bibnamefont {Jiang}},\ }\href
  {\doibase 10.1103/PhysRevB.92.245139} {\bibfield  {journal} {\bibinfo
  {journal} {Phys. Rev. B}\ }\textbf {\bibinfo {volume} {92}},\ \bibinfo
  {pages} {245139} (\bibinfo {year} {2015})}\BibitemShut {NoStop}%
\bibitem [{\citenamefont {Ortiz}\ and\ \citenamefont
  {Cobanera}(2016)}]{Cobanera}%
  \BibitemOpen
  \bibfield  {author} {\bibinfo {author} {\bibfnamefont {G.}~\bibnamefont
  {Ortiz}}\ and\ \bibinfo {author} {\bibfnamefont {E.}~\bibnamefont
  {Cobanera}},\ }\href {\doibase http://dx.doi.org/10.1016/j.aop.2016.05.020}
  {\bibfield  {journal} {\bibinfo  {journal} {Annals of Physics}\ }\textbf
  {\bibinfo {volume} {372}},\ \bibinfo {pages} {357 } (\bibinfo {year}
  {2016})}\BibitemShut {NoStop}%
\bibitem [{\citenamefont {Beenakker}(2015)}]{Beenakker1}%
  \BibitemOpen
  \bibfield  {author} {\bibinfo {author} {\bibfnamefont {C.~W.~J.}\
  \bibnamefont {Beenakker}},\ }\href {\doibase 10.1103/RevModPhys.87.1037}
  {\bibfield  {journal} {\bibinfo  {journal} {Rev. Mod. Phys.}\ }\textbf
  {\bibinfo {volume} {87}},\ \bibinfo {pages} {1037} (\bibinfo {year}
  {2015})}\BibitemShut {NoStop}%
\bibitem [{\citenamefont {Senthil}\ and\ \citenamefont
  {Fisher}(2000)}]{Senthil}%
  \BibitemOpen
  \bibfield  {author} {\bibinfo {author} {\bibfnamefont {T.}~\bibnamefont
  {Senthil}}\ and\ \bibinfo {author} {\bibfnamefont {M.~P.~A.}\ \bibnamefont
  {Fisher}},\ }\href {\doibase 10.1103/PhysRevB.61.9690} {\bibfield  {journal}
  {\bibinfo  {journal} {Phys. Rev. B}\ }\textbf {\bibinfo {volume} {61}},\
  \bibinfo {pages} {9690} (\bibinfo {year} {2000})}\BibitemShut {NoStop}%
\bibitem [{\citenamefont {Beenakker}(2014)}]{Beenakker2}%
  \BibitemOpen
  \bibfield  {author} {\bibinfo {author} {\bibfnamefont {C.~W.~J.}\
  \bibnamefont {Beenakker}},\ }\href {\doibase 10.1103/PhysRevLett.112.070604}
  {\bibfield  {journal} {\bibinfo  {journal} {Phys. Rev. Lett.}\ }\textbf
  {\bibinfo {volume} {112}},\ \bibinfo {pages} {070604} (\bibinfo {year}
  {2014})}\BibitemShut {NoStop}%
\bibitem [{\citenamefont {Chamon}\ \emph {et~al.}(2010)\citenamefont {Chamon},
  \citenamefont {Jackiw}, \citenamefont {Nishida}, \citenamefont {Pi},\ and\
  \citenamefont {Santos}}]{Chamon}%
  \BibitemOpen
  \bibfield  {author} {\bibinfo {author} {\bibfnamefont {C.}~\bibnamefont
  {Chamon}}, \bibinfo {author} {\bibfnamefont {R.}~\bibnamefont {Jackiw}},
  \bibinfo {author} {\bibfnamefont {Y.}~\bibnamefont {Nishida}}, \bibinfo
  {author} {\bibfnamefont {S.-Y.}\ \bibnamefont {Pi}}, \ and\ \bibinfo {author}
  {\bibfnamefont {L.}~\bibnamefont {Santos}},\ }\href {\doibase
  10.1103/PhysRevB.81.224515} {\bibfield  {journal} {\bibinfo  {journal} {Phys.
  Rev. B}\ }\textbf {\bibinfo {volume} {81}},\ \bibinfo {pages} {224515}
  (\bibinfo {year} {2010})}\BibitemShut {NoStop}%
\bibitem [{\citenamefont {Borstnik}\ \emph {et~al.}(2000)\citenamefont
  {Borstnik}, \citenamefont {Nielsen},\ and\ \citenamefont
  {Froggatt}}]{Nielsen1}%
  \BibitemOpen
  \bibfield  {author} {\bibinfo {author} {\bibfnamefont {N.~M.}\ \bibnamefont
  {Borstnik}}, \bibinfo {author} {\bibfnamefont {H.}~\bibnamefont {Nielsen}}, \
  and\ \bibinfo {author} {\bibfnamefont {C.}~\bibnamefont {Froggatt}},\
  }\href@noop {} {\  (\bibinfo {year} {2000})},\ \Eprint
  {http://arxiv.org/abs/arXiv:hep-th/0002048v1} {arXiv:hep-th/0002048v1}
  \BibitemShut {NoStop}%
\bibitem [{\citenamefont {Habara}\ \emph
  {et~al.}(2008{\natexlab{a}})\citenamefont {Habara}, \citenamefont {Nagatani},
  \citenamefont {Nielsen},\ and\ \citenamefont {Ninomiya}}]{Nielsen2}%
  \BibitemOpen
  \bibfield  {author} {\bibinfo {author} {\bibfnamefont {Y.}~\bibnamefont
  {Habara}}, \bibinfo {author} {\bibfnamefont {Y.}~\bibnamefont {Nagatani}},
  \bibinfo {author} {\bibfnamefont {H.~B.}\ \bibnamefont {Nielsen}}, \ and\
  \bibinfo {author} {\bibfnamefont {M.}~\bibnamefont {Ninomiya}},\ }\href
  {\doibase 10.1142/S0217751X08040342} {\bibfield  {journal} {\bibinfo
  {journal} {International Journal of Modern Physics A}\ }\textbf {\bibinfo
  {volume} {23}},\ \bibinfo {pages} {2733} (\bibinfo {year}
  {2008}{\natexlab{a}})}\BibitemShut {NoStop}%
\bibitem [{\citenamefont {Habara}\ \emph
  {et~al.}(2008{\natexlab{b}})\citenamefont {Habara}, \citenamefont {Nagatani},
  \citenamefont {Nielsen},\ and\ \citenamefont {Ninomiya}}]{Nielsen3}%
  \BibitemOpen
  \bibfield  {author} {\bibinfo {author} {\bibfnamefont {Y.}~\bibnamefont
  {Habara}}, \bibinfo {author} {\bibfnamefont {Y.}~\bibnamefont {Nagatani}},
  \bibinfo {author} {\bibfnamefont {H.~B.}\ \bibnamefont {Nielsen}}, \ and\
  \bibinfo {author} {\bibfnamefont {M.}~\bibnamefont {Ninomiya}},\ }\href
  {\doibase 10.1142/S0217751X08040354} {\bibfield  {journal} {\bibinfo
  {journal} {International Journal of Modern Physics A}\ }\textbf {\bibinfo
  {volume} {23}},\ \bibinfo {pages} {2771} (\bibinfo {year}
  {2008}{\natexlab{b}})}\BibitemShut {NoStop}%
\bibitem [{\citenamefont {Nielsen}\ and\ \citenamefont
  {Ninomiya}(2015)}]{Nielsen4}%
  \BibitemOpen
  \bibfield  {author} {\bibinfo {author} {\bibfnamefont {H.}~\bibnamefont
  {Nielsen}}\ and\ \bibinfo {author} {\bibfnamefont {M.}~\bibnamefont
  {Ninomiya}},\ }\href@noop {} {\  (\bibinfo {year} {2015})},\ \Eprint
  {http://arxiv.org/abs/arXiv:1510.03932v1} {arXiv:1510.03932v1} \BibitemShut
  {NoStop}%
\bibitem [{\citenamefont {Srivastava}\ \emph {et~al.}()\citenamefont
  {Srivastava}, \citenamefont {Widom},\ and\ \citenamefont
  {Swain}}]{Srivastava}%
  \BibitemOpen
  \bibfield  {author} {\bibinfo {author} {\bibfnamefont {Y.}~\bibnamefont
  {Srivastava}}, \bibinfo {author} {\bibfnamefont {A.}~\bibnamefont {Widom}}, \
  and\ \bibinfo {author} {\bibfnamefont {J.}~\bibnamefont {Swain}},\ }\href
  {https://arxiv.org/abs/hep-ph/9709434} {\ }\Eprint
  {http://arxiv.org/abs/arXiv:hep-ph/9709434} {arXiv:hep-ph/9709434}
  \BibitemShut {NoStop}%
\bibitem [{\citenamefont {Bedell}()}]{Bedell1}%
  \BibitemOpen
  \bibfield  {author} {\bibinfo {author} {\bibfnamefont {K.}~\bibnamefont
  {Bedell}},\ }\href@noop {} {\enquote {\bibinfo {title} {{\it Quantum
  Liquids}},}\ }\bibinfo {note} {Unpublished lecture notes}\BibitemShut
  {NoStop}%
\bibitem [{\citenamefont {Cowan}(2005)}]{Cowan}%
  \BibitemOpen
  \bibfield  {author} {\bibinfo {author} {\bibfnamefont {B.}~\bibnamefont
  {Cowan}},\ }\href@noop {} {\emph {\bibinfo {title} {Topics in Statistical
  Mechanics}}}\ (\bibinfo  {publisher} {Imperial College Press},\ \bibinfo
  {address} {United Kingdom},\ \bibinfo {year} {2005})\BibitemShut {NoStop}%
\bibitem [{\citenamefont {Heath}\ and\ \citenamefont {Bedell}(shed)}]{Heath}%
  \BibitemOpen
  \bibfield  {author} {\bibinfo {author} {\bibfnamefont {J.~T.}\ \bibnamefont
  {Heath}}\ and\ \bibinfo {author} {\bibfnamefont {K.~S.}\ \bibnamefont
  {Bedell}},\ }\href@noop {} {\  (\bibinfo {year} {to be
  published})}\BibitemShut {NoStop}%
\bibitem [{\citenamefont {Koepf}(2014)}]{Koepf}%
  \BibitemOpen
  \bibfield  {author} {\bibinfo {author} {\bibfnamefont {W.}~\bibnamefont
  {Koepf}},\ }\href@noop {} {\emph {\bibinfo {title} {Hypergeometric Summation:
  An Algorithmic Approach to Summation and Special Function Identities (Second
  Edition)}}}\ (\bibinfo  {publisher} {Springer},\ \bibinfo {address}
  {Berlin},\ \bibinfo {year} {2014})\BibitemShut {NoStop}%
\bibitem [{\citenamefont {Haldane}(1991)}]{Haldane}%
  \BibitemOpen
  \bibfield  {author} {\bibinfo {author} {\bibfnamefont {F.~D.~M.}\
  \bibnamefont {Haldane}},\ }\href {\doibase 10.1103/PhysRevLett.67.937}
  {\bibfield  {journal} {\bibinfo  {journal} {Phys. Rev. Lett.}\ }\textbf
  {\bibinfo {volume} {67}},\ \bibinfo {pages} {937} (\bibinfo {year}
  {1991})}\BibitemShut {NoStop}%
\bibitem [{\citenamefont {Wu}(1994)}]{Wu}%
  \BibitemOpen
  \bibfield  {author} {\bibinfo {author} {\bibfnamefont {Y.-S.}\ \bibnamefont
  {Wu}},\ }\href {\doibase 10.1103/PhysRevLett.73.922} {\bibfield  {journal}
  {\bibinfo  {journal} {Phys. Rev. Lett.}\ }\textbf {\bibinfo {volume} {73}},\
  \bibinfo {pages} {922} (\bibinfo {year} {1994})}\BibitemShut {NoStop}%
\bibitem [{\citenamefont {Khare}(2005)}]{Khare}%
  \BibitemOpen
  \bibfield  {author} {\bibinfo {author} {\bibfnamefont {A.}~\bibnamefont
  {Khare}},\ }\href@noop {} {\emph {\bibinfo {title} {Fractional Statistics and
  Quantum Theory (2nd edition)}}}\ (\bibinfo  {publisher} {World Scientific
  Publishing},\ \bibinfo {address} {Singapore},\ \bibinfo {year}
  {2005})\BibitemShut {NoStop}%
\bibitem [{\citenamefont {Green}(1953)}]{Green}%
  \BibitemOpen
  \bibfield  {author} {\bibinfo {author} {\bibfnamefont {H.~S.}\ \bibnamefont
  {Green}},\ }\href {\doibase 10.1103/PhysRev.90.270} {\bibfield  {journal}
  {\bibinfo  {journal} {Phys. Rev.}\ }\textbf {\bibinfo {volume} {90}},\
  \bibinfo {pages} {270} (\bibinfo {year} {1953})}\BibitemShut {NoStop}%
\bibitem [{\citenamefont {Lerda}(1992)}]{Lerda}%
  \BibitemOpen
  \bibfield  {author} {\bibinfo {author} {\bibfnamefont {A.}~\bibnamefont
  {Lerda}},\ }\href@noop {} {\emph {\bibinfo {title} {Anyons, Quantum Mechanics
  of Particles with Fractional Statistics}}}\ (\bibinfo  {publisher}
  {Spring-Verlag},\ \bibinfo {address} {Berlin},\ \bibinfo {year}
  {1992})\BibitemShut {NoStop}%
\bibitem [{\citenamefont {Murthy}\ and\ \citenamefont
  {Shankar}(2009)}]{Murthy}%
  \BibitemOpen
  \bibfield  {author} {\bibinfo {author} {\bibfnamefont {M.}~\bibnamefont
  {Murthy}}\ and\ \bibinfo {author} {\bibfnamefont {R.}~\bibnamefont
  {Shankar}},\ }\href
  {https://www.imsc.res.in/xmlui/bitstream/handle/123456789/334/MR120.pdf?sequence=1}
  {\emph {\bibinfo {title} {Fractional Statistics: From Pauli to Haldane}}},\
  \bibinfo {number} {The IMSc Report No. 120}\ (\bibinfo {year}
  {2009})\BibitemShut {NoStop}%
\bibitem [{\citenamefont {Rachidi}\ \emph {et~al.}(1997)\citenamefont
  {Rachidi}, \citenamefont {Saidi},\ and\ \citenamefont {Zerouaoui}}]{Rachidi}%
  \BibitemOpen
  \bibfield  {author} {\bibinfo {author} {\bibfnamefont {M.}~\bibnamefont
  {Rachidi}}, \bibinfo {author} {\bibfnamefont {E.}~\bibnamefont {Saidi}}, \
  and\ \bibinfo {author} {\bibfnamefont {J.}~\bibnamefont {Zerouaoui}},\ }\href
  {http://www.sciencedirect.com/science/article/pii/S0370269397008356}
  {\bibfield  {journal} {\bibinfo  {journal} {Phys. Lett. B}\ }\textbf
  {\bibinfo {volume} {409}},\ \bibinfo {pages} {349} (\bibinfo {year}
  {1997})}\BibitemShut {NoStop}%
\bibitem [{\citenamefont {Kaplan}(2017)}]{Kaplan}%
  \BibitemOpen
  \bibfield  {author} {\bibinfo {author} {\bibfnamefont {I.~G.}\ \bibnamefont
  {Kaplan}},\ }\href@noop {} {\emph {\bibinfo {title} {The Pauli Exclusion
  Principle; Origin, Verifications, and Applications}}}\ (\bibinfo  {publisher}
  {Wiley},\ \bibinfo {address} {United Kingdom},\ \bibinfo {year}
  {2017})\BibitemShut {NoStop}%
\bibitem [{\citenamefont {Graham}\ \emph {et~al.}(1988)\citenamefont {Graham},
  \citenamefont {Knuth},\ and\ \citenamefont {Patashnik}}]{Graham}%
  \BibitemOpen
  \bibfield  {author} {\bibinfo {author} {\bibfnamefont {R.~L.}\ \bibnamefont
  {Graham}}, \bibinfo {author} {\bibfnamefont {D.~E.}\ \bibnamefont {Knuth}}, \
  and\ \bibinfo {author} {\bibfnamefont {O.}~\bibnamefont {Patashnik}},\
  }\href@noop {} {\emph {\bibinfo {title} {Concrete Mathematics}}}\ (\bibinfo
  {publisher} {Addison-Wesley},\ \bibinfo {address} {Reading, MA},\ \bibinfo
  {year} {1988})\BibitemShut {NoStop}%
\bibitem [{\citenamefont {Knuth}(1997)}]{Knuth}%
  \BibitemOpen
  \bibfield  {author} {\bibinfo {author} {\bibfnamefont {D.~E.}\ \bibnamefont
  {Knuth}},\ }\href@noop {} {\emph {\bibinfo {title} {The Art of Computer
  Programming, Volume 1: Fundamental Algorithms (Third Edition)}}}\ (\bibinfo
  {publisher} {Addison-Wesley Professional},\ \bibinfo {address} {Reading,
  MA},\ \bibinfo {year} {1997})\BibitemShut {NoStop}%
\bibitem [{{\relax DLMF}()}]{NIST}%
  \BibitemOpen
  {\relax DLMF},\ \href {http://dlmf.nist.gov/} {\enquote {\bibinfo {title}
  {{\it NIST Digital Library of Mathematical Functions}},}\ }\bibinfo
  {howpublished} {http://dlmf.nist.gov/, Release 1.0.15 of 2017-06-01},\
  \bibinfo {note} {f.~W.~J. Olver, A.~B. {Olde Daalhuis}, D.~W. Lozier, B.~I.
  Schneider, R.~F. Boisvert, C.~W. Clark, B.~R. Miller and B.~V. Saunders,
  eds.}\BibitemShut {Stop}%
\bibitem [{\citenamefont {Pines}\ and\ \citenamefont
  {Nozi\'eres}(1989)}]{Pines}%
  \BibitemOpen
  \bibfield  {author} {\bibinfo {author} {\bibfnamefont {D.}~\bibnamefont
  {Pines}}\ and\ \bibinfo {author} {\bibfnamefont {P.}~\bibnamefont
  {Nozi\'eres}},\ }\href@noop {} {\emph {\bibinfo {title} {The Theory of
  Quantum Liquids, Vol. I}}}\ (\bibinfo  {publisher} {Westview Press},\
  \bibinfo {address} {Boulder, CO},\ \bibinfo {year} {1989})\BibitemShut
  {NoStop}%
\bibitem [{\citenamefont {Pauling}(1935)}]{Pauling}%
  \BibitemOpen
  \bibfield  {author} {\bibinfo {author} {\bibfnamefont {L.}~\bibnamefont
  {Pauling}},\ }\href {http://pubs.acs.org/doi/pdf/10.1021/ja01315a102}
  {\bibfield  {journal} {\bibinfo  {journal} {Journal of the American Chemical
  Society}\ }\textbf {\bibinfo {volume} {57}},\ \bibinfo {pages} {2680}
  (\bibinfo {year} {1935})}\BibitemShut {NoStop}%
\bibitem [{\citenamefont {Bramwell}\ and\ \citenamefont
  {Gingras}(2001)}]{Gingras}%
  \BibitemOpen
  \bibfield  {author} {\bibinfo {author} {\bibfnamefont {S.~T.}\ \bibnamefont
  {Bramwell}}\ and\ \bibinfo {author} {\bibfnamefont {M.~J.~P.}\ \bibnamefont
  {Gingras}},\ }\href {\doibase 10.1126/science.1064761} {\bibfield  {journal}
  {\bibinfo  {journal} {Science}\ }\textbf {\bibinfo {volume} {294}},\ \bibinfo
  {pages} {1495} (\bibinfo {year} {2001})}\BibitemShut {NoStop}%
\bibitem [{\citenamefont {Ferraro}\ \emph {et~al.}(2015)\citenamefont
  {Ferraro}, \citenamefont {Rech}, \citenamefont {Jonckheere},\ and\
  \citenamefont {Martin}}]{Ferraro}%
  \BibitemOpen
  \bibfield  {author} {\bibinfo {author} {\bibfnamefont {D.}~\bibnamefont
  {Ferraro}}, \bibinfo {author} {\bibfnamefont {J.}~\bibnamefont {Rech}},
  \bibinfo {author} {\bibfnamefont {T.}~\bibnamefont {Jonckheere}}, \ and\
  \bibinfo {author} {\bibfnamefont {T.}~\bibnamefont {Martin}},\ }\href
  {\doibase 10.1103/PhysRevB.91.075406} {\bibfield  {journal} {\bibinfo
  {journal} {Phys. Rev. B}\ }\textbf {\bibinfo {volume} {91}},\ \bibinfo
  {pages} {075406} (\bibinfo {year} {2015})}\BibitemShut {NoStop}%
\bibitem [{\citenamefont {DeMarco}\ and\ \citenamefont {Jin}(1999)}]{DeMarco1}%
  \BibitemOpen
  \bibfield  {author} {\bibinfo {author} {\bibfnamefont {B.}~\bibnamefont
  {DeMarco}}\ and\ \bibinfo {author} {\bibfnamefont {D.~S.}\ \bibnamefont
  {Jin}},\ }\href {\doibase 10.1126/science.285.5434.1703} {\bibfield
  {journal} {\bibinfo  {journal} {Science}\ }\textbf {\bibinfo {volume}
  {285}},\ \bibinfo {pages} {1703} (\bibinfo {year} {1999})}\BibitemShut
  {NoStop}%
\bibitem [{\citenamefont {DeMarco}(2001)}]{DeMarco2}%
  \BibitemOpen
  \bibfield  {author} {\bibinfo {author} {\bibfnamefont {B.}~\bibnamefont
  {DeMarco}},\ }\href
  {https://jila.colorado.edu/jin/sites/default/files/files/2001_demarco.pdf}
  {Ph.D. thesis},\ \bibinfo  {school} {University of Colorado} (\bibinfo {year}
  {2001})\BibitemShut {NoStop}%
\bibitem [{\citenamefont {Nayak}\ \emph {et~al.}(2008)\citenamefont {Nayak},
  \citenamefont {Simon}, \citenamefont {Stern}, \citenamefont {Freedman},\ and\
  \citenamefont {Das~Sarma}}]{Freedman}%
  \BibitemOpen
  \bibfield  {author} {\bibinfo {author} {\bibfnamefont {C.}~\bibnamefont
  {Nayak}}, \bibinfo {author} {\bibfnamefont {S.~H.}\ \bibnamefont {Simon}},
  \bibinfo {author} {\bibfnamefont {A.}~\bibnamefont {Stern}}, \bibinfo
  {author} {\bibfnamefont {M.}~\bibnamefont {Freedman}}, \ and\ \bibinfo
  {author} {\bibfnamefont {S.}~\bibnamefont {Das~Sarma}},\ }\href {\doibase
  10.1103/RevModPhys.80.1083} {\bibfield  {journal} {\bibinfo  {journal} {Rev.
  Mod. Phys.}\ }\textbf {\bibinfo {volume} {80}},\ \bibinfo {pages} {1083}
  (\bibinfo {year} {2008})}\BibitemShut {NoStop}%
\bibitem [{\citenamefont {Kempkes}\ \emph {et~al.}(2016)\citenamefont
  {Kempkes}, \citenamefont {Quelle},\ and\ \citenamefont {Smith}}]{Kempkes1}%
  \BibitemOpen
  \bibfield  {author} {\bibinfo {author} {\bibfnamefont {S.}~\bibnamefont
  {Kempkes}}, \bibinfo {author} {\bibfnamefont {A.}~\bibnamefont {Quelle}}, \
  and\ \bibinfo {author} {\bibfnamefont {C.~M.}\ \bibnamefont {Smith}},\ }\href
  {https://www.nature.com/articles/srep38530} {\bibfield  {journal} {\bibinfo
  {journal} {Scientific Reports}\ }\textbf {\bibinfo {volume} {6}},\ \bibinfo
  {pages} {38530} (\bibinfo {year} {2016})}\BibitemShut {NoStop}%
\bibitem [{\citenamefont {Kempkes}(2016)}]{Kempkes2}%
  \BibitemOpen
  \bibfield  {author} {\bibinfo {author} {\bibfnamefont {S.}~\bibnamefont
  {Kempkes}},\ }\href
  {https://web.science.uu.nl/ITF/Teaching/2016/2016Kempkes.pdf} {Ph.D.
  thesis},\ \bibinfo  {school} {Utrecht University} (\bibinfo {year}
  {2016})\BibitemShut {NoStop}%
\bibitem [{\citenamefont {Quelle}\ \emph {et~al.}(2016)\citenamefont {Quelle},
  \citenamefont {Cobanera},\ and\ \citenamefont {Smith}}]{Kempkes3}%
  \BibitemOpen
  \bibfield  {author} {\bibinfo {author} {\bibfnamefont {A.}~\bibnamefont
  {Quelle}}, \bibinfo {author} {\bibfnamefont {E.}~\bibnamefont {Cobanera}}, \
  and\ \bibinfo {author} {\bibfnamefont {C.~M.}\ \bibnamefont {Smith}},\ }\href
  {\doibase 10.1103/PhysRevB.94.075133} {\bibfield  {journal} {\bibinfo
  {journal} {Phys. Rev. B}\ }\textbf {\bibinfo {volume} {94}},\ \bibinfo
  {pages} {075133} (\bibinfo {year} {2016})}\BibitemShut {NoStop}%
\bibitem [{\citenamefont {Hill}(1994)}]{Hill}%
  \BibitemOpen
  \bibfield  {author} {\bibinfo {author} {\bibfnamefont {T.~L.}\ \bibnamefont
  {Hill}},\ }\href@noop {} {\emph {\bibinfo {title} {The thermodynamics of
  small systems (Second Edition)}}}\ (\bibinfo  {publisher} {Dover
  Publishing},\ \bibinfo {address} {New York},\ \bibinfo {year}
  {1994})\BibitemShut {NoStop}%
\bibitem [{\citenamefont {Erten}\ \emph {et~al.}(2017)\citenamefont {Erten},
  \citenamefont {Chang}, \citenamefont {Coleman},\ and\ \citenamefont
  {Tsvelik}}]{Erten}%
  \BibitemOpen
  \bibfield  {author} {\bibinfo {author} {\bibfnamefont {O.}~\bibnamefont
  {Erten}}, \bibinfo {author} {\bibfnamefont {P.-Y.}\ \bibnamefont {Chang}},
  \bibinfo {author} {\bibfnamefont {P.}~\bibnamefont {Coleman}}, \ and\
  \bibinfo {author} {\bibfnamefont {A.~M.}\ \bibnamefont {Tsvelik}},\ }\href
  {\doibase 10.1103/PhysRevLett.119.057603} {\bibfield  {journal} {\bibinfo
  {journal} {Phys. Rev. Lett.}\ }\textbf {\bibinfo {volume} {119}},\ \bibinfo
  {pages} {057603} (\bibinfo {year} {2017})}\BibitemShut {NoStop}%
\bibitem [{\citenamefont {Eder}\ \emph {et~al.}(1998)\citenamefont {Eder},
  \citenamefont {Rogojanu},\ and\ \citenamefont {Sawatzky}}]{Eder}%
  \BibitemOpen
  \bibfield  {author} {\bibinfo {author} {\bibfnamefont {R.}~\bibnamefont
  {Eder}}, \bibinfo {author} {\bibfnamefont {O.}~\bibnamefont {Rogojanu}}, \
  and\ \bibinfo {author} {\bibfnamefont {G.}~\bibnamefont {Sawatzky}},\ }\href
  {\doibase 10.1103/PhysRevB.58.7599} {\bibfield  {journal} {\bibinfo
  {journal} {Phys. Rev. B}\ }\textbf {\bibinfo {volume} {58}},\ \bibinfo
  {pages} {7599} (\bibinfo {year} {1998})}\BibitemShut {NoStop}%
\bibitem [{\citenamefont {\"Ostlund}(2007)}]{Ostlund}%
  \BibitemOpen
  \bibfield  {author} {\bibinfo {author} {\bibfnamefont {S.}~\bibnamefont
  {\"Ostlund}},\ }\href {\doibase 10.1103/PhysRevB.76.153101} {\bibfield
  {journal} {\bibinfo  {journal} {Phys. Rev. B}\ }\textbf {\bibinfo {volume}
  {76}},\ \bibinfo {pages} {153101} (\bibinfo {year} {2007})}\BibitemShut
  {NoStop}%
\bibitem [{\citenamefont {Li}\ \emph {et~al.}(2014)\citenamefont {Li} \emph
  {et~al.}}]{Li}%
  \BibitemOpen
  \bibfield  {author} {\bibinfo {author} {\bibfnamefont {G.}~\bibnamefont {Li}}
  \emph {et~al.},\ }\href {\doibase 10.1126/science.1250366} {\bibfield
  {journal} {\bibinfo  {journal} {Science}\ }\textbf {\bibinfo {volume}
  {346}},\ \bibinfo {pages} {1208} (\bibinfo {year} {2014})}\BibitemShut
  {NoStop}%
\bibitem [{\citenamefont {Aronson}\ \emph {et~al.}(1999)\citenamefont
  {Aronson}, \citenamefont {Sarrao}, \citenamefont {Fisk}, \citenamefont
  {Whitton},\ and\ \citenamefont {Brandt}}]{Aronson}%
  \BibitemOpen
  \bibfield  {author} {\bibinfo {author} {\bibfnamefont {M.~C.}\ \bibnamefont
  {Aronson}}, \bibinfo {author} {\bibfnamefont {J.~L.}\ \bibnamefont {Sarrao}},
  \bibinfo {author} {\bibfnamefont {Z.}~\bibnamefont {Fisk}}, \bibinfo {author}
  {\bibfnamefont {M.}~\bibnamefont {Whitton}}, \ and\ \bibinfo {author}
  {\bibfnamefont {B.~L.}\ \bibnamefont {Brandt}},\ }\href {\doibase
  10.1103/PhysRevB.59.4720} {\bibfield  {journal} {\bibinfo  {journal} {Phys.
  Rev. B}\ }\textbf {\bibinfo {volume} {59}},\ \bibinfo {pages} {4720}
  (\bibinfo {year} {1999})}\BibitemShut {NoStop}%
\bibitem [{\citenamefont {Yin}\ \emph {et~al.}(2015)\citenamefont {Yin} \emph
  {et~al.}}]{Ding1}%
  \BibitemOpen
  \bibfield  {author} {\bibinfo {author} {\bibfnamefont {J.-X.}\ \bibnamefont
  {Yin}} \emph {et~al.},\ }\href
  {http://www.nature.com/nphys/journal/v11/n7/full/nphys3371.html} {\bibfield
  {journal} {\bibinfo  {journal} {Nature Physics}\ }\textbf {\bibinfo {volume}
  {11}},\ \bibinfo {pages} {543} (\bibinfo {year} {2015})}\BibitemShut
  {NoStop}%
\bibitem [{\citenamefont {Zhang}\ \emph {et~al.}()\citenamefont {Zhang} \emph
  {et~al.}}]{Ding2}%
  \BibitemOpen
  \bibfield  {author} {\bibinfo {author} {\bibfnamefont {P.}~\bibnamefont
  {Zhang}} \emph {et~al.},\ }\href {https://arxiv.org/abs/1706.05163} {\
  }\Eprint {http://arxiv.org/abs/arXiv:1706.05163v1} {arXiv:1706.05163v1}
  \BibitemShut {NoStop}%
\bibitem [{\citenamefont {Sasaki}\ \emph {et~al.}(2011)\citenamefont {Sasaki}
  \emph {et~al.}}]{Sasaki}%
  \BibitemOpen
  \bibfield  {author} {\bibinfo {author} {\bibfnamefont {S.}~\bibnamefont
  {Sasaki}} \emph {et~al.},\ }\href {\doibase 10.1103/PhysRevLett.107.217001}
  {\bibfield  {journal} {\bibinfo  {journal} {Phys. Rev. Lett.}\ }\textbf
  {\bibinfo {volume} {107}},\ \bibinfo {pages} {217001} (\bibinfo {year}
  {2011})}\BibitemShut {NoStop}%
\bibitem [{\citenamefont {Levy}\ \emph {et~al.}(2013)\citenamefont {Levy},
  \citenamefont {Zhang}, \citenamefont {Ha}, \citenamefont {Sharifi},
  \citenamefont {Talin}, \citenamefont {Kuk},\ and\ \citenamefont
  {Stroscio}}]{Levy}%
  \BibitemOpen
  \bibfield  {author} {\bibinfo {author} {\bibfnamefont {N.}~\bibnamefont
  {Levy}}, \bibinfo {author} {\bibfnamefont {T.}~\bibnamefont {Zhang}},
  \bibinfo {author} {\bibfnamefont {J.}~\bibnamefont {Ha}}, \bibinfo {author}
  {\bibfnamefont {F.}~\bibnamefont {Sharifi}}, \bibinfo {author} {\bibfnamefont
  {A.~A.}\ \bibnamefont {Talin}}, \bibinfo {author} {\bibfnamefont
  {Y.}~\bibnamefont {Kuk}}, \ and\ \bibinfo {author} {\bibfnamefont {J.~A.}\
  \bibnamefont {Stroscio}},\ }\href {\doibase 10.1103/PhysRevLett.110.117001}
  {\bibfield  {journal} {\bibinfo  {journal} {Phys. Rev. Lett.}\ }\textbf
  {\bibinfo {volume} {110}},\ \bibinfo {pages} {117001} (\bibinfo {year}
  {2013})}\BibitemShut {NoStop}%
\bibitem [{\citenamefont {Kitaev}(2006)}]{Kitaev_honey}%
  \BibitemOpen
  \bibfield  {author} {\bibinfo {author} {\bibfnamefont {A.}~\bibnamefont
  {Kitaev}},\ }\href {\doibase http://dx.doi.org/10.1016/j.aop.2005.10.005}
  {\bibfield  {journal} {\bibinfo  {journal} {Annals of Physics}\ }\textbf
  {\bibinfo {volume} {321}},\ \bibinfo {pages} {2 } (\bibinfo {year} {2006})},\
  \bibinfo {note} {\,January Special Issue}\BibitemShut {NoStop}%
\bibitem [{\citenamefont {Do}\ \emph {et~al.}()\citenamefont {Do} \emph
  {et~al.}}]{Do}%
  \BibitemOpen
  \bibfield  {author} {\bibinfo {author} {\bibfnamefont {S.-H.}\ \bibnamefont
  {Do}} \emph {et~al.},\ }\href {https://arxiv.org/abs/1703.01081} {\ }\Eprint
  {http://arxiv.org/abs/arXiv:1703.01081v1} {arXiv:1703.01081v1} \BibitemShut
  {NoStop}%
\bibitem [{\citenamefont {Glamazda}\ \emph {et~al.}(2016)\citenamefont
  {Glamazda} \emph {et~al.}}]{Glamazda}%
  \BibitemOpen
  \bibfield  {author} {\bibinfo {author} {\bibfnamefont {A.}~\bibnamefont
  {Glamazda}} \emph {et~al.},\ }\href
  {https://www.nature.com/articles/ncomms12286} {\bibfield  {journal} {\bibinfo
   {journal} {Nature Communications}\ }\textbf {\bibinfo {volume} {7}},\
  \bibinfo {pages} {12286} (\bibinfo {year} {2016})}\BibitemShut {NoStop}%
\bibitem [{\citenamefont {Hermanns}\ and\ \citenamefont
  {Trebst}(2014)}]{Trebst}%
  \BibitemOpen
  \bibfield  {author} {\bibinfo {author} {\bibfnamefont {M.}~\bibnamefont
  {Hermanns}}\ and\ \bibinfo {author} {\bibfnamefont {S.}~\bibnamefont
  {Trebst}},\ }\href {\doibase 10.1103/PhysRevB.89.235102} {\bibfield
  {journal} {\bibinfo  {journal} {Phys. Rev. B}\ }\textbf {\bibinfo {volume}
  {89}},\ \bibinfo {pages} {235102} (\bibinfo {year} {2014})}\BibitemShut
  {NoStop}%
\bibitem [{\citenamefont {Nasu}\ \emph {et~al.}(2015)\citenamefont {Nasu},
  \citenamefont {Udagawa},\ and\ \citenamefont {Motome}}]{Nasu}%
  \BibitemOpen
  \bibfield  {author} {\bibinfo {author} {\bibfnamefont {J.}~\bibnamefont
  {Nasu}}, \bibinfo {author} {\bibfnamefont {M.}~\bibnamefont {Udagawa}}, \
  and\ \bibinfo {author} {\bibfnamefont {Y.}~\bibnamefont {Motome}},\ }\href
  {\doibase 10.1103/PhysRevB.92.115122} {\bibfield  {journal} {\bibinfo
  {journal} {Phys. Rev. B}\ }\textbf {\bibinfo {volume} {92}},\ \bibinfo
  {pages} {115122} (\bibinfo {year} {2015})}\BibitemShut {NoStop}%
\bibitem [{\citenamefont {Sandilands}\ \emph {et~al.}(2015)\citenamefont
  {Sandilands}, \citenamefont {Tian}, \citenamefont {Plumb}, \citenamefont
  {Kim},\ and\ \citenamefont {Burch}}]{Burch}%
  \BibitemOpen
  \bibfield  {author} {\bibinfo {author} {\bibfnamefont {L.~J.}\ \bibnamefont
  {Sandilands}}, \bibinfo {author} {\bibfnamefont {Y.}~\bibnamefont {Tian}},
  \bibinfo {author} {\bibfnamefont {K.~W.}\ \bibnamefont {Plumb}}, \bibinfo
  {author} {\bibfnamefont {Y.-J.}\ \bibnamefont {Kim}}, \ and\ \bibinfo
  {author} {\bibfnamefont {K.~S.}\ \bibnamefont {Burch}},\ }\href {\doibase
  10.1103/PhysRevLett.114.147201} {\bibfield  {journal} {\bibinfo  {journal}
  {Phys. Rev. Lett.}\ }\textbf {\bibinfo {volume} {114}},\ \bibinfo {pages}
  {147201} (\bibinfo {year} {2015})}\BibitemShut {NoStop}%
\bibitem [{\citenamefont {Janka}(1993)}]{Janka1}%
  \BibitemOpen
  \bibfield  {author} {\bibinfo {author} {\bibfnamefont {H.-T.}\ \bibnamefont
  {Janka}},\ }in\ \href
  {http://www.iaea.org/inis/collection/NCLCollectionStore/_Public/28/026/28026738.pdf}
  {\emph {\bibinfo {booktitle} {Proc. Vulcano Workshop 1992 Frontier Objects in
  Astrophysics and Particle Physics}}},\ Vol.~\bibinfo {volume} {40},\ \bibinfo
  {editor} {edited by\ \bibinfo {editor} {\bibfnamefont {F.}~\bibnamefont
  {Giovannelli}}\ and\ \bibinfo {editor} {\bibfnamefont {G.}~\bibnamefont
  {Mannocchi}}}\ (\bibinfo {address} {Societa Italiana di Fisica, Bologna},\
  \bibinfo {year} {1993})\ pp.\ \bibinfo {pages} {345--374}\BibitemShut
  {NoStop}%
\bibitem [{\citenamefont {Zuber}(2012)}]{Zuber}%
  \BibitemOpen
  \bibfield  {author} {\bibinfo {author} {\bibfnamefont {K.}~\bibnamefont
  {Zuber}},\ }\href@noop {} {\emph {\bibinfo {title} {Neutrino Physics (Second
  Edition)}}}\ (\bibinfo  {publisher} {CRC Press, Taylor \& Francis Group},\
  \bibinfo {address} {Boca Raton, FL},\ \bibinfo {year} {2012})\BibitemShut
  {NoStop}%
\bibitem [{\citenamefont {M.}\ \emph {et~al.}(1987)\citenamefont {M.} \emph
  {et~al.}}]{Aglietta}%
  \BibitemOpen
  \bibfield  {author} {\bibinfo {author} {\bibfnamefont {A.}~\bibnamefont {M.}}
  \emph {et~al.},\ }\href
  {http://iopscience.iop.org/article/10.1209/0295-5075/3/12/011/meta}
  {\bibfield  {journal} {\bibinfo  {journal} {Europhys. Lett.}\ }\textbf
  {\bibinfo {volume} {3}} (\bibinfo {year} {1987})},\ \bibinfo {note} {(with
  Frejus Collaboration)}\BibitemShut {NoStop}%
\bibitem [{\citenamefont {Alekseev}\ \emph {et~al.}(1987)\citenamefont
  {Alekseev}, \citenamefont {Alekseeva}, \citenamefont {Volchenko},\ and\
  \citenamefont {Krivosheina}}]{Alekseev}%
  \BibitemOpen
  \bibfield  {author} {\bibinfo {author} {\bibfnamefont {E.}~\bibnamefont
  {Alekseev}}, \bibinfo {author} {\bibfnamefont {L.}~\bibnamefont {Alekseeva}},
  \bibinfo {author} {\bibfnamefont {V.~I.}\ \bibnamefont {Volchenko}}, \ and\
  \bibinfo {author} {\bibfnamefont {I.~V.}\ \bibnamefont {Krivosheina}},\
  }\href {http://www.jetpletters.ac.ru/ps/1245/article_18825.shtml} {\bibfield
  {journal} {\bibinfo  {journal} {JETP Letters}\ }\textbf {\bibinfo {volume}
  {45}},\ \bibinfo {pages} {589} (\bibinfo {year} {1987})}\BibitemShut
  {NoStop}%
\bibitem [{\citenamefont {Bionta}\ \emph {et~al.}(1987)\citenamefont {Bionta}
  \emph {et~al.}}]{Bionta}%
  \BibitemOpen
  \bibfield  {author} {\bibinfo {author} {\bibfnamefont {R.~M.}\ \bibnamefont
  {Bionta}} \emph {et~al.},\ }\href {\doibase 10.1103/PhysRevLett.58.1494}
  {\bibfield  {journal} {\bibinfo  {journal} {Phys. Rev. Lett.}\ }\textbf
  {\bibinfo {volume} {58}},\ \bibinfo {pages} {1494} (\bibinfo {year}
  {1987})},\ \bibinfo {note} {(with IMB Collaboration)}\BibitemShut {NoStop}%
\bibitem [{\citenamefont {Hirata}\ \emph {et~al.}(1987)\citenamefont {Hirata}
  \emph {et~al.}}]{Hirata}%
  \BibitemOpen
  \bibfield  {author} {\bibinfo {author} {\bibfnamefont {K.}~\bibnamefont
  {Hirata}} \emph {et~al.},\ }\href {\doibase 10.1103/PhysRevLett.58.1490}
  {\bibfield  {journal} {\bibinfo  {journal} {Phys. Rev. Lett.}\ }\textbf
  {\bibinfo {volume} {58}},\ \bibinfo {pages} {1490} (\bibinfo {year}
  {1987})},\ \bibinfo {note} {(with Kamiokande Collaberation)}\BibitemShut
  {NoStop}%
\bibitem [{\citenamefont {Y\"uksel}\ and\ \citenamefont
  {Beacom}(2007)}]{Yuksel}%
  \BibitemOpen
  \bibfield  {author} {\bibinfo {author} {\bibfnamefont {H.}~\bibnamefont
  {Y\"uksel}}\ and\ \bibinfo {author} {\bibfnamefont {J.~F.}\ \bibnamefont
  {Beacom}},\ }\href {\doibase 10.1103/PhysRevD.76.083007} {\bibfield
  {journal} {\bibinfo  {journal} {Phys. Rev. D}\ }\textbf {\bibinfo {volume}
  {76}},\ \bibinfo {pages} {083007} (\bibinfo {year} {2007})}\BibitemShut
  {NoStop}%
\bibitem [{\citenamefont {Costantini}\ \emph {et~al.}(2007)\citenamefont
  {Costantini}, \citenamefont {Ianni}, \citenamefont {Pagliaroli},\ and\
  \citenamefont {Vissani}}]{Costantini}%
  \BibitemOpen
  \bibfield  {author} {\bibinfo {author} {\bibfnamefont {M.~L.}\ \bibnamefont
  {Costantini}}, \bibinfo {author} {\bibfnamefont {A.}~\bibnamefont {Ianni}},
  \bibinfo {author} {\bibfnamefont {G.}~\bibnamefont {Pagliaroli}}, \ and\
  \bibinfo {author} {\bibfnamefont {F.}~\bibnamefont {Vissani}},\ }\href
  {http://stacks.iop.org/1475-7516/2007/i=05/a=014} {\bibfield  {journal}
  {\bibinfo  {journal} {Journal of Cosmology and Astroparticle Physics}\
  }\textbf {\bibinfo {volume} {2007}},\ \bibinfo {pages} {014} (\bibinfo {year}
  {2007})}\BibitemShut {NoStop}%
\bibitem [{\citenamefont {Dolgov}\ and\ \citenamefont
  {Smirnov}(2005)}]{Dolgov2}%
  \BibitemOpen
  \bibfield  {author} {\bibinfo {author} {\bibfnamefont {A.}~\bibnamefont
  {Dolgov}}\ and\ \bibinfo {author} {\bibfnamefont {A.}~\bibnamefont
  {Smirnov}},\ }\href {\doibase
  http://dx.doi.org/10.1016/j.physletb.2005.06.035} {\bibfield  {journal}
  {\bibinfo  {journal} {Physics Letters B}\ }\textbf {\bibinfo {volume}
  {621}},\ \bibinfo {pages} {1 } (\bibinfo {year} {2005})}\BibitemShut
  {NoStop}%
\bibitem [{\citenamefont {Choubey}\ and\ \citenamefont {Kar}(2006)}]{Choubey}%
  \BibitemOpen
  \bibfield  {author} {\bibinfo {author} {\bibfnamefont {S.}~\bibnamefont
  {Choubey}}\ and\ \bibinfo {author} {\bibfnamefont {K.}~\bibnamefont {Kar}},\
  }\href {\doibase http://dx.doi.org/10.1016/j.physletb.2006.01.041} {\bibfield
   {journal} {\bibinfo  {journal} {Physics Letters B}\ }\textbf {\bibinfo
  {volume} {634}},\ \bibinfo {pages} {14 } (\bibinfo {year}
  {2006})}\BibitemShut {NoStop}%
\bibitem [{\citenamefont {Dolgov}(2008)}]{Dolgov3}%
  \BibitemOpen
  \bibfield  {author} {\bibinfo {author} {\bibfnamefont {A.~D.}\ \bibnamefont
  {Dolgov}},\ }\href {https://doi.org/10.1134/S1063778808120181} {\bibfield
  {journal} {\bibinfo  {journal} {Physics of Atomic Nuclei}\ }\textbf {\bibinfo
  {volume} {71}},\ \bibinfo {pages} {2152} (\bibinfo {year}
  {2008})}\BibitemShut {NoStop}%
\bibitem [{\citenamefont {Faessler}\ \emph {et~al.}(2017)\citenamefont
  {Faessler}, \citenamefont {Hod\'ak}, \citenamefont {Kovalenko},\ and\
  \citenamefont {Simkovic}}]{Faessler}%
  \BibitemOpen
  \bibfield  {author} {\bibinfo {author} {\bibfnamefont {A.}~\bibnamefont
  {Faessler}}, \bibinfo {author} {\bibfnamefont {R.}~\bibnamefont {Hod\'ak}},
  \bibinfo {author} {\bibfnamefont {S.}~\bibnamefont {Kovalenko}}, \ and\
  \bibinfo {author} {\bibfnamefont {F.}~\bibnamefont {Simkovic}},\ }\href
  {\doibase 10.1142/S0218301317400080} {\bibfield  {journal} {\bibinfo
  {journal} {International Journal of Modern Physics E}\ }\textbf {\bibinfo
  {volume} {26}},\ \bibinfo {pages} {1740008} (\bibinfo {year}
  {2017})}\BibitemShut {NoStop}%
\bibitem [{\citenamefont {Yanagisawa}(2014)}]{Chiaki}%
  \BibitemOpen
  \bibfield  {author} {\bibinfo {author} {\bibfnamefont {C.}~\bibnamefont
  {Yanagisawa}},\ }\href
  {http://journal.frontiersin.org/article/10.3389/fphy.2014.00030} {\bibfield
  {journal} {\bibinfo  {journal} {Frontiers in Physics}\ }\textbf {\bibinfo
  {volume} {2}},\ \bibinfo {pages} {30} (\bibinfo {year} {2014})}\BibitemShut
  {NoStop}%
\bibitem [{\citenamefont {Betts}\ \emph {et~al.}()\citenamefont {Betts} \emph
  {et~al.}}]{Betts}%
  \BibitemOpen
  \bibfield  {author} {\bibinfo {author} {\bibfnamefont {S.}~\bibnamefont
  {Betts}} \emph {et~al.},\ }\href {https://arxiv.org/abs/1307.4738} {\
  }\Eprint {http://arxiv.org/abs/arXiv:1307.4738v2} {arXiv:1307.4738v2}
  \BibitemShut {NoStop}%
\bibitem [{\citenamefont {Bhattacharjee}(1997)}]{Bhattacharjee}%
  \BibitemOpen
  \bibfield  {author} {\bibinfo {author} {\bibfnamefont {P.}~\bibnamefont
  {Bhattacharjee}},\ }\href
  {http://adsabs.harvard.edu/full/1997JApA...18..263B} {\bibfield  {journal}
  {\bibinfo  {journal} {J. Astrophys. Astr.}\ }\textbf {\bibinfo {volume}
  {18}},\ \bibinfo {pages} {263} (\bibinfo {year} {1997})}\BibitemShut
  {NoStop}%
\bibitem [{\citenamefont {Dolgov}\ \emph {et~al.}(2005)\citenamefont {Dolgov},
  \citenamefont {Hansen},\ and\ \citenamefont {Smirnov}}]{Dolgov1}%
  \BibitemOpen
  \bibfield  {author} {\bibinfo {author} {\bibfnamefont {A.~D.}\ \bibnamefont
  {Dolgov}}, \bibinfo {author} {\bibfnamefont {S.~H.}\ \bibnamefont {Hansen}},
  \ and\ \bibinfo {author} {\bibfnamefont {A.~Y.}\ \bibnamefont {Smirnov}},\
  }\href {http://stacks.iop.org/1475-7516/2005/i=06/a=004} {\bibfield
  {journal} {\bibinfo  {journal} {Journal of Cosmology and Astroparticle
  Physics}\ }\textbf {\bibinfo {volume} {2005}},\ \bibinfo {pages} {004}
  (\bibinfo {year} {2005})}\BibitemShut {NoStop}%
\bibitem [{\citenamefont {Kuno}(2009)}]{Kuno}%
  \BibitemOpen
  \bibfield  {author} {\bibinfo {author} {\bibfnamefont {Y.}~\bibnamefont
  {Kuno}},\ }in\ \href@noop {} {\emph {\bibinfo {booktitle} {Neutrinos in
  Particle Physics, Astrophysics, and Cosmology}}},\ \bibinfo {editor} {edited
  by\ \bibinfo {editor} {\bibfnamefont {F.}~\bibnamefont {Soler}}, \bibinfo
  {editor} {\bibfnamefont {C.~D.}\ \bibnamefont {Froggatt}}, \ and\ \bibinfo
  {editor} {\bibfnamefont {F.}~\bibnamefont {Muheim}}}\ (\bibinfo  {publisher}
  {CRC Press, Taylor \& Francis Group},\ \bibinfo {address} {Boca Raton, FL},\
  \bibinfo {year} {2009})\BibitemShut {NoStop}%
\end{thebibliography}%

\end{document}